\documentclass[twocolumn]{article}

\usepackage{style}

\usepackage[utf8]{inputenc} 
\usepackage[T1]{fontenc}    
\usepackage[square,numbers,comma,sort]{natbib}
\usepackage{amsthm,amsfonts,amsmath,amssymb}       
\usepackage{nicefrac}       
\usepackage{microtype}      
\usepackage[dvipsnames]{color,xcolor}         
\usepackage{lineno}
\usepackage[labelfont=bf,font=bf]{caption}
\usepackage[labelfont=bf,font=bf]{subcaption}
\usepackage{multirow,tabularx,longtable,booktabs}
\usepackage{setspace}
\usepackage{mathrsfs}
\usepackage{pdfpages}
\usepackage{float}
\usepackage{algorithm}
\usepackage{algorithmic}
\usepackage{enumitem}
\usepackage{totcount}
\usepackage{url}            
\usepackage{booktabs}       
\usepackage{hyperref}
\usepackage{authblk}
\usepackage{siunitx}

\hypersetup{colorlinks=true,citecolor=black,filecolor=black,linkcolor=black,urlcolor=black,menucolor=black,
            pdftitle={Time Series Information Visualization -- A Review of Approaches and Tools},
            pdfsubject={Time Series Information Visualization},
            pdfauthor={Ortigossa, et al.},
            pdfkeywords={},
}

\title{Time Series Information Visualization -- A Review of Approaches and Tools}

\author[1,2]{Evandro S. Ortigossa}
\author[3]{Fábio F. Dias}
\author[4]{Diego C. Nascimento}
\author[1]{Luis Gustavo Nonato}
\affil[1]{Institute of Mathematics and Computer Science, University of S{\~a}o Paulo (ICMC-USP), S{\~a}o Carlos 13566-590, Brazil}
\affil[2]{Department of Computer Science and Applied Mathematics, Weizmann Institute of Science, Rehovot 7610001, Israel}
\affil[3]{Tandon School of Engineering, New York University, Brooklyn, NY 11201, USA}
\affil[4]{NEOMA Business School, 76825 Mont-Saint-Aignan, France}

\affil[ ]{\texttt{evortigosa@usp.br, ffd2011@nyu.edu, diego.nascimento@neoma-bs.fr, gnonato@icmc.usp.br}}

\newtotcounter{citnum} 
\def\oldbibitem{} \let\oldbibitem=\bibitem
\def\bibitem{\stepcounter{citnum}\oldbibitem}

\begin{document}

\maketitle

\begin{abstract}
    Time series data are prevalent across various domains and often encompass large datasets containing multiple time-dependent features in each sample. 
    Exploring time-varying data is critical for data science practitioners aiming to understand dynamic behaviors and discover periodic patterns and trends.
    However, the analysis of such data often requires sophisticated procedures and tools. 
    Information visualization is a communication channel that leverages human perceptual abilities to transform abstract data into visual representations. 
    Visualization techniques have been successfully applied in the context of time series to enhance interpretability by graphically representing the temporal evolution of data. 
    The challenge for information visualization developers lies in integrating a wide range of analytical tools into rich visualization systems that can summarize complex datasets while clearly describing the impacts of the temporal component. 
    Such systems enable data scientists to turn raw data into understandable and potentially useful knowledge. 
    This review examines techniques and approaches designed for handling time series data, guiding users through knowledge discovery processes based on visual analysis. 
    We also provide readers with theoretical insights and design guidelines for considering when developing comprehensive information visualization approaches for time series, with a particular focus on time series with multiple features. 
    As a result, we highlight the challenges and future research directions to address open questions in the visualization of time-dependent data.\vspace{1pt}

    \textbf{Keywords:} InfoVis, \MakeLowercase{Visual data analysis, User interfaces, Interactivity, graphical properties}.
\end{abstract}

\section{Introduction} 
\label{sec1}
    Time series data are frequently found in a wide variety of applications. Exploration, analysis, and understanding of their dynamic behaviors are valuable for data science but require complex processing. Furthermore, time series instances can have multiple correlated and uncorrelated attributes (dimensions), varying over time~\cite{Shneiderman1996}, which hampers data representation even in systems specifically designed for handling multidimensional data. 

    Almost half of human neural tissue is directly and indirectly associated with vision, enabling sophisticated abilities related to pattern recognition, particularly when visually stimulated~\cite{hoffman1998}. Therefore, visualization emerges as a natural strategy for representing complex data. The area of Information Visualization (InfoVis) leverages human cognitive abilities to transform abstract data patterns into comprehensible visual representations.
    It reduces cognitive effort to identify, interpret, and extract patterns from raw data by depicting them as graphical elements~\cite{Ward2010}, facilitating data filtering and smoothing (modeling).
    Monitoring multiple features over time is an important but challenging task, and InfoVis has developed innovative techniques incorporating multi-view user interfaces to facilitate information extraction~\cite{Heer2012}. Such tools also leverage human visual capabilities to generate knowledge about large datasets in an interactive way~\cite{Badam2016}.

    Traditional visualization methods, such as line and bar charts, have long represented temporal data in several applications. 
    However, line-based metaphors are not scalable for treating multiple series, often leading to visual cluttering and occlusion problems~\cite{Heer2009}. 
    Specialists have applied advanced InfoVis approaches to analyze data from different contexts (e.g., business planning, social networks, climate, pollution, finances, or criminal cases)~\cite{Kehrer2013,Silva2000,Aigner2011,Thakur2010}. These approaches help the identification of patterns and anomalies to support informed decision-making procedures across both unidimensional and multidimensional datasets~\cite{McLachlan2008}. 
    By graphically representing information through computer systems, these methods enable alternative visions for describing complex data structures~\cite{Ward:2015}. 
    However, a significant gap persists between the current capacity of devices to generate and store data and the ability of visual-analytical tools to effectively process, organize, and display the extracted information~\cite{Keim:2002,Alexandrina:2019}.

    The design of efficient visualization tools capable of representing multiple time series depends on the specific characteristics of the application domain~\cite{byron2008}. This review discusses multiple visualization approaches and techniques employed to extract information from time series data, emphasizing the conceptual, practical, and aesthetic considerations of developing comprehensive tools for managing time-oriented data graphically. Our review also introduces readers to new concepts and modern InfoVis applications, with a particular focus on multidimensional time-series analysis and recent techniques based on user-interaction strategies that dynamically connect data scientists with complex data.

    In summary, the main contributions of this research are:

    \begin{itemize}[itemsep=2pt,topsep=0pt]
        \item A comprehensive discussion of aesthetic principles and graphical properties that developers must consider when designing efficient InfoVis systems.

        \item A presentation of critical challenges in visualizing time series, especially the multidimensional ones, and tools that can be used to interact with time-varying data.

        \item An in-depth review of approaches and applications for time series InfoVis. All codes and data used to generate the charts presented here are available online.
    \end{itemize}

    \subsection{Organization}
    \label{sec1:organization}

    The remainder of the paper is organized as follows: 
    Section~\ref{sec2} presents the research methodology adopted for collecting literature; 
    Section~\ref{sec3} presents previous reviews covering different aspects of visualization for time series;
    Section~\ref{sec4} addresses the ground theory to understand the time series domain and its applications, which includes formal definitions (\ref{sec4:time_series}), dimensionality (\ref{sec4:sub:multid}), missing data (\ref{sec4:sub:missing_data}), modeling approaches (\ref{sec4:sub:modeling}), an introduction to the visualization pipeline (\ref{sec4:vis_ts}), taxonomies and guidelines (\ref{sec4:sub:tipos_vis}), graphical properties (\ref{sec4:sub:propriedades}), interactivity (\ref{sec4:sub:interacao}), and large-scale datasets (\ref{sec4:sub:big_data}); 
    Section~\ref{sec5} describes visualization methods and approaches designed for time series InfoVis; 
    Section~\ref{sec6} extends the previous section and presents composed interfaces applied to multidimensional time series in the context of text and topics (\ref{sec6:sub:text}), graphs (\ref{sec6:sub:graphs}), financial data (\ref{sec6:sub:financial}), environmental monitoring (\ref{sec6:sub:air_qual}), audio and images (\ref{sec6:sub:audio}), and event sequences (\ref{sec6:sub:event});
    Section~\ref{sec7} discusses software tools for time series InfoVis (\ref{sec7:sub:software}), open challenges (\ref{sec7:sub:desafio}), and suggests future research directions (\ref{sec7:sub:future}); finally, 
    Section~\ref{sec8} provides our final remarks.

\section{Method of the Systematic Review}
\label{sec2}
    We analyzed the literature to understand how visual techniques have been used in time series analyses over the past years. 
    The systematic review process follows \citet{Moher:2009}, establishing objective criteria to define the scope of the relevant literature and the appropriate report on the findings.    
    From the available literature, we identified and classified key contributions to the time series field, clarifying current trends and practices, and indicating research opportunities.

    \subsection{Literature Search Procedure}
    \label{sec2:procedure}
        The procedure combines four databases (IEEE Xplore Digital Library, Association for Computing Machinery (ACM) Digital Library, DBLP, and Elsevier's Scopus) to search time-oriented InfoVis. Such databases are well-regarded for their comprehensive collections of peer-reviewed articles in computer science.
        In association with these data, we used Google Scholar, Elsevier's ScienceDirect, and Web of Science as search engines.

        We queried about multidimensional time series and InfoVis in publications' titles, abstracts, and keywords, selecting studies from (but not restricted to) the 2005--2024 period. This period covers most of the recently published literature, including seminal studies. In this context, the search outcomes were evaluated using the following criteria to select a set of publications for review:

        \begin{itemize}[itemsep=2pt,parsep=0.1cm]
            \item Papers published as articles in peer-reviewed journals and conference proceedings from leading venues on visualization, and available online in English. Besides, we also used arXiv preprints, theses, and books;

            \item Studies explicitly addressing time series representation. Articles that only cite InfoVis in keywords, do not explain the methodology employed, or the time-oriented data context were not included.
        \end{itemize}

        Papers that did not satisfy at least one of the selection criteria were excluded. The queries returned 297 publications, and duplicates were removed after a second filtering based on abstract and introduction readings. 
        For each selected paper, we reviewed the main content and did not consider the appendices or supplementary material.
        A total of 232 studies were then read in full and, after a careful review of their contents, \total{citnum} references are reported in this review.

\section{Previous Visualization Reviews}
\label{sec3}
    Research on data analysis and mining has introduced various approaches and methods for information visualization (InfoVis). Several researchers have proposed comprehensive discussions on the visualization environment, defining categorizations and practical aspects of techniques. In this section, we present an overview of previous works, from more general surveys on visualization that also discussed time series to those focused on the different aspects of time series InfoVis.

    \citet{Oliveira2003} presented a foundational study on data mining and exploratory strategies supported by visualization methods, introducing the concept of visual data mining.
    \citet{Kehrer2013} conducted a broad survey of methods for visualizing data that hold more diverse aspects, including multidimensionality, multisource, and multitype elements. They introduced the term multifaceted scientific data and addressed several techniques for dealing with such data.

    \citet{liu2014survey} presented an introductory overview of multiple approaches for InfoVis, focusing on data representation using interactive methods, in which users are active participants in refining the exploratory process of visual data.
    \citet{cui2019visual} differentiated the multiple and closely related research areas of visualization, highlighting the synergy between visualization and interactive data exploration to facilitate knowledge discovery.
    Recently, \citet{shakeel2022comprehensive} elaborated a survey on data visualization, categorizing its research tasks. The paper also analyzed the advantages and limitations of diverse programming tools and platforms used to generate visualizations. 

    A common subject in the visualization literature concerns large data collections and how to represent them effectively. In this context, \citet{liu2016visualizing} focused their investigation on high-dimensional data, proposing a taxonomy of techniques for multidimensional visualization. \citet{veras2019discriminability} evaluated the perceptual effectiveness of visualization methods when faced with data scalability, i.e., the challenges associated with distinguishing variations in information when representing large datasets.

    When visualizing large datasets, the volume of information presented to users is a critical consideration since human working memory has limitations.
    \citet{cockburn2009review} thoroughly reviewed visual exploration techniques based on focus+context approaches, allowing users to work with focused and contextual views by selectively presenting content to reduce information load.

    Previous works have addressed foundational concepts on visualization as a broad area encompassing a wide range of data. However, the specialized literature on temporal data analysis, modeling, and visualization techniques is also extensive, providing designers with valuable content on mining
    ~\cite{esling2012time}, clustering
    ~\cite{aghabozorgi2015time}, and aggregation~\cite{nguyen2020exploring} concepts and approaches that can be applied to the development pipeline of visualization tools devoted to exploring time series. \citet{clustering2019} reviewed a wide range of supervised and unsupervised machine learning algorithms within visual analytics frameworks. They demonstrated how integrating interactive visualization methods can assist analysts in refining classification and clustering tasks.
    
    Representing and understanding the evolution of time-dependent attributes imposes particular challenges on visualization developers and data scientists.
    \citet{Aigner2007,aigner2008} conducted remarkable surveys on the visualization of time-oriented data and related interactive techniques to capture the dynamic behavior of time-dependent features. They provided a categorization based on the nature of the temporal dimension with important considerations for designing visual systems, such as data granularity (e.g., discrete-time and intervals) or periodicity (e.g., linear-time, cyclic, and branching).
    Similar discussions can be found in classical surveys~\cite{Silva2000} and books~\cite{shumway2006,Aigner2011} devoted to time series visualization.

    Recently, \citet{Fang2020} presented a structured overview of time series visualization, characterizing data into linear and cyclic, point and interval, and sequential and branching types. The authors also reviewed multiple visualization methods, both static and animated, as well as interactive approaches.

    \citet{ANDRIENKO2003} provided an early but comprehensive taxonomy of spatiotemporal visualization techniques, categorizing existing methods by data type, visual metaphor, and analytical tasks. However, the authors underestimated real‑time interaction challenges.
    Recent works reviewed the complexities of visualizing spatiotemporal data~\cite{Bach2017,wang2017spatial,andrienko2017visual}.
    \citet{bai2020time} presented a systematic review of time-varying volumetric data frequently used in scientific visualization to understand the temporal evolution of complex phenomena such as fluid dynamics, cosmology, and climatology.

    Despite the rich literature, effectively visualizing multidimensional time series remains a significant challenge. This work focuses specifically on solutions for time series information visualization (InfoVis). We provide an updated discussion on techniques and tools for interactive visual analysis that were previously unavailable or only recently emerged. Therefore, this study provides developers with a comprehensive overview, supporting the design of tools that enable data scientists to better understand time-series data and visually transform information into knowledge. In addition, we have made all the code and data used to render our visualizations publicly available in an online repository\footnote{Available on GitHub at {\url{https://github.com/evortigosa/InfoVis}}}.

\section{Ground Theory}
\label{sec4}
    This section provides an overview of the time series definitions that should be considered in the development of visualization tools. The aim is to offer a comprehensive review of time series InfoVis methods and approaches and to address important characteristics and concepts for multidimensional time-oriented data (or model) visualization.

    \subsection{Time Series Notation and Nomenclature}
    \label{sec4:time_series}
        \begin{figure*}[!t]
            \centering
            \includegraphics[width=0.8\linewidth]{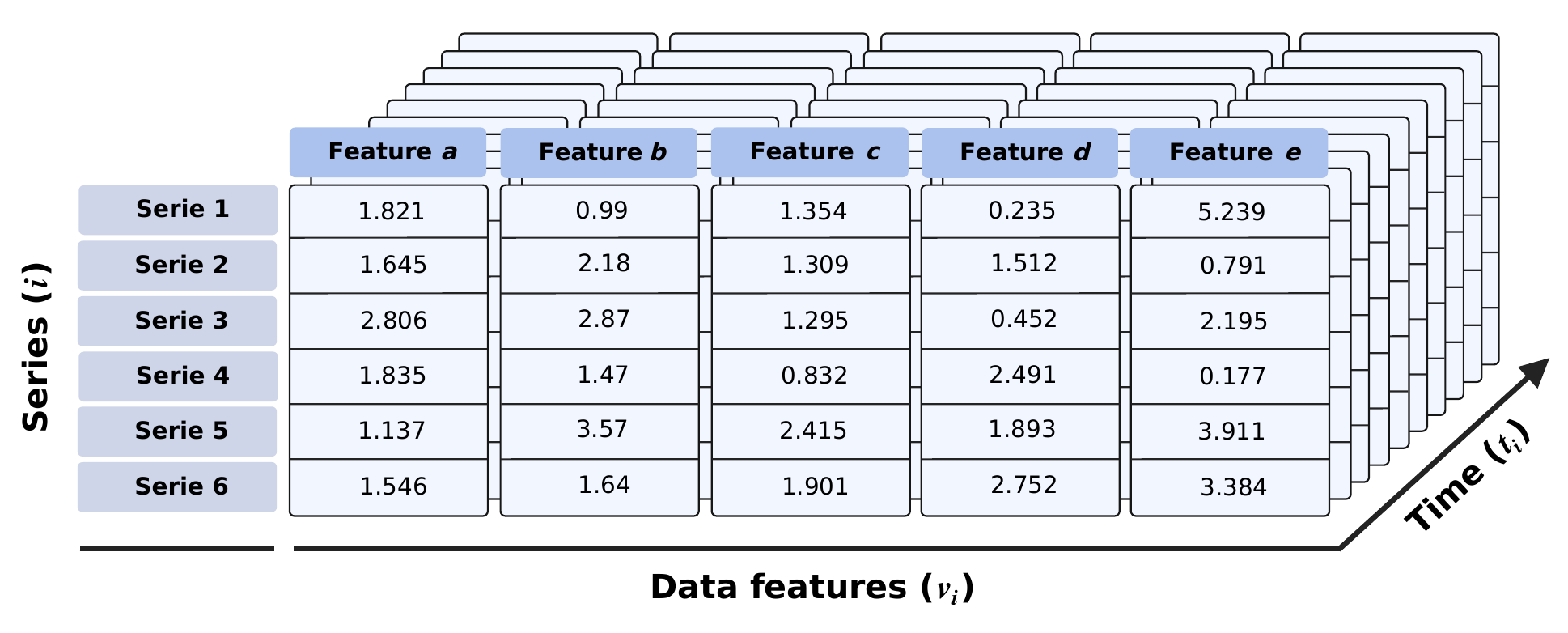}
            \caption{Multidimensional time series can be conceptually represented as a data cube structure with samples varying over time slices and having one or more features. This cube abstraction can be extended to higher dimensions by adding spatial features (in terms of increasing algebraic space) and different temporal frequency bands.}
            \label{fig:data_cube}                 
        \end{figure*}
    
        Time-series data are generated and found in a wide range of real-world application contexts, such as finance, health, meteorology, astronomy, news, remote sensing and monitoring, scientific experimental simulations, image collections, and project planning and management. From a practical point of view, each data context requires specific tools tailored to support the information discovery process~\cite{Aigner2007}. Formally, a time series dataset can be expressed as
        
        \begin{equation*}
        \label{eq:time_serie}
            D = \big\{(\mathbf{t}_1, \mathbf{v}_1), (\mathbf{t}_2, \mathbf{v}_2),\dotsc, (\mathbf{t}_n, \mathbf{v}_n)\big\}
        \end{equation*}
        
       \noindent being $t_i$ the instant of time and $\mathbf{v}_i = f \big \langle \mathbf{t}_i \big \rangle$ the time-dependent vector (also called an instance or sample) that can bear properties of different types (e.g., an integer, a real, a categorical, an image, or a binary sequence)~\cite{Weber01}. 
        The term ``\textit{time series}'' is frequently used to refer to time-oriented data comprising numerical values, while time-oriented data consisting of categorical values is sometimes termed \textit{event sequences}~\cite{bernard2024ivesa}.
        To ensure terminological clarity and prevent misunderstandings, we refer to time-dependent data of any type as simple ``\textit{time series}''.

        A time series dataset is considered unidimensional when it contains only a single qualitative or quantitative feature, also known as an attribute or variable.
        Although the temporal evolution of ordered pairs $(\mathbf{t}_i, \mathbf{v}_i)$ can be represented in a Cartesian coordinate space, each instance of $\mathbf{v}_i$ can also bear a subset carrying an arbitrary number of features, expressed as $\mathbf{v}_i = \{v_{i1}, v_{i2},\dotsc, v_{ik}\}$. 
        These $k$ elements define the feature space of the time-dependent instances, thus leading to a multidimensional time series dataset, since each instance can be interpreted as a point in the $m$-dimensional space where $m=k+1$ is the total number of elements in $(\mathbf{t}_i, \mathbf{v}_i)$, that is, the $k$ time-dependent features and the time component. 
        Consequently, families of time-varying curves for each feature of time-dependent instances can be produced in the $m$-dimensional space~\cite{Konyha2012}.
        
        Throughout this text, we reserve the term ``parameter'' to identify any adjustable element used to configure a data modeling or graphical object.

	\subsection{Multidimensionality}
	\label{sec4:sub:multid}
        An important definition must be clarified before addressing the fundamentals of time series visualization. The literature reports terms such as ``multidimensional,'' ``high-dimensional,'' ``multivariate,'' or even ``multivalued'' to describe datasets where each instance (time-dependent vectors) comprises a subset of elements. However, these terms are not used consistently~\cite{Fua1999}. By convention, we adopt the term ``multidimensional'', equivalent to multivariate, to denote datasets composed of multiple features, which may or may not be dependent on each other at each time interval. 

        Due to the growing capacity of current devices and sensors for data collection, storage, and transmission, time series datasets tend to be large, not only in the number of instances or time intervals but also in the number of attributes (dimensions) associated with each instance, making time series analysis a complex and cumbersome task~\cite{aigner2008, Javed2010}. Their difficult interpretation is closely related to data dimensionality, i.e., an adequate representation of the many elements linked to each time interval. 
        
        Figure~\ref{fig:data_cube} illustrates the structure of a multidimensional time series dataset conceptually represented as a data cube.
        Consider, for example, a simulation procedure that involves parametric optimization, where a set of features is given as input for multiple simulation methods at each time step. More specifically, different vectors of input features are received and processed by different algorithms, and then multiple series of output vectors are produced. 
        By recording both inputs and outputs at each time step, we obtain a snapshot of the simulation. Iterating this procedure over subsequent time instants for modeling features and possible adjustments to the simulation parameters generates a sequence of such snapshots that contains all stimuli and outcomes processed over time.
        When organizing these data as rows (series of outputs), columns (features), and depth (temporal evolution), we have a multidimensional time series that resembles a cube-like temporal structure, as shown in Figure~\ref{fig:data_cube}, which can be extended by adding spatial features or stacking temporal layers into the data structure.

        As the dimensionality of the ``time cube'' increases, representing and interpreting its content becomes increasingly challenging~\cite{Nonato2018}, as it may present time-varying groups of relational structures. This increased complexity complicates the analytical tasks related to extracting useful information from the data. In particular, analyzing trends and interdependencies among time-varying features is at the core of multidimensional time series mining~\cite{Gruendl2016}.

	\subsection{Fragmented and Intermittent Time Series}
	\label{sec4:sub:missing_data}

        Real-world time series often contain noise and interruptions due to sensor failure, communication loss, or irregular sampling intervals. These interruptions produce missing observations with distinct statistical characteristics that can obscure the underlying patterns~\cite{donders2006gentle}, which require explicit treatment to avoid biased inference while preserving the integrity of the temporal dynamic.

        Approaches for handling fragmented time series fall into two broad categories. 
        First, classical statistical methods, such as linear or spline interpolation, ARIMA and Kalman filters, and state-space methods, estimate missing data points when the missing spans are short and the underlying dynamics are locally smooth~\cite{lepot2017interpolation,pratama2016review}. 
        For example, \citet{Alexandrina:2019} employed a spline interpolation approach to represent fragmented points in an interactive system that dynamically adjusts the visualizations.
        Second, modern data-driven machine learning techniques directly learn temporal and cross-series dependencies and can impute long or structured gaps~\cite{wang2024deep}. For example, \citet{cao2018brits} proposed a bidirectional recurrent model that propagates information in forward and backward directions, demonstrating effectiveness in reconstructing missing extended segments.
        
        From the visualization perspective, designers must balance continuity and transparency, i.e., communicating data imputation. Best practices can include: exposing the missingness pattern and provenance (so analysts know which values were measured versus inferred)~\cite{song2018s}; rendering imputed values together with explicit uncertainty encodings, such as error bars or translucent bands~\cite{hullman2019authors}; and providing interactive controls that allow users to switch between raw and imputed views, to view per-segment confidence intervals, and to inspect the imputation method and parameters on demand.

	\subsection{Overlook Through Time Series Modeling}
	\label{sec4:sub:modeling}
        Before presenting the visual tools for time series InfoVis, it is essential to mention that graphical representations can extend beyond raw empirical dependent (observational) dynamics. For example, visualizations can also be applied to components derived from model outputs or filter states, such as dynamics associated with time-varying parameters or phase evolutions. In this context, time-series modeling falls into two broad categories: time-domain and frequency-domain approaches, with their goodness-of-fit results supported by visualization.

        Time-domain methods decompose a time series/signals into interpretable estimated/learned components (e.g., trend, seasonality, cycles, and stochastic volatility), combined additively or multiplicatively in equations indexed over time. For instance, let us suppose a univariate ($i=1$) collection of observations indexed over time ($v_i$), such as
        
        \begin{equation}
            v_i^{(t)} = f(v_i^{(1)},v_i^{(2)},\dots,v_i^{(t-1)}) 
        \end{equation}

        \noindent where the $f(\cdot)$ function can be a combination of elements like a simple explanation of the current event as a contribution of the previous one ($\mathbf{\phi}$), plus a random noise ($\epsilon$),
        
        \begin{equation}
            v_i^{(t)} = f(v_i^{(1)},v_i^{(2)},\dots,v_i^{(t-1)}|\mathbf{\phi}_t) = \sum_{j=1}^{t-1} \phi_j v_i^{(t-j)} + \epsilon^{(t)}
        \end{equation}
        
        \noindent or more complex structures (components) like integration of trend ($\beta_t$) and season ($s_t$) components,
        
        \begin{equation}
            v_i^{(t)} = f(v_i^{(1)},v_i^{(2)},\dots,v_i^{(t-1)}|\beta_t,s_t)
        \end{equation}
        
        \noindent where those elements are a time-dependent window combined additively or multiplicatively. For the sake of illustration, the season component may carry information lagged $r$-steps (and may also incorporate other components in its structure, through $h$-functions, such as trend or change points), measured by $\alpha^{(t)}$,
        
        \begin{equation}
            s_t =  \sum_{w=0}^r [\alpha_w^{(t)} h_w(v_i^{(t-w)}|\beta_t)] \text{   or   } \prod_{w=0}^r [\alpha_w^{(t)} h_w(v_i^{(t-w)}|\beta_t)].
        \end{equation}

        These breakdown components and how these elements are constructed may explicitly describe a combination of elements, such as level, trend, seasonal, and residual (through the addition or multiplication of parameters). Classical statistical frameworks~\cite{hyndman2018forecasting} include (S)ARIMA, GARCH, Holt-Winters, SEATS, X-11 and variations, STL, ETS, (dynamic) regression (e.g., Vector Autoregressive and dynamic factor models), TBATS~\cite{de2011forecasting}, hidden Markov models, harmonic regression, and multivariate singular spectrum analysis (mSSA). 
        Modern machine learning literature presents several tree-based ensembles, such as XGBoost~\cite{XGBoost}, and popular artificial neural network architectures (ANNs, based on CNNs, LSTMs, GNNs, or LLMs) that are designed to capture complex nonlinear dependencies, such as Prophet~\cite{prophet}, DeepAR~\cite{salinas2020deepar}, N-BEATS~\cite{oreshkin2019n}, EfficientNet-based~\cite{abbaspourazad2023large}, Temporal Fusion Transformer~\cite{lim2021temporal}, and Vision Transformer-based~\cite{narayanswamy2024scaling}.
        
        From a frequency domain perspective, the same concept of series decomposition applies to sinusoidal functions. It is seen as an infinite summation replaced by an integral, focusing on a function  mapped from a set of integers, i.e., considering a univariate time series
        
        \begin{equation}
            v^{(t)} = \int_{\omega=0}^{\pi} [cos(\omega t)dA(\omega ) + sin(\omega t)dB(\omega)],
        \end{equation}

        \noindent where the combination of cosine and sine functions is alongside functions $A(\cdot)$ and $B(\cdot)$ in terms of differentials indexed on the continuous parameter $\omega$, easily extended from $[0, \pi]$ to $[-\pi, \pi]$ domain, allowing the existence of the spectral distribution/density function. There are standard methods of frequency transformation (FT), such as the discrete Fourier transform (DFT) or discrete cosine transform (DCT), which can be used to detect periodicities and frequency content.

        Real-world time series often arise as wave-like patterns in natural and engineered experiments, embedding concepts from physics such as energy transfer, phase shifts, and oscillatory dynamics. In this context, wavelet transforms decompose non-stationary series into time-frequency components, localizing energy transfer and phase changes that Fourier transforms can overlook, enabling time series visualization as oscillatory waves. \citet{cazelles2007time} applied the continuous wavelet transform (CWT) to generate decompositions that identify transient, non-stationary oscillatory behavior in epidemiological time series. \citet{priyadarshini2025energy} used wavelet packet decomposition (WPD) to detect and visualize voltage disturbances in a computationally efficient power quality monitoring approach. 

        In practice, time-domain components can be visualized through coordinated views, supporting the dynamic exploration of each element. Frequency domain visualizations often focus on spectral plots of dominant periods, where the number of observations between seasonal repetitions (the ``frequency'') guides interpretation.
        Recent studies have integrated spectral coefficients and ANN models to uncover deeper frequency-based patterns~\cite{gong2021ast,yi2023survey}.

        \subsection{Time Series and Information Visualization}
        \label{sec4:vis_ts}
        \begin{figure*}[!t]
            \centering
            \includegraphics[width=0.75\linewidth]{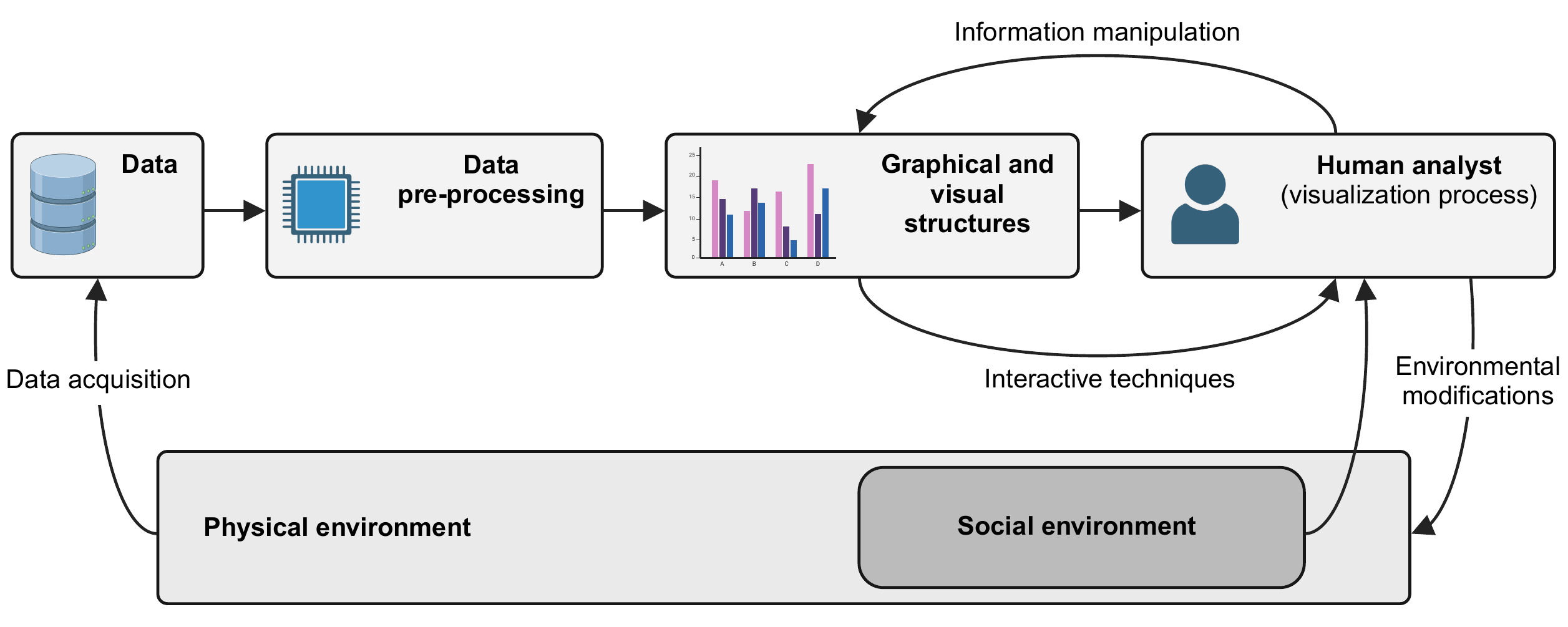}
            \caption{Schematic diagram of the interactive information visualization process. Visualization is closely related to data knowledge discovery, as it supports data interpretation through graphical tools.}
            \label{fig:vis_stages}            
        \end{figure*}
        
        The vast volume of time series data introduces analytical challenges that require sophisticated approaches to manage and interpret large volumes of content~\cite{bernard2024ivesa}. 
        Data mining and InfoVis share the goal of improving data interpretation; InfoVis specifically supports this process through interactive visual representations and tools that reveal useful patterns~\cite{Oliveira2003}. Visual mining is the union of techniques from both research areas by embedding visual strategies into the knowledge discovery process~\cite{Wong1999}.

        Figure~\ref{fig:vis_stages} illustrates the basic steps of the visualization pipeline. 
        The process begins with data acquisition, where raw data are collected, recorded, and organized in structured formats.
        Data acquisition is often the most time-consuming step, as it typically involves collecting data directly from the real-world environment. The physical environment serves as the data source, and the social context determines how conventions from the environment are interpreted~\cite{Ware2004}. 
        At first, raw data do not convey insights into the source environment because of their complexity of direct interpretation. At this stage, extensive pre-processing and transformation are applied to extract relevant attributes (by manual selection or automated methods).

        Next, graphical algorithms are developed to model the pre-processed content. 
        These algorithms leverage visual properties and graphical symbols to map sequences and generate adjustable representations that dynamically capture temporal changes~\cite{Card1999}. This step transforms numerical and categorical attributes into visual formats that enable monitoring of temporal evolution.

        The final stage involves user interaction. Using an input device, such as a computer mouse, analysts select intervals of interest to explore through comparisons among different data views, filtering out noise, and converting the visualized information into actionable knowledge. Although information tends to be cumulative, knowledge is selective. Only the information displayed on a computer screen may not be sufficient for making a decision, and an excess of information can clutter the analysis. Therefore, when performing an analysis, the analysts must know what to do with the discovered information, since it is the foundation of knowledge.

        The connection of analysts with an interactive information space drives the cognitive knowledge discovery process, helping visual perception of previously unknown patterns essential for informed decision-making. Using the knowledge generated through selective information, users can understand and, if necessary, modify the environment (data source).

        For example, the classical map created by Charles Joseph Minard in 1869 is one of the earliest effective multidimensional time-series representations. Minard used a data map to describe the flow of French army soldiers during Napoleon's disastrous 1812 war against Russia. According to \citet{tufte2007}, Minard's chart is probably one of the best statistical representations ever made, since it simultaneously integrates spatial, temporal, and quantitative variables (e.g., troop numbers and temperature), effectively summarizing the conflict's evolution. Minard's chart uses a graphical approach currently known as the Sankey diagram, which is discussed further in Section~\ref{sec5:sub:sankey}.

    \subsection{Time Series InfoVis Taxonomies and Guidelines}
    \label{sec4:sub:tipos_vis}
        \begin{figure*}[!t]
            \centering
            \begin{subfigure}[b]{0.71\linewidth}
                \centering
                \includegraphics[width=1\textwidth]{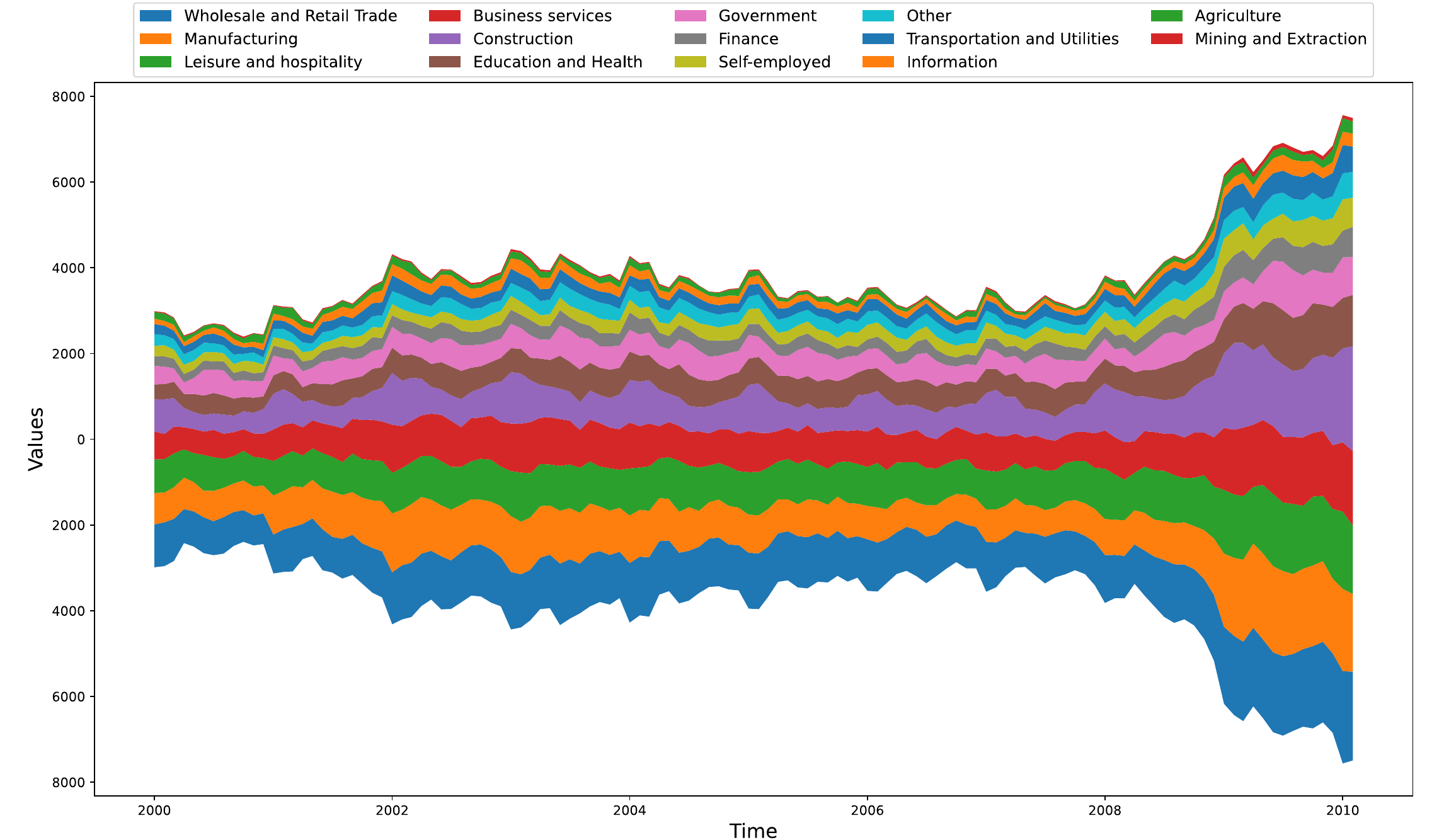}
                \caption{{Stream graph} shares the chart space by stacking the time features one on top of the other.}
                \label{fig:sub:space_share}
            \end{subfigure}\\ \vspace{0.5cm}
            \begin{subfigure}[b]{0.71\linewidth}
                \centering
                \includegraphics[width=1\textwidth]{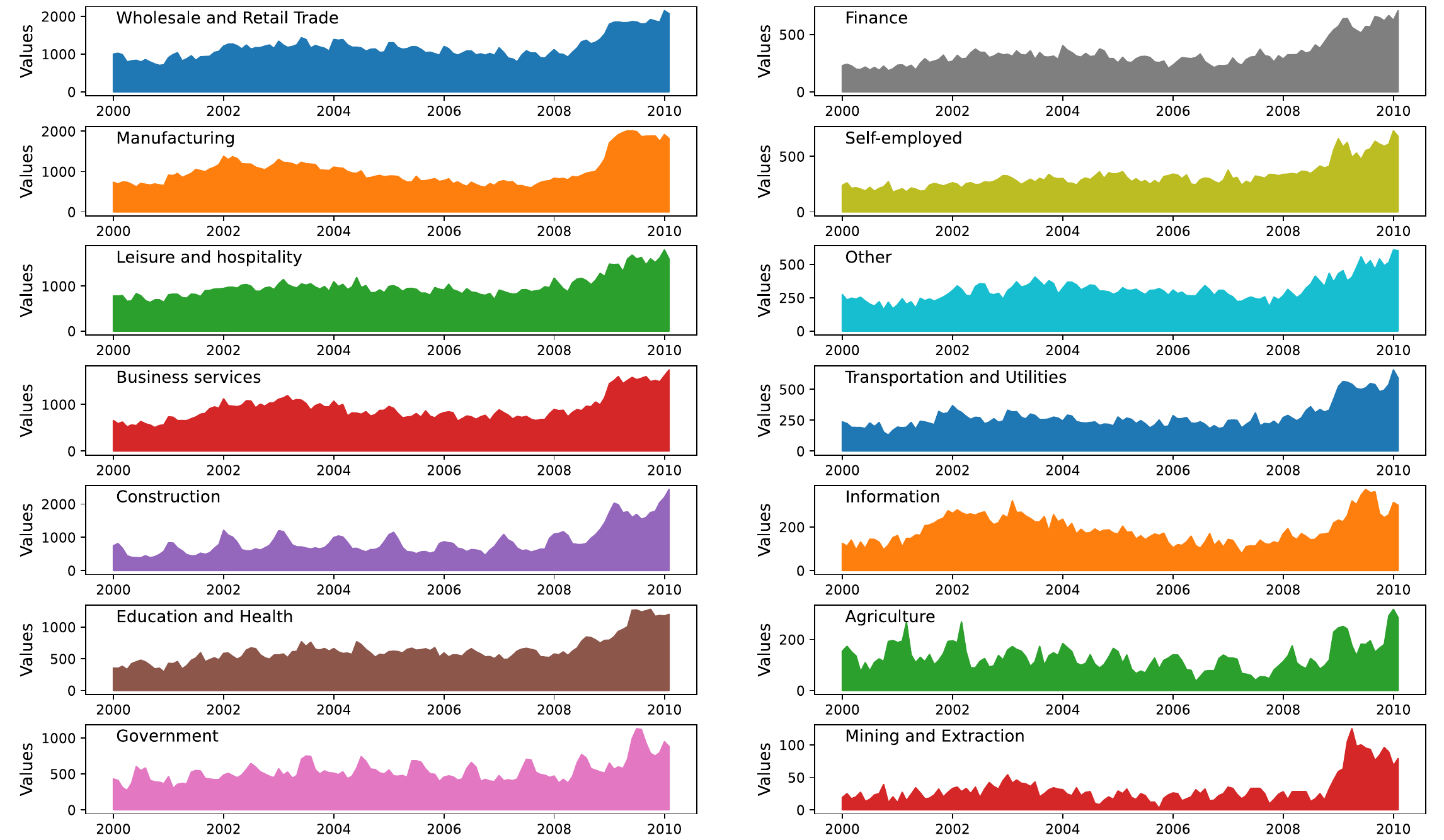}
                \caption{Small multiple views split the time-dependent features in their charts.}
                \label{fig:sub:space_split}            		
            \end{subfigure} \vspace{0.1cm}
            \caption{Two graphical representations of the unemployment rate in the United States, across the economic sectors, from 2000 to 2010.}
            \label{fig:space_share_split}          	
        \end{figure*}
        
        Since the representation of multidimensional time series is a persistent challenge, \citet{aigner2008} proposed a framework that classifies the different approaches of visual temporal-data analysis, while \citet{Javed2010Perception} categorized visualization techniques for multiple series into two main categories, namely shared-space and split-space. 
        
        In the \textbf{shared-space} approach, multiple time series are plotted within the same graphical area~\cite{Perin2013}. A visualization in which data features show overlapping curves on the same axes can facilitate direct comparisons among values, since the resultant chart is a visual summarization of multiple values in a single and aggregated view~\cite{Heer2009}. However, the amount of information to be simultaneously represented without clutter is limited. Displaying more than four curves turns the readability significantly challenging, causing cognitive overload, especially when the number of data dimensions demands the use of color codes with similar contrasts~\cite{Javed2010Perception}.

        In \textbf{split-space} techniques, each series holds its own space (or chart) -- usually small multiples (views) -- and the dimensions of a multidimensional time series with different data types should be divided accordingly. Depending on the context, the charts share similar structures (e.g., size and type of metaphor), and side-by-side positioning can enhance the estimation of temporal information, minimizing the overloads generated by a single and complex visualization, which may be cognitively tiring~\cite{McLachlan2008}. On the other hand, the accuracy of information estimation may decrease, as displaying multiple and simultaneous charts requires each of them to be constructed in smaller portions of space, which can lead to occlusions and reduced details~\cite{Milos2011}.

        Figure~\ref{fig:space_share_split} shows examples of the shared- and split-space visualization systems. The charts provide information on unemployment rates in the United States, with attributes related to economic sectors restricted to the years 2000--2010\footnote{\url{https://www.bls.gov/}, visited on June 2023}. 
        Figure~\ref{fig:sub:space_share} illustrates a {Stream graph} (see Section~\ref{sec5:sub:stacked}), a shared-space technique in which time-dependent attributes are rendered as stacked layers, one on top of the other, arranged along a central time axis~\cite{byron2008}. Furthermore, Figure~\ref{fig:sub:space_split} presents the same time-dependent features holding their own space in small area charts, using the same graphical metaphor for all charts. Notice that the aspect ratio of the charts is homogeneous~\cite{Hao2005}.
        
        As examples of these categories, \citet{Wattenberg2005} identified techniques based on the \textbf{space-filling} concept. Typically, the information area that fills a chart occupies only a small portion of the total area between the axes (see Figure~\ref{fig:space_share_split}). In contrast, space-filling strategies make use of all (or virtually all) available space on a screen and are considered efficient regarding space use~\cite{jugel2014m4}. Although several space-filling techniques are devoted to hierarchical data structures~\cite{Heer2010,Stasko2000}, the growing demand for multidimensional time series visualizations in limited spaces has motivated the development of systems that apply the space-filling concept~\cite{Perin2013,Hao2005}.

        \citet{Card1999} classified visualization systems that interconnect multiple visualization approaches to investigate complex, conceptual, and/or heterogeneous entities as \textbf{hybrid approaches}. These systems employ at least two distinct approaches to improve the expressiveness capacity of visualization information transmission~\cite{Oliveira2003}. In this context, ``different views'' refer to distinct views that emphasize different aspects of a dataset and promote their comprehension by the user~\cite{Baldonado2000}.

        Modern multi-view systems exploit data diversity by integrating different views and graphical elements to build a direct visual connection between the features and the time axis. Such integration supports the exploratory process based on interactive browsing, a critical component for building a successful visual analysis~\cite{Aigner2007}. By partitioning the data into multiple interconnected views, exploratory strategies based on the ``divide and conquer'' approach can be applied, thus generating manageable segments of information that reduce the memory effort required for the user to assimilate the information on screen~\cite{Baldonado2000}.

        The spatial placement of interface elements must be consistent toward minimizing the user's need to continually switch contexts~\cite{Baldonado2000}, or browse views that are distant from each other. The design must ensure that the user's attention remains focused on the right places. As an example, \citet{Gruendl2016} proposed a hybrid interface with four information groups: a parameter setup panel, a parallel coordinates panel, a time series repository, and a collection of charts displaying interactively selected variables. This structure enables users to explore multiple views, segmenting time-dependent features according to different perspectives and data granularities in a compact design.

        User interaction and data presentation are critical components for an effective data mining process~\cite{VanWijk1999}. Hybrid systems are highly dependent on interaction techniques, requiring coordination mechanisms that synchronize data exploration across multiple graphical elements. In other words, actions performed in one view must be fluidly propagated to all other views. Coordinating multiple views requires sophisticated management strategies, which makes the design of hybrid interactive interfaces particularly challenging~\cite{Baldonado2000}. Therefore, designers must carefully address attention to coding errors while developing such environments, since errors can be propagated across interconnected components and be difficult to fix due to the intricacy of the graphical elements.

        Presenting multiple levels of detail simultaneously provides users with a high-density context of information, thus increasing the analytical power over the data by enabling exploration from different perspectives~\cite{Sadana2016}. Multi-view systems improve exploratory performance by reducing the time required to access complex information, since visual comparisons are cognitively easier for the user than those relying only on working memory. This design supports the discovery of unknown relationships in the data through a linked and coordinated environment where complex information is divided into multiple and simplified views~\cite{North1997}. In this context, multiple views should be arranged based on perceptual design principles and interconnected via interaction tools, so that hybrid visualization interfaces with coordinated views are emerging as a modern solution for exploring complex multidimensional time series.

        Visual metaphors related to time series are numerous and, therefore, other authors have categorized representation techniques based on alternative properties related to the geometrical aspects of the graphical elements, iconography, and information access, among other visual structuring properties (layout). Comprehensive taxonomies for visualization techniques can be found in \citet{Oliveira2003} and \citet{Grinstein2001}. Additionally, \citet{Heer2009} analyzed space-efficient methods for multidimensional time series, and \citet{Roberts2007} and \citet{Baldonado2000} provided overviews of design fundamentals and requirements for multi-view systems. 
        
    \subsection{Graphical Perception and Visual Properties}
    \label{sec4:sub:propriedades}
        Visualization aims to effectively communicate information to human users~\cite{borland2007rainbow}. Therefore, an InfoVis tool must transform complex data into comprehensible information by leveraging human graphical abilities through the exploration of visual properties to enhance the clarity and accuracy of information transmission.
        Several studies have addressed the limitations of electronic device displays for providing information in a two-dimensional domain. Researchers have created tools for visualizing datasets with multiple attributes that cannot be mapped on a Cartesian plane or even on three-dimensional spatial coordinate systems~\cite{Chalmers1996}.
        This section addresses aesthetic principles to consider when designing InfoVis tools that enhance users' comprehension and avoid distractions during analysis.
        
        \citet{veras2019discriminability} defined discriminability as the ability to differentiate the graphical elements that human vision can perceive. Increasing the number of data curves represented in the same space might drastically decrease the capacity of individualized information perception; thus, it becomes necessary to inspect the relationships among the data series individually. More specifically, individual contributions must be recognized, that is, series that positively or negatively influence the overall data trend~\cite{Wang2016}. The comprehensive exploration of multiple curves in a line graph simultaneously is challenging, mainly due to issues such as overlays and visual disorder~\cite{Perin2013}. Although alternative metaphors (e.g., bars) can summarize large volumes of data, they typically require more space per single element. In addition to assessing correlation dynamics among families of data curves, analysts must understand inter-series relationships by individualizing them at different scales and across distinct time intervals. 

        Following, visual scalability is defined as the ability of visualization systems to effectively represent large datasets, supporting increasing volumes of information in terms of the number of instances and dimensions~\cite{Eick2002}. This aspect should always be considered when developing visual approaches, as human working memory can handle and interpret a limited number of cognitive entities at once~\cite{miller1956magical,cowan2010magical,Javed2010Perception}. Occlusion refers to the obstruction of graphical symbols caused by overlapping elements, and it is a recurring scalability issue when handling large sets. In this sense, effectively representing thousands of records within the limited visual space of computer screens remains one of the major challenges in the InfoVis domain~\cite{Shneiderman2008}.
        
\begin{figure}[!ht]
            \centering
            \begin{subfigure}[b]{1\linewidth}
                \centering
                \includegraphics[width=1\textwidth]{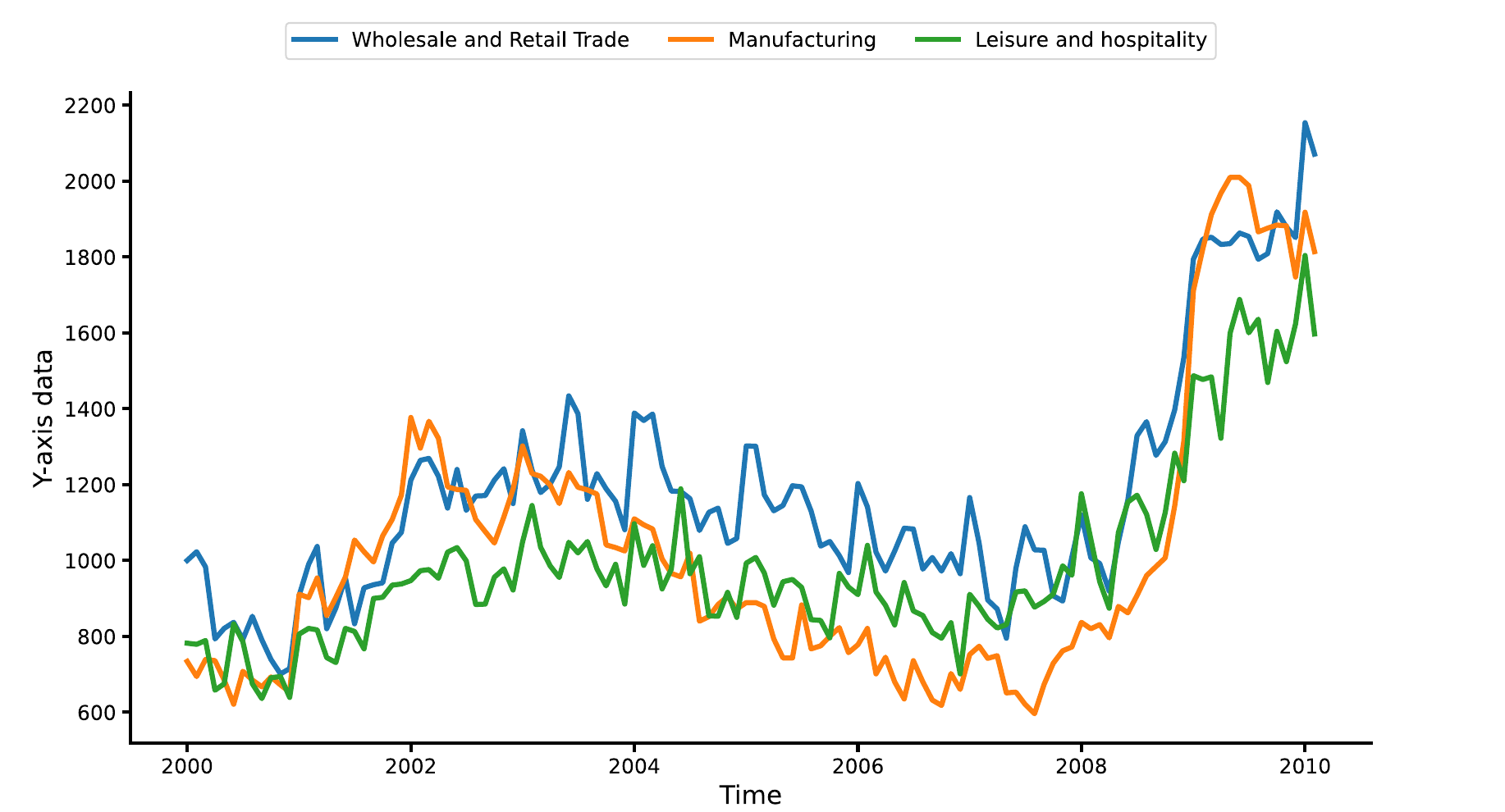}
                \caption{Three features.}
                \label{fig:sub:3series}
            \end{subfigure}\\ \vspace{0.4cm}
            \begin{subfigure}[b]{1\linewidth}
                \centering
                \includegraphics[width=1\textwidth]{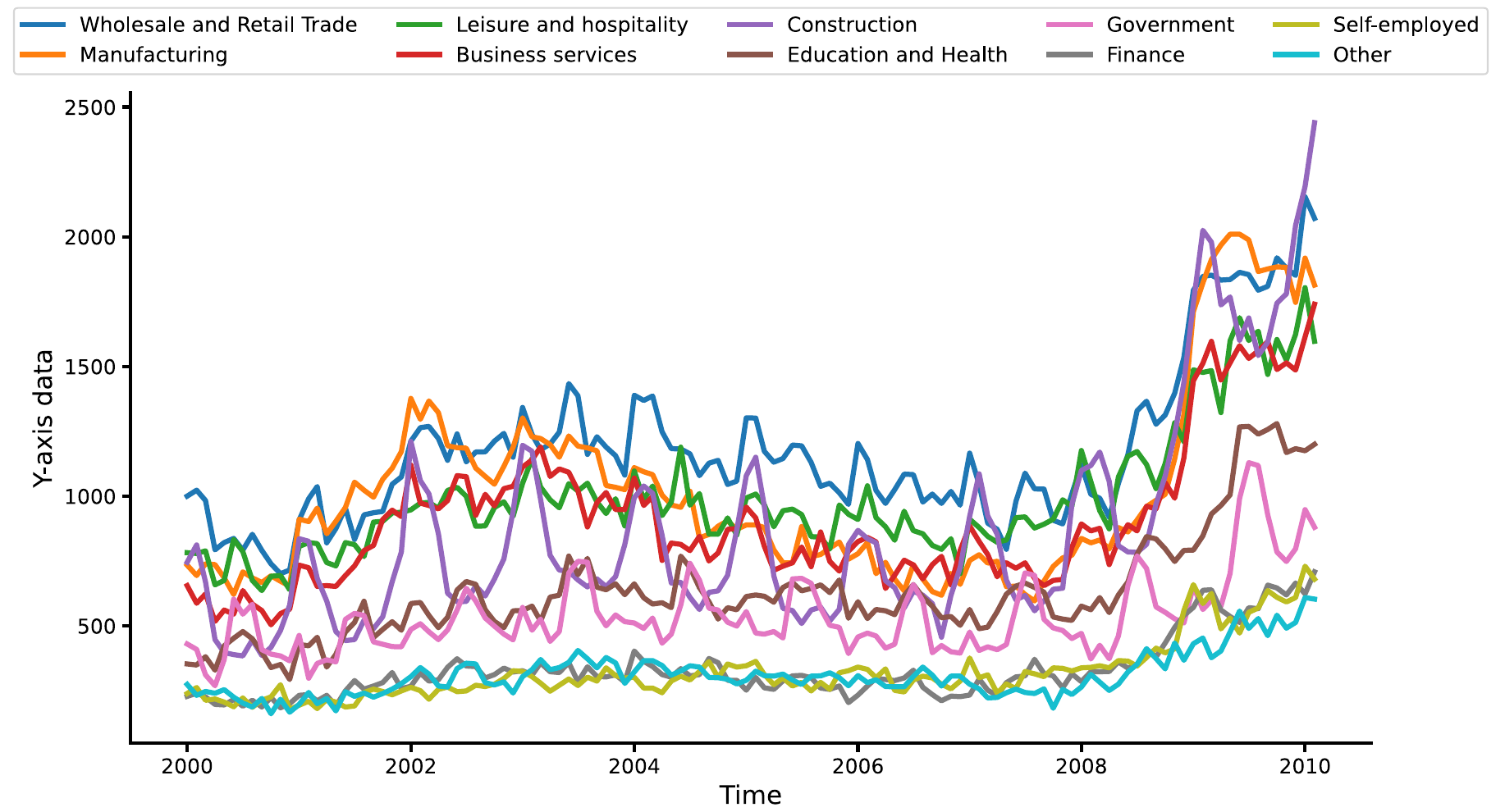}
                \caption{Ten features.}
                \label{fig:sub:10series}            		
            \end{subfigure}
            \caption{Two visualizations with different numbers of features related to unemployment rates in the United States per industrial sector from 2000 to 2010.}
            \label{fig:multi_series}
        \end{figure}

        Figure~\ref{fig:multi_series} illustrates the ``curse of dimensionality''~\cite{Nonato2018} in line charts: As the number of time-varying features increases from three (Figure~\ref{fig:sub:3series}) to ten (Figure~\ref{fig:sub:10series}), individual legibility decreases, resulting in visual disorder and clutter. The charts in Figure~\ref{fig:multi_series} use the same unemployment rate data presented in Figure~\ref{fig:space_share_split}. In a related study, \citet{gogolou2018comparing} evaluated how three variants of line-based charts (line chart itself, horizon graphs~\cite{Heer2009,Javed2010Perception,Perin2013}, and color fields~\cite{saito2005two,Albers2014}) influence the similarity perception abilities of humans in dynamic time series analyses tasks. They compared such charts with automatic similarity measures, highlighting good performances of horizon graphs.

        \begin{figure*}[!ht]
            \centering
            \includegraphics[width=0.8\linewidth]{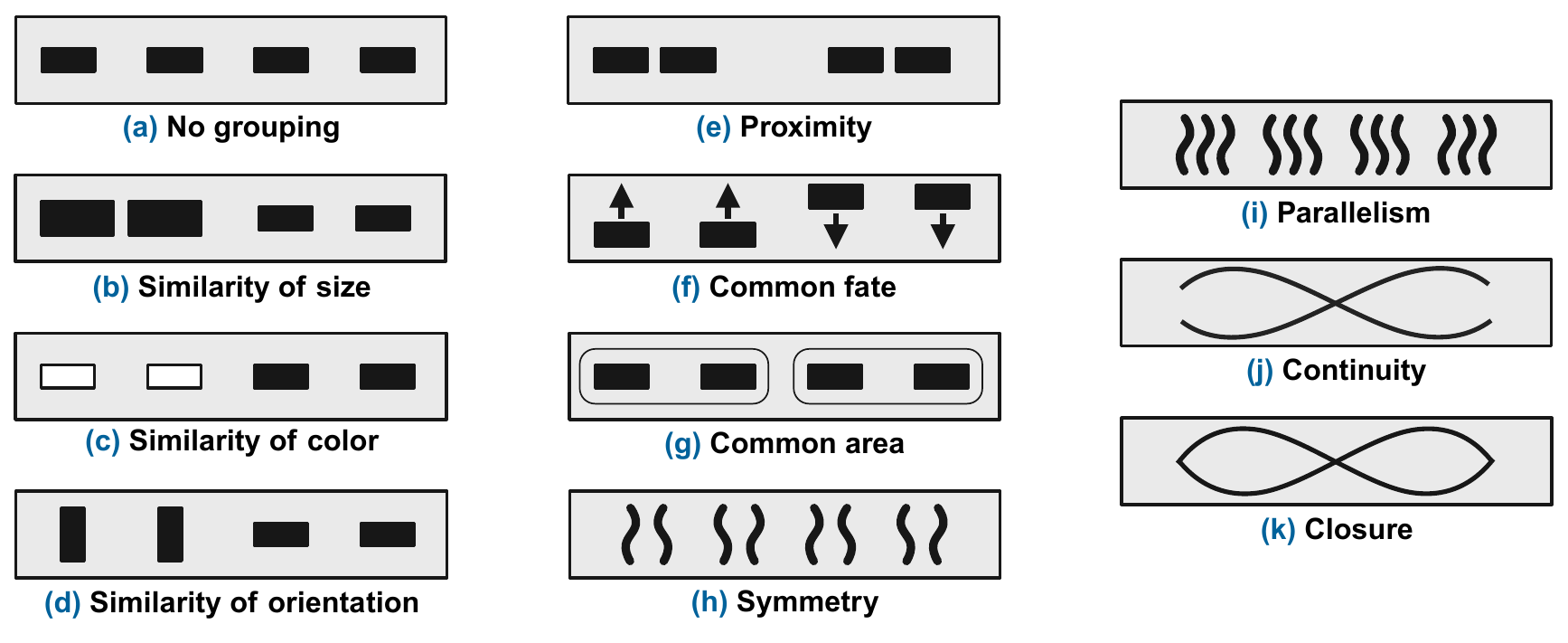}
            \caption{Illustration of some Gestalt theory's grouping principles. Different data instances can be visually perceived as belonging to the same group when sharing similar graphical encodings.}
            \label{fig:gestalt}                
        \end{figure*}
        
        \citet{Cleveland1984} defined visual perception as a users' ability to interpret graphic encodings to decode and understand the information. Perceiving scenes and objects through transmitted information is not merely a passive observation procedure. Research on graphical perception for reducing cognitive load in data visualization had been conducted before computers transformed graphics into interactive tools. In the early 20th century, the influential \textit{Gestalt} theory emerged from studies of fundamental laws in behavioral psychology on elemental perceptual organization and grouping.
        According to Gestalt theory, the brain is a pattern and discrepancy detector that identifies, distinguishes, and groups graphical elements based on shared characteristics to form more complex perceptual constructions~\cite{ChangD2007,Wagemans2012}. Exploring the brain's grouping capacity is valuable for developers designing visualization tools that integrate powerful communication elements to help users discover patterns within large amounts of information.

        Key perceptual properties that can guide the graphical encoding include: ($i$) groups of units sharing similar goals (e.g., common fate or proximity); ($ii$) groups of elements with similar shape characteristics (e.g., size, color, orientation, symmetry, or parallelism); ($iii$) groups of objects related by the same action; ($iv$) smooth or curving elements perceived as more visually friendly; and ($v$) groups of elements enclosed within well-delimited boundaries. Figure~\ref{fig:gestalt} illustrates aesthetic principles based on Gestalt theory.

        A comprehensive overview of the historical origins of the theory, as well as a detailed discussion of organization and visual perception by grouping objects, can be found in \citet{Wagemans2012}. Additionally, \citet{Wagemans2012b} explored recent formulations related to Gestalt principles related to efficient information structuring in visual systems from neurological research on complex perceptual states.

        Moreover, \citet{Aigner2012} found that visualization tools exploring graphical properties demonstrate better capabilities for information transmission when applied to complex tasks involving multiple analyses, as observed when such tools are contrasted to visualization techniques that do not explore graphical properties. Thus, modern visualization systems should consider aspects of visual perception, such as space management, the use of familiar symbols and shapes related to the application domain, data orientation, and visual consistency among different elements under visualization; these aspects should deal with the visual disorder generated by large amounts of information displayed simultaneously~\cite{Lohse1991,Javed2010Perception}.

        \citet{Hao2005} analyzed the influence of alignment and aspect ratios on multidimensional time series visualizations, and claimed that elements placed at the top of the screen are cognitively perceived as more important. Similarly, elements on the left side are also considered important (where the labels and other descriptive information related to the $Y$-axis variables are usually placed). Spatial placement reveals the order of the data and emerges as one of the most expressive perceptual principles in visualization~\cite{McLachlan2008}. 
        
        Besides, size is a critical information channel, with larger elements typically perceived as more important than smaller ones. Color is another highly influential graphical property. Well-defined color contrasts facilitate accurate reading; that is, adjacent objects representing different variables are more easily distinguished when encoded with high-contrasting color schemes, whereas the efforts required for the individualization of such objects tend to increase when the color contrasting difference is smoother. Moreover, large graphical areas, such as backgrounds, should display low saturation levels (e.g., white or lighter shades) to provide a clear contrast with the vivid colors used for symbols~\cite{Ware2004}. 
        Figure~\ref{fig:tooltip} presents an example where size and color communicate different data attributes.
        
        A colormap, or color encoding, is a sequential arrangement of colors used to represent data values in a visualization~\cite{ware2023rainbow}. One of the most common design mistakes in InfoVis is overusing the rainbow colormap, which is probably the most popular in visualization applications. According to \cite{borland2007rainbow}, the rainbow colormap is a poor choice because it can confuse viewers for three primary reasons: (\textit{i}) it is isoluminant for large segments, and the lack of monotonic luminance can obscure or distort data shapes; (\textit{ii}) it is not perceptually uniform, which can result in misleading interpretations of data variations; and (\textit{iii}) it is based on non-data-dependent gradients, creating colored regions with artificial divisions within the data. In addition, rainbow maps are difficult to make accessible, especially for viewers with color vision deficiencies~\cite{ware2023rainbow}.
        
        However, \citet{ware2023rainbow} presented a study that demonstrated scenarios where improved rainbow colormap designs (e.g., multihued maps~\cite{liu2018somewhere}) could be useful, increasing accuracy in visualization tasks (if carefully used). It is important to notice that the choice of suitable color schemes depends on the context. \cite{borland2007rainbow} suggested more suitable colormaps than rainbow-based. We recommend checking ColorBrewer~\cite{harrower2003colorbrewer}, a valuable guide to help visualization designers select color encodings that align with perceptual properties.

        Following, investigations into cognitive perception principles for time series visualization, assess the extent to which different graphical properties influence the analysts' abilities to compare multiple series effectively~\cite{Albers2014}. By carefully considering these graphical properties, InfoVis developers can optimize the interaction between data, visualization, and analytical tasks, simplifying the communication process that supports users' engagement with the information space. \citet{Adnan2016} and \citet{Albers2014} conducted comprehensive research on graphical perception and compiled ``good design practices'' directions for visualization systems. Similarly, \citet{Javed2010Perception} systematically investigated the performance of several chart designs, documenting critical graphical perception factors and providing recommendations for selecting appropriate visualization methods for multidimensional time series.

        As discussed above, the human cognitive system is a powerful machine; however, it has significant limitations. Even in a visualization interface that explores advanced graphical properties, excessive simultaneous information can overload users and become unworkable, leading to visual clutter and reduced clarity~\cite{Javed2010Perception}. In such cases, visualization systems should incorporate exploratory methods that support selections and reorganizations of the content on the screen at varying levels of granularity. 
        
        We refer to the books of~\citet{Munzner:2014} and~\citet{Ware2004} for comprehensive studies on design principles in information visualization. The following section explores interactive techniques designed to address human visual limitations in dense visualizations.

    \subsection{Interactive Exploratory Techniques}
    \label{sec4:sub:interacao}
        Modern visualization systems essentially comprise two main components: visual representations (i.e., data mapping in graphical structures) and interactive techniques (i.e., communication channels between users and systems)~\cite{Yi2007}. However, few studies have evaluated the impact of interactive techniques in InfoVis systems related to visual perception improvements. Most of the available literature focuses only on classifying and developing interactivity~\cite{Elmqvist2011}. Historically, the InfoVis community has concentrated on developing graphical presentation methods, while the Human-Computer Interaction (HCI) community has addressed interaction and design of components that influence user experience while browsing in data mining tools~\cite{Adnan2016}.

        Meanwhile, datasets are increasingly dense, and traditional time series visualization approaches often suffer from visual clutter and occlusion of details (see Section~\ref{sec4:sub:propriedades}). These methods often rely on page scrolling for users to access the information contained in all data points~\cite{Hao2005}. Scrolling through pages loaded with diversified data types depends on the user's working memory, and linking different pieces of information that are not visible simultaneously can lead to a fragmented overview of the data.

        On the other hand, digital devices have evolved in terms of graphical quality and computational power, thus supporting the development of comprehensive \textbf{multi-view systems} and refined user interaction techniques, which are crucial for coordinating and integrating the information context on the screen. These interaction mechanisms allow users to select points of interest on one chart, which are then highlighted across other linked views. Multiple levels of abstraction and detail support dynamic exploration of data patterns and anomalies~\cite{Heer2010}. In this sense, direct interaction with the information space can reduce data complexity and control the flow of information under analysis.
        Moreover, through iterative knowledge acquisition and hypothesis refinement, users can discover complex relations within dense datasets, including accessing information on demand, without the need to visualize all details at once~\cite{cui2019visual}.

        \begin{table*}[!t]
            \vspace{4pt}
            \caption{Conceptual elements of design for visual data exploration.}
            \vspace{2pt}
            \label{tab:concepts}
            \small
            \centering
            \begin{tabular}{>{\raggedright}p{4.0cm}|p{11.0cm}}
                \toprule[0.8pt]
                \textbf{Design Concept} & \textbf{Description} \\
                \midrule[0.8pt]
                Multi-view systems & Simultaneously display multiple coordinated visualizations to support comparison, cross-referencing, and exploration across different aspects of time‑series data without forcing context switches. \\
                \midrule
                Overview first, zoom and filter, then details-on-demand & A visual exploration paradigm where users start with a high‑level temporal summary to identify macro‑patterns, refine intervals of interest through zooming and filtering, and finally access detailed data or segments selectively. \\
                \midrule
                Focus+context & A visual exploration paradigm that combines localized detail (focus) regions within a broader contextual view. Helps users maintain orientation while inspecting temporal subsequences or anomalies in time series data. \\
                \midrule
                Fluidity & Smooth, continuous transitions and immediate visual feedback during interactive operations. Improves user experience and preserves cognitive continuity during exploration of temporal structures. \\
                \midrule
                Intuitive design & Interface designs leveraging elements and interactions that align with user expectations, requiring minimal instruction for a quick understanding of temporal trends and interactions across views. \\
                \bottomrule[0.8pt]
            \end{tabular}
        \end{table*}

        To address the challenge of handling massive datasets with effective interaction mechanisms, \citet{Shneiderman1996} introduced the \textbf{``overview first, zoom and filter, then details-on-demand''} paradigm, one of the most remarkable and widely recognized theoretical concepts in visual data exploration. This approach assumes that some patterns and trends can be observed only when data is analyzed from a broad, contextual perspective. For example, regions of a dataset containing local peaks/minima may be easily misinterpreted as global maxima or minima, inducing analysts to draw conclusions based on apparently adequate but incorrect information. In an InfoVis application, the overview provides a broad context of the dataset, presenting an overall picture of the entire entity under visualization. However, the number of items can easily scale to thousands when large datasets, such as multidimensional time series, are visualized, which hinders the effective detailing of individual elements~\cite{Craft2005}.

        After identifying the information of interest, users can select it to verify its specific characteristics. \textbf{``Details on demand''} refers to the selection of an individual object or a group of objects for point-by-point detailing, allowing users to selectively adjust the depth of detail, hiding, or revealing the information~\cite{Shneiderman1996}. Additionally, \textbf{zoom and filter} are exploration operations that reduce visualization complexity by removing non-essential information from the visualization and reorganizing the on-screen representation~\cite{Craft2005}. Reducing the initial set of elements helps users individually investigate the details of selected data more effectively~\cite{Shneiderman1996}.

        After a detailed overview of the original concept application, \citet{Craft2005} defined zoom as a filtering method through data browsing. This approach supports users in controlling content with no change in the overall representational context, filtering out undesired items from the current view. According to the authors, the dynamics of the details-on-demand strategy is an abstraction process that adheres to the Principle of Selective Omission~\cite{Resnikoff1989}. The cognitive system processes sensory information into manageable stages, i.e., complex data tends to be reduced to manageable portions through organization or format simplification. As a result, the brain processes the available information quickly, allowing rapid responses to changes in the environment.

        In addition, the \textbf{``overview first, details-on-demand''} strategy fits when applied to multi-view systems, since it aims to reduce visual complexity by revealing details of the selected items without changing the overall representational context. 
        After selecting an element group or a specific time interval, the corresponding data views are updated to display detailed information about the chosen items. This action is particularly important for steering the discovery of links and seasonal patterns between two (or more) data dimensions, building associations of the identified information with the remaining parts of the dataset~\cite{Craft2005}.

        \citet{cockburn2009review} summarized interface strategies such as \textbf{``focus+context''}, which integrate focal details and contextual visualization in a multi-level composition where all parts are visible in a single view. This principle relies on magnification tools (e.g., lenses~\cite{furnas2006fisheye}) to distribute content across multiple coordinated views where all focused elements are visible while also displaying the surrounding context. On the other hand, in the details-on-demand strategy, zooming operations alter the scale of a view, relying on users' spatial memory to compare multiple pieces of information~\cite{satriadi2020maps}.

        In this context, \citet{Kothur2015} integrated cross-correlation windows with a four-view design, while \citet{Milos2011} employed a focus+context technique associated with a compact tool for visually analyzing multiple events within temporal data. In their approach, a distortion lens was used to verify areas with high data density. \textbf{Distortion lenses}~\cite{furnas2006fisheye} are interactive techniques that magnify data; their magnification level and spatial extent are configurable so that they operate as local zoom tools, typically activated and positioned using the mouse cursor. 
        \citet{Zhao2011b} extended the interactive capabilities of the lenses for temporal data exploration, allowing an arbitrary number of series to be used as lenses' inputs. They integrated supplementary interactive tools to compare and statistically analyze regions of interest across one or multiple series.
        For an overview of lens abstractions, \citet{Tominski2017} identified related studies and discussed their properties and approaches within a general visualization context.

        \begin{table*}[!t]
            \vspace{4pt}
            \caption{Interactive tools and techniques for visual data exploration.}
            \vspace{2pt}
            \label{tab:tools}
            \small
            \centering
            \begin{tabular}{>{\raggedright}p{4.0cm}|p{11.0cm}}
                \toprule[0.8pt]
                \textbf{Exploratory tool} & \textbf{Description} \\
                \midrule[0.8pt]
                Zoom & Interactive scaling of a selected time interval. Expands the temporal axis to reveal intra‑interval detail, allowing the inspection of fine-grained variations in the data. \\
                \midrule
                Filter & Dynamic inclusion or exclusion of series, temporal ranges, or categories, reducing clutter and helping users focus on meaningful patterns. \\
                \midrule
                Distortion lenses & Local magnification tools that locally enlarge dense areas around a cursor position without losing surrounding context. \\
                \midrule
                Brush selection & Manipulation tool to select temporal segments by directly dragging over plots. Enables comparison, highlighting, or linking selected time intervals across coordinated views. \\
                \midrule
                Tooltips & On‑hover text pop‑ups displaying timestamped values, metadata, or analytic annotations. Enable value inspection without requiring persistent visual changes. \\
                \midrule
                Animations & Smooth transitions between visual states or time intervals that guide users through changes. Animations support mental mappings of evolving data, reducing jump disorientation. \\
                \bottomrule[0.8pt]
            \end{tabular}
        \end{table*}

        \citet{Aigner2007} defined the \textbf{coordination of multiple views} as the proper way to propagate an interaction from one view to all other dependent views. An important matter arises regarding the coordination among multiple views in interactive tools, i.e., the difficulty in precisely tracking changes when a delay occurs between interaction and transition (adjustment in the views' content after an interaction)~\cite{Heer2007}. User interactions should result in immediate responses from the interface, which promptly update all coordinated elements for a seamless interactive exploration process. However, these visual changes must be smooth and not abrupt~\cite{McLachlan2008}, i.e., modern InfoVis tools should work based on fluid interactions. 

        \citet{Elmqvist2011} described the \textbf{fluidity} of visualization systems as an intangible concept characterized by smooth and continuous effects that allow charts to quickly react to user commands. Due to its intangible nature, developing tools with a fluid design is not a trivial task. The authors argued fluidity is a powerful abstraction that can transform the exploratory process into a comprehensive and enjoyable experience, helping users compose an evolutionary mental model of the information flow. Furthermore, they also elaborated guidelines for building InfoVis systems that integrate responsive interactions with graphical representations. 
        As the previous paragraphs present many different and dense concepts, Table~\ref{tab:concepts} summarizes elements that developers can explore when creating modern visualization tools.

        From a different perspective, the interactive information search involves exploring data features to test and refine hypotheses. If the visualization tool takes too long to render the new layout after an interaction, users may lose track of their initial expectations~\cite{Nonato2018}. Several interactive techniques can improve the user experience in visualization systems, ranging from simple methods, such as highlighting (which acts on visual control by capturing the user's attention to specific data elements~\cite{Adnan2016}) to more sophisticated animations that guide viewers through transition events.

        Following this idea, \citet{Heer2007} claimed that \textbf{animation} is a promising strategy to enhance users' perception of object changes across charts with correlated information. Through supervised experiments, they evaluated the advantages of animated transitions during operations such as resizing axes and values rescaling, toward reducing occlusions. However, animation effects do not automatically guarantee effective exploratory performance, with \citet{Heer2007} recommending that graphical developers avoid overly complex animations. The authors warned that animations in statistical charts require careful studies, since inappropriate transitions might induce analysts to misinterpretations or errors when the actions violate the visual semantics of the data. Furthermore, they suggest that animations are preferable for making screen updates more user-friendly, recommending transitions that take one to three seconds as a response time. 

        \citet{Sadana2016} designed interactive systems featuring multiple coordinated views optimized specifically for tablets, which have become very popular touch-screen mobile devices with no input tools such as a keyboard or mouse. Despite their considerable computational power, tablets and smartphones impose more space constraints than a regular computer since their screens range roughly from four to thirteen inches. User interaction on these devices is addressed directly through the user's touch, which may lead to inaccuracies during data selection tasks, for example. However, due to their widespread use, InfoVis systems can exploit their resources to engage a wider audience. \citet{Sadana2016} also proposed a requirement analysis for multi-view interfaces in addition to interaction techniques considered suitable for data selection and view coordination on small devices.

\begin{figure}[!ht]
            \centering
            \includegraphics[width=1\linewidth]{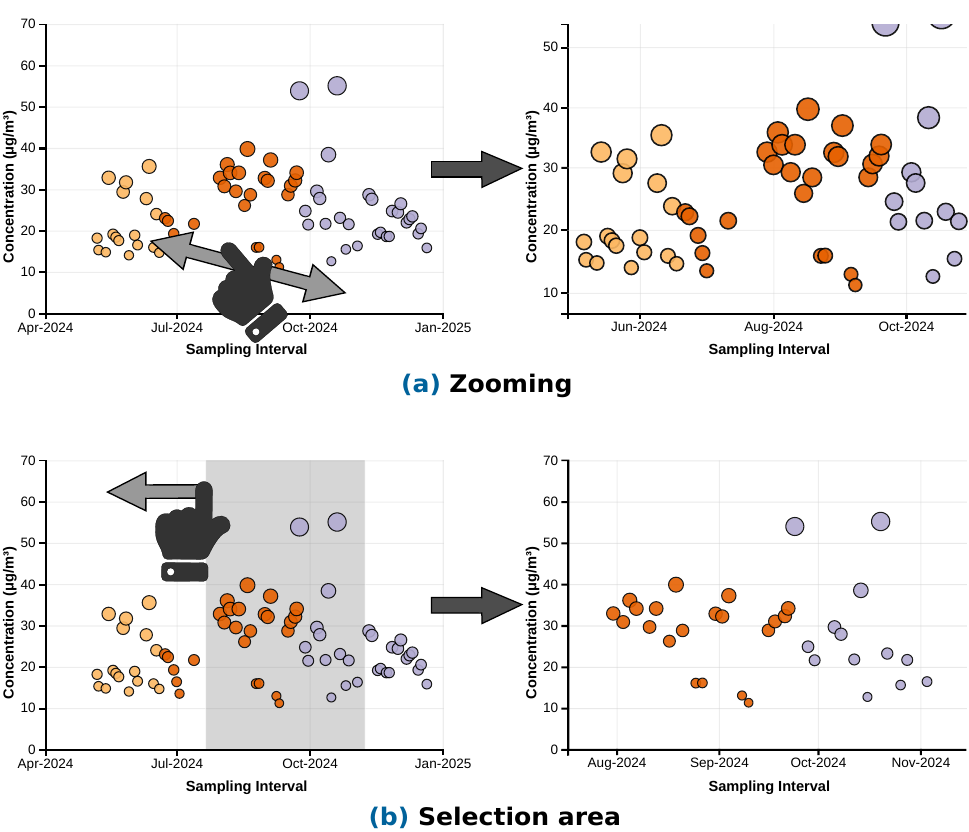}
            \caption{Interactive visualization tools for exploring data presented on touch-screen displays, such as tablets and smartphones.}
            \label{fig:tablet}                
        \end{figure}

        Figure~\ref{fig:tablet} illustrates interaction commands for touch-screen mobile devices with graphics based on visualizations developed by~\citet{Alexandrina:2019}. For example, to zoom, the user should place two fingers on the screen and then move them apart (Figure~\ref{fig:tablet}a). To define a selection area, the user touches a region in the view and drags horizontally across the view until the open area encompasses the items of interest (Figure~\ref{fig:tablet}b).

        \begin{table*}[!t]
            \vspace{4pt}
            \caption{User requirements and exploration challenges in time series visualization, and their mitigation using design concepts and interactive tools.}
            \vspace{2pt}
            \label{tab:challenges}
            \small
            \centering
            \begin{tabular}{>{\raggedright}p{4.0cm}|p{11.0cm}}
                \toprule[0.8pt]
                \textbf{Requirement / Challenge} & \textbf{Description} \\
                \midrule[0.8pt]
                Cognitive load &  
                Users' working memory is limited to a handful of items; high-density plots can overwhelm analysts, requiring abstraction, progressive disclosure, and coordinated views to distribute information processing effectively. \\
                \midrule
                Screen constraints &  
                Display area, resolution, and aspect ratio (especially on mobile or embedded devices) restrict the number and size of simultaneous views, requiring adaptive layouts, responsive controls, and aggregation techniques. \\
                \midrule
                Latency and responsiveness &  
                Interaction delays (network, rendering, data retrieval) degrade the exploratory loop; maintaining short response times and smooth transitions is critical to keep users' mental models aligned with visual updates. \\
                \bottomrule[0.8pt]
            \end{tabular}
        \end{table*}

        Interactive techniques are integral to current analytical-exploratory processes, with virtually all modern visualization tools supporting some degree of data exploration through dynamic mappings~\cite{Oliveira2003}. 
        \citet{Pike2009} described interactivity as a science and identified challenges for developing effective interactive tools. They argued that interactive tools should act naturally, with interactivity-dependent objects performing obvious-like actions without drawing the user's attention to the interactive operation mechanisms (even the most sophisticated ones, as they can overload the user's attention). They also claimed that \textbf{a good design should be intuitive}. Analysts must easily recognize elements across different views of a graphical interface during interactions and understand the relationships between current and previous data states~\cite{Heer2007}.

\begin{figure}[!ht]
            \centering
            \includegraphics[width=0.95\linewidth]{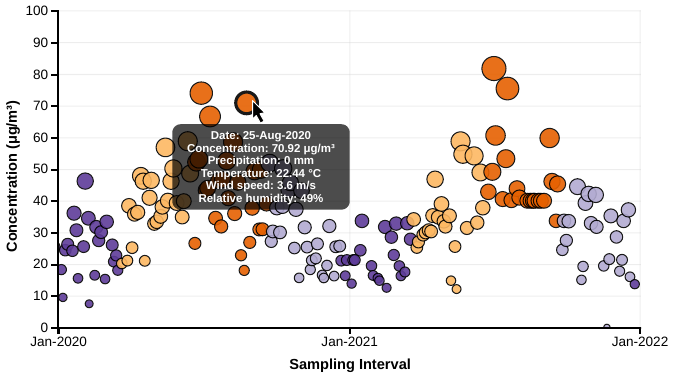}
            \caption{Accessing multiple information from a time point with an interactive tooltip~\cite{Alexandrina:2019}.}
            \label{fig:tooltip}                
        \end{figure}

        To add more information to a visualization, 
        \citet{Adnan2016} investigated the use of \textbf{tooltips}, which are text messages displayed when a user hovers over a visual element with the mouse. Such approaches help users identify data values and characteristics. The authors found that tooltips can enhance navigation in time series visualizations without loss of efficiency or analytical accuracy. Their results showed that some users prefer simple textual instructions about points of interest over more complex interactions with a high learning curve. Figure~\ref{fig:tooltip} presents an example of interactivity using tooltips~\cite{Alexandrina:2019}.

        In addition, \citet{Goethem2017} applied \textbf{brush selection} as an interactive technique to facilitate trend detection in time series data. Brushing allows users to select subsets of data by interacting with graphical representations directly: toward defining a selection area using brushing, the user clicks at a starting point (where the target objects start), drags the cursor until reaching a termination point (comprising all desired target objects), and finally releases the mouse button to complete the selection process. After the release moment, the selected area is highlighted and the corresponding data can be transferred to another verification method (see Figure~\ref{fig:tablet}b). For an easy overview, Table~\ref{tab:tools} summarizes the interactive tools discussed in this section.

        Generally speaking, visual analysis focuses on the connection between data presentation and human cognition; simply creating new graphical metaphors does not ensure information perception~\cite{Pike2009}. Instead, graphical development should be integrated with interactive approaches for assisting users in discovering patterns and trends on demand in multidimensional time series data. Transitions between the states of visualization interfaces can be smoothly animated toward an oriented and continuous analytical environment. Table~\ref{tab:challenges} summarizes the main challenges faced by modern visualization systems in managing large, multidimensional time series datasets.
        
        Complementing the other studies presented in this section, 
        \citet{Yi2007} outlined a taxonomy for interactivity objectives in InfoVis, while \citet{McLachlan2008} designed development guidelines for interactive-based systems with multiple views, in which time series can be side-by-side compared on the same chart, under different levels of detail. \citet{Perin2013} compared static charts with interactive approaches to evaluate the effectiveness of alternative representations applied to multiple time series and found that even simple interactive techniques applied to unify different views can substantially improve exploratory tasks. Finally, \citet{wu2021diel} proposed a declarative relational framework that incorporates asynchronous events to support interactive visualization.

	\subsection{Large-Scale and Real-Time Time Series}
	\label{sec4:sub:big_data}

        The proliferation of big data has generated extreme-scale time series containing millions or even billions of samples, requiring techniques capable of handling such a large volume of data while keeping computational efficiency, perceptual clarity, and interactivity. Two complementary classes of methods are commonly used in this context: \textbf{aggregation/downsampling} that reduces raw samples to a compact representation and \textbf{multi-view systems} that progressively refine visual detail on demand (see Section~\ref{sec4:sub:interacao}).

        A straightforward aggregation is temporal binning, where data points are grouped into fixed intervals (e.g., hourly or daily bins) and summarized using averages, medians, or densities, which reduces data volume while preserving high-level trends~\cite{chan2025tivy}. For example, imMens~\cite{liu2013immens} employed binned aggregation for the real-time querying of massive datasets, allowing subsecond response times for interactive explorations, while TiVy~\cite{chan2025tivy} provides visual summaries through adaptive binning and summarization.
        Similarly, curve-preserving downsampling algorithms select representative points to conserve salient visual features. In this context, M4~\cite{jugel2014m4} performs query-level aggregation to produce error-bounded line visualizations at very high reduction rates.

        Multidimensional (or even high-dimensional) time series are ubiquitous in Industry 4.0, and alternative Symbolic Data Analysis (SDA) representations for time series have emerged. \citet{nascimento2020dynamic} summarized more than 1.5 billion 256-channel electroencephalogram (i.e., 256 time series) observations into 15 million symbolic intervals (approximately 0.98\% of the original size) through dynamic linear modeling, discovering temporal patterns beyond simple average trends. Moreover, questions of dynamic accommodation, as observed in time series intervals in neuroscience research, were revealed through visualizations of brain activity patterns using an interactive graph that highlighted the group communication of the brain's Regions of Interest (ROIs) across time, also obtained through quantile regression on symbolic intervals. Given their flexibility and speed in data convergence, SDA-based models can incorporate dynamic events related to complex time-dependent data, demonstrating computational efficiency to summarize, detect outliers, and visualize high-dimensional time series.

        In real-time contexts, systems must process data in (near) constant time. Progressive partitioning and multidimensional pattern extraction incrementally update aggregations through operations that bin data on-the-fly~\cite{liu2018tpflow}, while deep learning-based analytics approaches integrate visual analytics with neural networks to handle massive datasets by extracting embeddings for clustered visualization~\cite{rodriguez2023deepvats}. Refer to \citet{ulmer2023survey} for a comprehensive overview of progressive visualization and visual analytics.

\section{Visual Approaches and Methods}
\label{sec5}
    Despite the multiple taxonomies addressed in the previous section, categorizing visualization approaches nowadays is challenging because time-varying data have become more extensive in size and dimensionality. As a result, solutions designed to manage such complex data have also become more sophisticated, often integrating multiple conceptual frameworks.
    
    To better organize our discussion on InfoVis methods devoted to time series, we have partitioned the presentation of approaches into two sections. This section provides an overview of the different InfoVis techniques for time-oriented data by grouping them according to specific (or closely related) visual metaphors, while highlighting the main strengths and limitations of each approach. In contrast, the following section (Section~\ref{sec6}) addresses interfaces composed of multiple concepts and tools, grouping them according to their respective application domains.

    To begin with, visualizing similarities in time series is essential in many application domains~\cite{Fu2011}, and several approaches deal with data organization. For example, some tools create visualizations that group data according to global similarities, relying on clustering methods~\cite{ziegler2010} or grid-adaptive layouts~\cite{Hao2007}, helping users visually compare different time series directly. Additionally, temporal data can also be recursively mapped onto ordered subareas of the screen according to the global importance of data points~\cite{Hao2005}, or transformed into thumbnail representations for comparisons~\cite{Kumar2005}. 
    The following subsections cover several of these variations within the context of time series InfoVis.

    \subsection{Spiral and Radial Charts}
    \label{sec5:sub:spiral}

        To identify cyclical trends in time series, \citet{Weber01} introduced spiral graphs, a space-filling metaphor that supports the detection of previously unknown periodic behaviors such as cyclical trends or recurring events by arranging the temporal axis as a spiral based on concentric rings. Later, \citet{Zhao2008} refined this concept by presenting a two-dimensional (2D) radial graph illustrating a set of daily human activities (allocating one ring per day), in addition to a central three-dimensional (3D) representation of the 2D chart, communicating the magnitude of data patterns over time. 
        
        Spiral charts provide continuous visualizations for multiple overlapping cycles and are flexible to enable visualization at different data granularities (e.g., days, hours, minutes, weeks, or months). However, their effective use depends on careful parameterization (e.g., data aggregation to reduce noise), which can be a challenging task in contexts where there is no previous knowledge of the data properties~\cite{Andrienko2010}. Moreover, spiral metaphors can offer limited advantages when the time series has long periodic cycles.

\begin{figure}[!ht]
            \centering
            \includegraphics[width=1\linewidth]{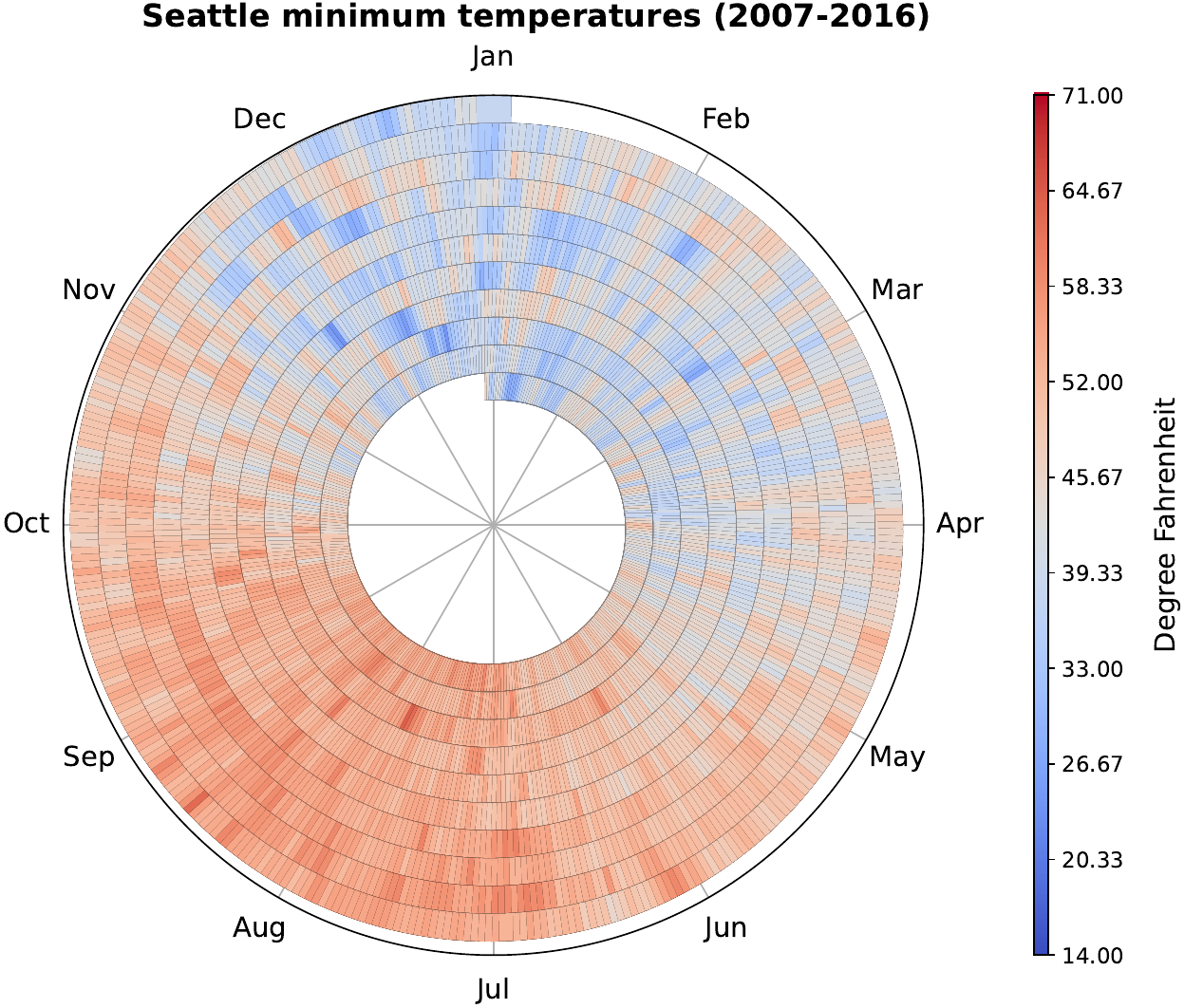}
            \caption{Spiral-based chart with daily minimum temperature in Seattle, WA, USA, recorded from 2007 (inner-most) to 2016 (outer-most).}
            \label{fig:ring}
        \end{figure}

         As an example, Figure~\ref{fig:ring} shows a ringmap, which is a spiral-based chart with a central space, depicting ten years (2007--2016) of daily minimum temperatures from Seattle, WA, USA~\cite{seattle_data}. As we can see, the spiral approach is able to effectively highlight clear seasonal patterns of lower temperatures during the winter months (December to March). 

        \begin{figure*}[!t]
            \centering
            \includegraphics[width=0.9\linewidth]{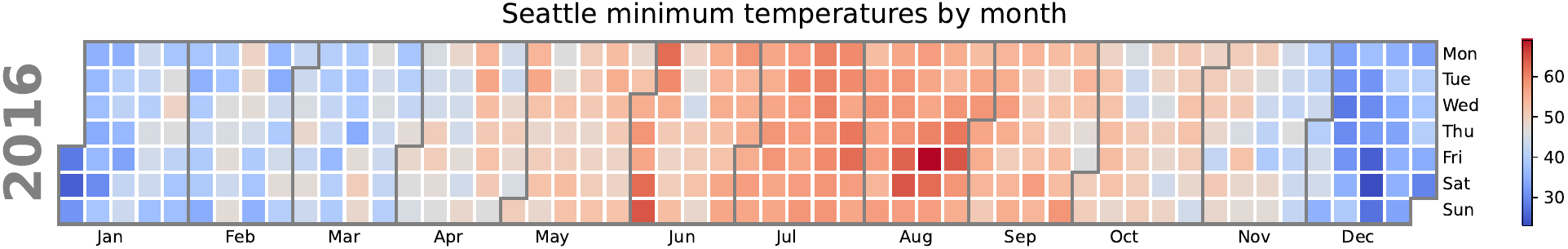}
            \caption{Calendar map with daily minimum temperature in Seattle, WA, USA, recorded in 2016.}
            \label{fig:calendar}
        \end{figure*}

        Following these ideas, \citet{Zhao2011} developed KronoMiner, a tool for dynamic exploration of time series that attempts to be flexible, supporting both complex analytical tasks and domain independence. It is based on a radial display that enables direct data manipulation through interactive selection, which simultaneously highlights and organizes distinct segments of the series. Data details are revealed using the MagicAnalytics Lens, a technique that performs zooming and compares parts of time series in real-time. 
        KronoMiner has multiple interconnected components that support its radial structure, providing multiple levels of detail, including views with different levels of granularity (from general overviews to more detailed views), tooltip messages, and pop-up menus. Although KronoMiner deals with temporal data in a generalized way, manipulating different types of data, the tool has limitations in handling large data volumes and lacks analytical features for more in-depth quantitative data analysis.

    \subsection{Calendar View and Heatmaps}
	\label{sec5:sub:calendar}
        Spiral charts are visually engaging, but they can add a cognitive layer to be interpreted once they require users to mentally map the spiral layout back to a linear one.
        Calendar-based charts aggregate temporal data and present it within a calendar-like grid. 
        In addition to enabling a detailed overview through fine-grained presentations (daily data), users can recognize underlying patterns in a temporal chart organized in a familiar calendar shape. 
        However, calendar views are restricted in their granularity and usually require more display space compared to approaches such as the spiral-based.
        As an example, Figure~\ref{fig:calendar} also shows temperature data from Seattle for the year 2016, but now presented in a calendar view.
        As we can see, it is also straightforward to identify seasonal patterns throughout the year in this visualization.

        In this context, \citet{VanWijk1999} developed a tool integrating two main components: a calendar and a multiple-line chart. Initially, the interface segments the dataset into daily sequences, and a clustering algorithm groups series with similar patterns (objects or data units) according to proximity measures between pairs. Each day on the calendar is color-coded according to its day cluster, and the line chart shows curves representing each time series group defined by the clustering process. 
        The approach was applied to a one-year dataset regarding the daily frequency of employees in a research laboratory in the Netherlands. 
        Although the tool provides a detailed overview of the information, it deals with data bearing periodicity characteristics that fit into a calendar-like format. Thus, it can have limitations when working with time series that lack daily or weekly periodic patterns~\cite{Lin2005}. 

\begin{figure}[!ht]
            \centering
            \includegraphics[width=1\linewidth]{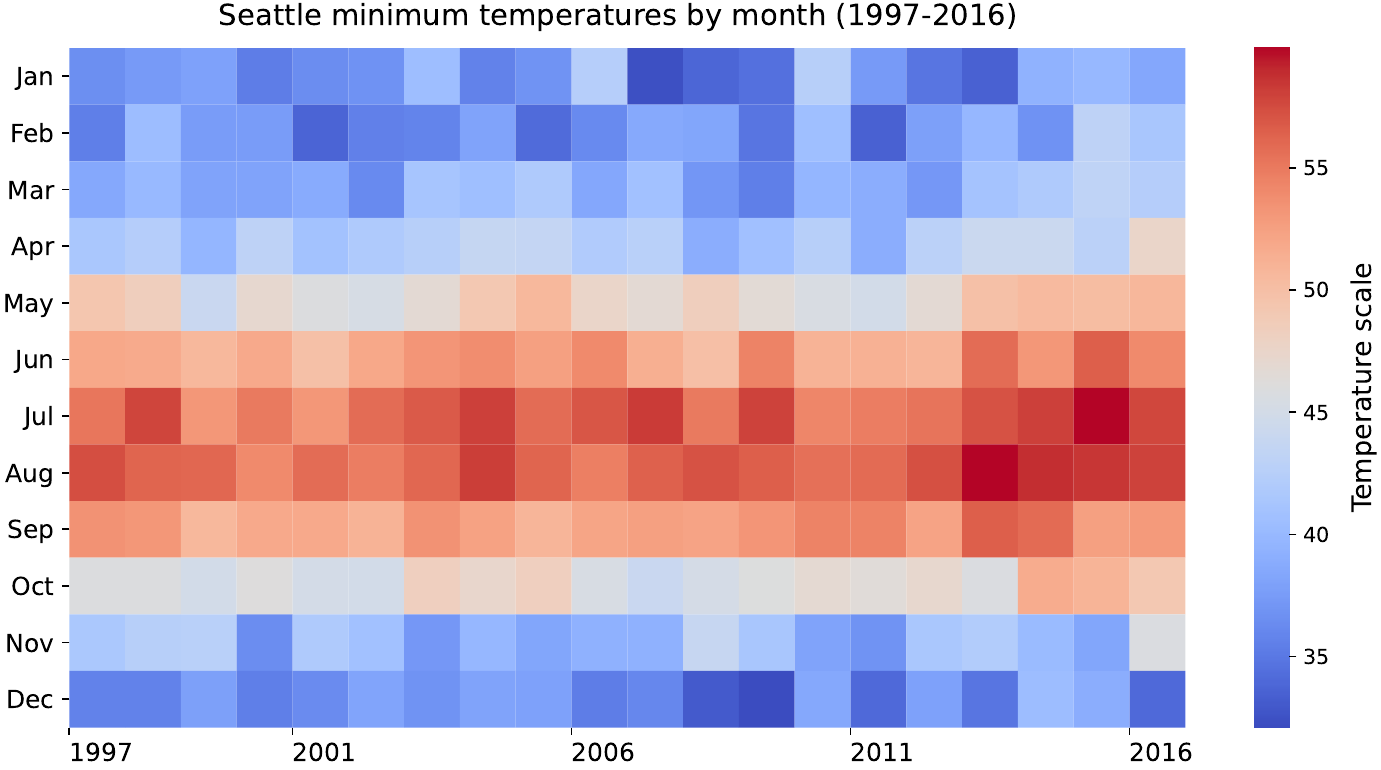}
            \caption{Heatmap with the average of the monthly minimum temperature in Seattle, WA, USA, recorded from 1997 to 2016.}
            \label{fig:heat}
        \end{figure}

        Calendar views are closely related to heatmaps~\cite{Grinstein2001}, a widely used space-filling visualization technique that provides a compact design to represent large amounts of data. Heatmaps are typically arranged as matrix layouts, relying on color encodings to represent magnitude values. The one-dimensional applications of heatmaps are often referred to as colorfields~\cite{Albers2014,gogolou2018comparing}. 
        Figure~\ref{fig:heat} presents the temperatures of Seattle in a twenty-year timeframe (1997--2016). We aggregated daily measures into monthly averages and visualized them in a heatmap.

        \citet{iram2019connecting} explored time series related to energy consumption to identify utilization patterns. They organized the data into a date $\times$ time heatmap, where color is used to encode consumption rates. 
        Despite the easily accessible information on energy consumption behavior over time, the fine-grained organization and lack of interactive tools limit the visualization to a few attributes.
        Similarly, \citet{zhang2021time} converted energy time series data from heating stations into heatmaps to identify temperature setbacks; however, their proposal differs by using convolutional neural networks (CNNs) fed with heatmap images for automated pattern recognition. Such a solution is particularly interesting because it leverages a visualization method to feed machine learning models for automated time-series analysis instead of only providing humans with charts for manual analysis.
        
        Whereas conventional heatmaps render data in individual cells, \citet{pham2020contimap} evolved the visualization of extensive time series data by developing a continuous abstraction that sorts and groups multiple series based on proximity similarities, resulting in a heatmap arranged with similar cells next to each other. Although such continuous visualizations achieve high levels of generalization, they depend on additional exploratory tools to access detailed information.
        \citet{Hao2005} presented a matrix-based visualization method that allocates space proportionally to the degree of interest in specific data subintervals.
        Data+Shift~\cite{palmeiro2022data} supports data scientists when investigating changes in real-time data used in machine learning tasks. A covariate detection algorithm is applied to measure distributional shifts over time, which are then visualized using a heatmap-based interface that enables users to conduct detailed investigations on distribution changes through feature selections.
        
        \citet{andrienko2023episodes} investigated population mobility during the COVID-19 pandemic with heatmaps to represent the evolution of daily mortality indicators due to COVID-19 in different countries. 
        Although heatmaps can handle large datasets, they often become cluttered and hard to interpret when dealing with excessive data density. Such scalability challenges can be mitigated by incorporating clustering and hierarchical methods or by providing different temporal resolutions to compress the data. 
        Additionally, given that heatmaps are 2D visualizations, it is critical to enhance them with interactive tools or integrate them with other visualization metaphors to explore multiple features.
        In certain contexts, 3D views can also be applied to improve the analytical capabilities of heatmaps. 
        For example, \citet{chu2023spatio} proposed a framework for visualizing space-time-varying wind fields. They used heatmaps to represent experimental results, capturing both velocity and trajectory of wind particles through a combination of color encoding and symbolic representations.

        Similarly to the spiral approaches discussed above, calendars and heatmaps can leverage interactive tools such as zooming, filtering, and tooltips (see Section~\ref{sec4:sub:interacao}) to allow a more detailed exploration of specific time intervals.

    \subsection{Time, Space, and Maps}
	\label{sec5:sub:maps}
        Maps are a usual choice to visualize geospatial data; however, how can temporal information be introduced to a map? \citet{Boyandin2011} designed Flowstrates, a visualization system that analyzes temporal variations in georeferenced datasets containing data on origin-destination flows between multiple locations. It utilizes two lateral, geographic-based maps and a central heatmap to separately depict spatial and temporal dimensions, allowing for independent exploration of each aspect. 
        Specifically, each row of the heatmap defines an origin-destination pair, and its cells are color-encoded to denote the magnitude of flows in each period. 
        The heatmap is surrounded on both sides by an origin map (left) and a destination map (right), with both maps linked to the heatmap to provide coordinated exploration of spatial and temporal patterns.
        
        In case studies, \citet{Boyandin2011} demonstrated Flowstrates with data on the displacement of Ethiopian refugees, revealing that Somalia was historically the country with the most expressive destination flow. 
        They observed that after 1988, Ethiopian migration to Somalia drastically decreased, following an increase in reverse migratory flow. 
        Such a reduction is attributed to the 1986 outbreak of the Somali Civil War, whose peak of violence was between 1991 and 1993, when the main destination of Ethiopian migrants shifted to Sudan and Kenya. 
        Flowstrates provides an effective visual summary of spatiotemporal patterns; however, it must deal with occlusion issues and has limitations in transmitting an accurate perception of real travel distances.

        Similarly, \citet{andrienko2023s} used heatmaps as mosaic glyphs on maps to represent temporal patterns of daily flight counts from airports. Relying on multidimensional projection embeddings~\cite{Maaten2008,eu2022projections} and introducing a time periodization approach, the authors investigated complex dynamic phenomena related to temporal and geographical patterns. However, the approach is limited to the existence of periods and can suffer when handling high-dimensional time series.
        
        \citet{Thakur2010} introduced a visualization system with 3D interactive maps, where temporal data are represented by geographically distributed polygons with disc-based shapes. Each disc corresponds to a specific time interval, and its color and size represent the evolution of the data features normalized according to their values.
        Although quite interactive, the system suffers from occlusion in its graphical elements, a common problem in 3D visualizations. The authors added a 2D linked view representing a regional classification to mitigate such occlusions. The tool was demonstrated on unemployment rate data from North Carolina, USA, where the stacked elements describe county-level trends over time. The users can then interactively select stacks to inspect and compare individual or subgroups of information.

        Three-dimensional visualization approaches often use space-time cube abstractions to represent the spatial evolution of data over time. In a space-time cube abstraction, spatial components are represented on the horizontal plane, whereas temporal components are placed on the vertical plane, thus creating an information volume. \citet{Bach2014} reviewed the literature on temporal data visualization, focusing on operations related to the space-time cube, a study that was later extended by \citet{Bach2017}. Furthermore, \citet{Andrienko2013} extensively reviewed tools and methods for space-time representations, particularly from the perspective of tracking the displacements and movements of entities. 

	\subsection{Scatter Plots and Boxplots}
	\label{sec5:sub:scatter}
        Scatter plots~\cite{becker1987} are well-known graphical tools that map data instances to Cartesian coordinates to identify trends, patterns, and correlations between two variables. Each instance is represented as a point, and multiple instances or groups can be differentiated within a scatter plot by using distinct symbols or color encodings. Such characteristics make scatter plots space-efficient approaches, as only a few pixels are necessary to visualize each point (see Figure~\ref{fig:tablet}). Despite their wide usage, scatter plots have significant limitations on visual scalability when handling multidimensional data (similar to line charts' limitations discussed above). As the density of points increases, occlusions and overlaps become significant issues, especially when the density of points exceeds the screen resolution~\cite{jugel2014m4}, reducing the legibility of individual data points.
        
        In the context of multidimensional time series, the limitations of scatter plots become more pronounced because of the difficulty in simultaneously tracking the temporal dynamics (e.g., seasonality) of multiple variables. In such cases, revealing interaction effects among multiple series can be challenging, which limits the depth of temporal analysis based on standard scatter plots. 
        Several strategies can be applied to mitigate those limitations and allow for more effective and insightful temporal analysis based on scatter plots, including color encoding and size variations for data points, incorporating interactive exploratory tools (e.g., filtering and zooming, see Section~\ref{sec4:sub:interacao}), and arranging small multiples to track different dimensions across multiple scatter plots~\cite{Alexandrina:2019}. Extending scatter plots into three dimensions can be an alternative, although it can introduce further challenges related to 3D perception and occlusion (see Section~\ref{sec5:sub:maps}).

        \citet{feng2010matching} introduced a data aggregation methodology for scatter plots (and other visual metaphors) that generates density plots through the blurring of visual marks, mitigating visual clutter, and enhancing preattentive pattern identification. The solution highlights regions of high certainty and relies on interactive techniques for data selection, which can help analysts identify multidimensional uncertainty.

        According to \citet{wang2017line}, the choice between line charts and scatter plots for time series visualization depends on data characteristics, such as noise. For datasets with a high frequency of outliers, scatter plots can depict temporal trends more clearly, since excessive clutter is introduced in line charts when data variance increases. The authors proposed design guidelines and an algorithm to automatically select between line graphs and scatter plots in time series contexts.

\begin{figure}[!ht]
            \centering
            \includegraphics[width=1\linewidth]{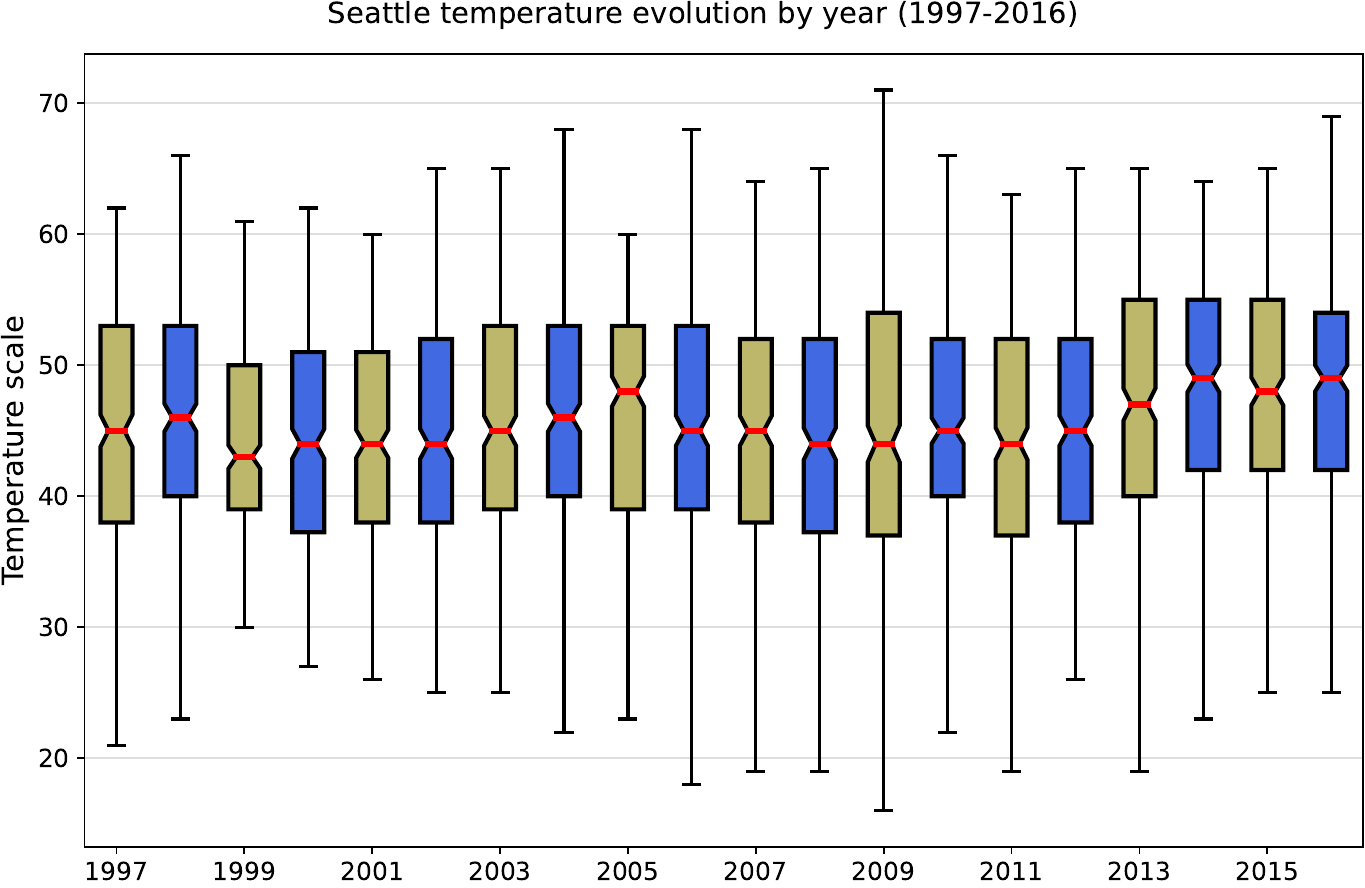}
            \caption{Boxplot describing the temperature variation of Seattle, WA, USA, over a twenty-year timeframe, with information measured daily and aggregated by years.}
            \label{fig:bplot}
        \end{figure}

        On the other hand, boxplots provide a compact statistical summary of data distributions, including medians, quartiles, central tendency measures, and outliers. They use box-shaped symbols with extension lines called ``whiskers'' to represent the range of data aggregated over predetermined time intervals (e.g., daily, weekly, monthly). Thus, boxplots are indicated to provide a high-level statistical summary of time series data and compare different intervals, instead of depicting the continuous temporal evolution of individual data points. As an example, Figure~\ref{fig:bplot} shows a temporal boxplot summarizing the statistical components of Seattle’s temperature variations, with daily measurements aggregated in annual intervals.

        At a high level, scatter plots and boxplots provide complementary perspectives that together can enrich the analysis of complex multidimensional time series. Scatter plots are typically applied to investigate fine-grained pointwise relationships (detailed context), whereas boxplots summarize statistical distribution measures across intervals (overview context).

	\subsection{Parallel Coordinates}
	\label{sec5:sub:par_coords}
        Parallel coordinates are a classical technique applied to multidimensional data visualization. However, in contrast to spirals or calendar-based approaches, parallel coordinates are rarely seen in the context of time-series data.
        Their modern development began in the 1990s and was especially noticed by the visualization community after the seminal work conducted by \citet{Inselberg1990}. Since then, they have been applied in a wide range of research aimed at representing complex multidimensional data. 
        
        Parallel coordinates break the conventional single orthogonal $Y$-axis representation paradigm by arranging multiple vertical axes evenly spaced and horizontally distributed on the screen (i.e., parallel), with each axis (or coordinate) corresponding to a different data attribute. Adjacent axes are linked by straight-line segments that intersect each axis at the position corresponding to the data instance's measurement for that attribute. The trajectory of each segment throughout all parallel axes describes the profile of each data instance across all dimensions.

        The approach offers advantages in comparing attributes with different magnitudes and in identifying patterns and data trends. In theory, representing many data dimensions simultaneously is not a major concern for parallel coordinates; however, in practice, visualizing high-dimensional datasets using such an approach can lead to significant element occlusion and visual clutter, which can impair the user's ability to perceive attribute correlations~\cite{Fua1999}.

        Several researchers have addressed efforts to mitigate these limitations. \citet{Fua1999} and \citet{feng2010matching} improved the technique with interactive aggregation methods to structure data into different abstraction levels to reduce density and visual clutter, while \citet{roberts2018smart} implemented interactive brushing techniques to allow focused exploration of subsets.
        Even with many advances, parallel coordinates have traditionally struggled to reveal temporal data variations. 
        As an example, \citet{Tominski2004} attempted to deal with the temporal problem by reconfiguring parallel coordinates into a radial layout with a central temporal axis and the other feature axes around it. However, the approach suffers from occlusions and has limitations in providing data overviews.

        In contrast, \citet{Gruendl2016} introduced an innovative ``pseudo-perspective'' strategy by incorporating a temporal panel into the parallel coordinates layout; the panel can be opened between two or more axes and moved in and out, simulating a time-varying effect. 
        Specifically, the temporal panel shows the series' evolution by mapping time onto screen depth, enabling analysts to observe attribute fluctuations over time. 
        In addition, the system provides interactive tools to assist in comparing individual or grouped time series.
        For a comprehensive overview of parallel coordinates in visualization, we refer to \citet{heinrich2013state}.

	\subsection{Sankey Diagrams and Flow Maps}
	\label{sec5:sub:sankey}
        A Sankey diagram is a type of node-link chart in which the links represent flows between entities (nodes), and the width of each link is proportional to the magnitude of the flow~\cite{schmidt2008sankey}. As a flow-based visualization, Sankey diagrams can be naturally extended to represent time series data, providing an understanding of how relative magnitudes change over time.

        \citet{cuba2015research} aggregated the time component by year, representing each year as a distinct node. In contrast, \citet{lupton2017hybrid} proposed a Sankey solution in which the time-dependent attribute is represented as a single node, distributing the values along that time node. Both approaches yield layouts similar to parallel coordinates, which can limit the effective perception of temporal evolution when dealing with fine-grained time data. 
        
\begin{figure}[!ht]
            \centering
            \includegraphics[width=1\linewidth]{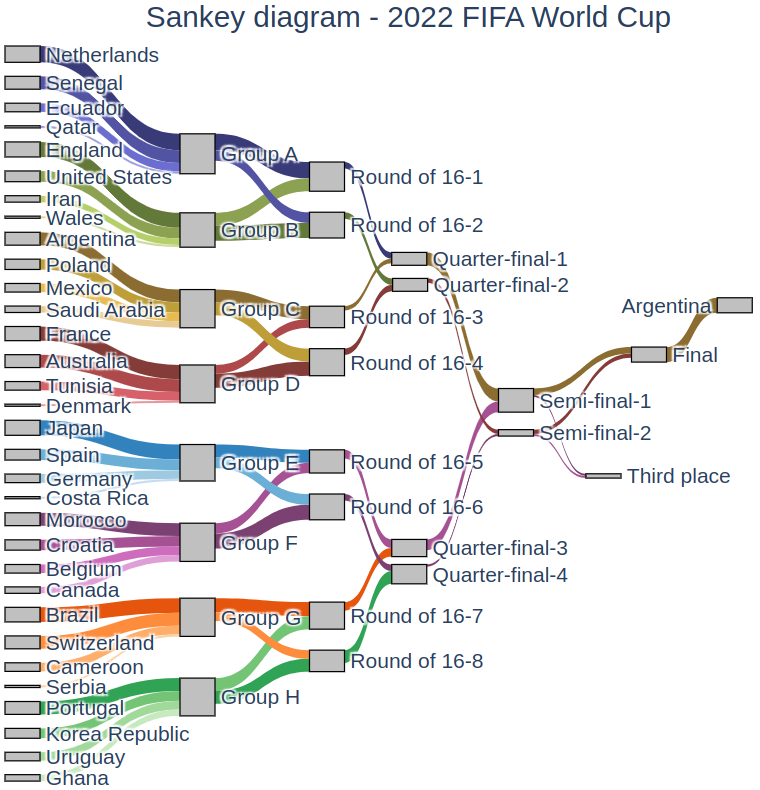}
            \caption{Sankey diagram representing the evolution of the 2022 FIFA World Cup. The event took place in Qatar from November 20 to December 18, 2022, with 32 participating national football teams.}
            \label{fig:sankey}
        \end{figure}

        \citet{gilch2018elo} simulated the probabilities that each team would advance through different stages of the 2018 FIFA World Cup, presenting the flow of all teams throughout the tournament using a Sankey diagram. Although the approach aims to illustrate probability flows instead of temporal evolution, it cleverly outlines the tournament stages and simultaneously reflects the progression of each team. Figure~\ref{fig:sankey} displays our version of a World Cup Sankey diagram, but for the 2022 championship. 
        In this diagram, we used the 2022 FIFA World Cup dataset~\cite{fifa_data}, where link widths encode the number of goals scored by each team between the tournament stages, and the size of the boxes with country names represents their points during the group stage.

        Despite their long history in visualization tasks, the design of effective layouts based on Sankey diagrams remains a challenge. Node placement is critical for achieving a clear and interpretable design, as Sankey diagrams are prone to clutter when the number of entities grows. In particular, minimizing link crossings is essential to avoid visual clutter, which can be challenging for highly interconnected nodes and result in unreadable charts. Simplifying or even omitting weak flow connections can reduce information overload, along with exploring color encodings to distinguish flows and node categories, and incorporating interactive tools such as zooming and selections for details-on-demand (see Section~\ref{sec4:sub:interacao}), can enhance the readability.

        Flow maps share similarities with Sankey diagrams but use links to represent the movement of objects between origin-destination locations. Flow maps are also prone to clutter when applied to large datasets, in addition to long flow links that can overlap and occlude shorter ones. To mitigate such limitations, \citet{zeng2014visualizing} designed enhanced flow maps specifically to visualize the temporal information of passenger mobility in public transportation systems. The solution allows users to track multiple metrics, such as riding, waiting, and transfer times, by interactively selecting route segments (pairs of origin-destination links). Given that public transportation systems generate complex spatiotemporal data, we refer to \citet{andrienko2017visual} for a comprehensive review of methods.
        
    \subsection{Stacked Graphs}
	\label{sec5:sub:stacked}
        A stacked graph (also known as a ``stacked area chart'') is a technique for visualizing multidimensional data, which stacks individual attributes on top of each other to share the visual space. The result is a visual summarization that provides an aggregated view of the data flow~\cite{Heer2009}. Several graphical primitives can be used to compose a stacked graph; for example, bars can be applied to represent categorical attributes, while lines or areas are suitable for visualizing time series. Stacked graphs emphasize summed or aggregated values in contexts where that information is more important than or equal to individual contributions.

        \citet{Wattenberg2005b} introduced NameVoyager, a web-based tool that works with animated stacked graphs to explore demographic data. It displays historical trends in baby names, placing temporal variations on the horizontal axis and stacking the popularity of individual names through the vertical axis~\cite{Elmqvist2011}. The tool updates the view to each key pressed during the user's searches. According to the author, the immediate reaction saves the user from having to click on a ``submit'' button and helps demonstrate how data mining is executed in real-time. Each attribute (baby names) is represented as alphabetically ordered colored strips. A specific name stack can be selected through mouse interactions, triggering smooth animations and a fluid user experience~\cite{Elmqvist2011}. NameVoyager has reached popularity due to its responsive approach, similar to a game~\cite{Wattenberg2005b}.

\begin{figure}[!ht]
            \centering
            \includegraphics[width=1\linewidth]{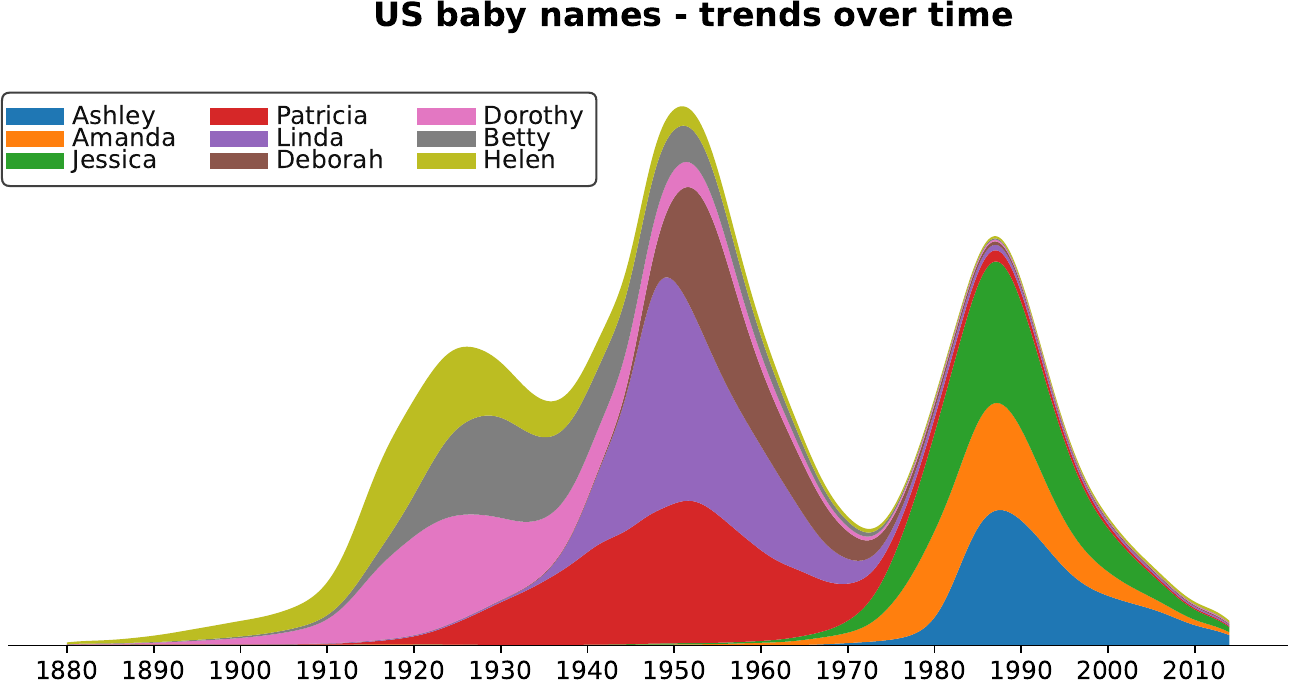}
            \caption{Stacked graph depicting part of the frequency of babies' names in the US from 1880 to 2014.}
            \label{fig:stacked}
        \end{figure}

        Figure~\ref{fig:stacked} shows an example based on the US Baby Names dataset~\cite{usnames_data}, which tracks trends in the popularity of names in the US. Such data came from the US Social Security records and comprises thousands of names and their evolution from 1880 to 2014. We selected a reduced set of names to avoid visual clutter.

        \citet{Havre2000} developed the ThemeRiver prototype, one of the first applications designed explicitly for visualizing the evolution of multiple time series. Enhanced by \citet{Havre2002}, ThemeRiver produces a fluid layout with soft curves. Unlike conventional stacked graph approaches, ThemeRiver centers its stacked layers around a horizontal baseline, allowing the time axis to show information flows in both directions (upper and lower dispersion)~\cite{byron2008}.

        As a case study, \citet{Havre2002} illustrated ThemeRiver by applying it to a collection of Fidel Castro speeches from 1959 to 1961. 
        The visualizations highlight some of Castro's most frequently discussed topics, relating them to history by overlaying a chronological sequence of events related to the speeches' period above the chart. 
        The design provides an enriched flow view that reveals general trends and temporal changes in the content of historical speeches, working as a ``timeline'' instead of only pointing outliers~\cite{chang2007}.

        The Stream Graph, introduced by \citet{byron2008}, refines ThemeRiver by optimizing the layer ordering to improve readability. The approach became popular when The New York Times (NYT) used a stream graph to compare the box-office revenues of movies released in 2007 with Oscar nominations. 
        The NYT visualization highlighted \emph{La Vie En Rose}, a movie that won an Oscar for Best Actress, but attracted a modest audience compared to other high-grossing films such as \emph{Shrek the Third}, demonstrating that the public and the Academy sometimes have divergent opinions. 
        In particular, the NYT rotated its chart $90$ degrees, placing the time evolution vertically and the gross earnings of the movies on the horizontal axis. According to \citet{byron2008}, this rotated layout was specifically applied to fit the chart in the print media's column format. An online version of the visualization is provided in \citet{Bloch2008} but with a non-rotated layout.

\begin{figure}[!ht]
            \centering
            \includegraphics[width=0.85\linewidth]{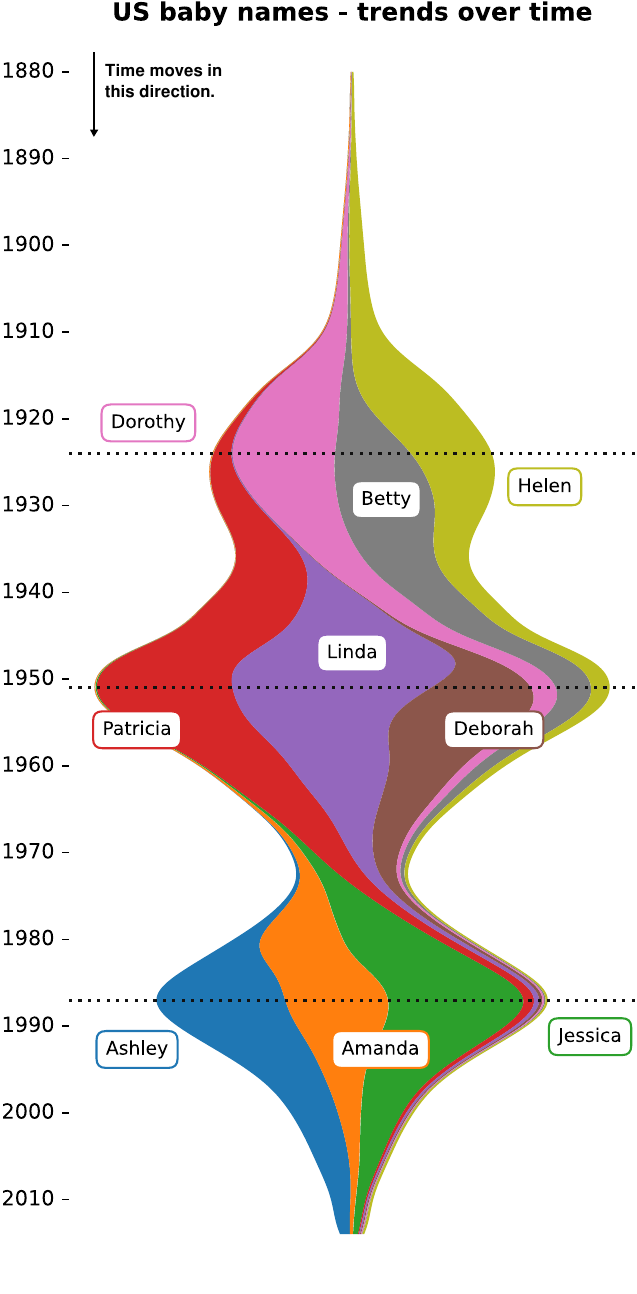}
            \caption{Stream graph representing name tendencies between 1880 and 2014. Vertical positioning creates a fluid design that fits a large chart in a two-column page layout.}
            \label{fig:stream_names}
        \end{figure}

        Unfortunately, the NYT's interactive visualization is no longer available, but a static copy remains accessible in the NYT archives~\cite{nyt_stream}. To demonstrate the tool, we rendered a stream graph following the NYT chart's design using the US Baby Names dataset described above. Figure~\ref{fig:stream_names} shows our version, which fits nicely in a two-column layout and allows users to track magnitude variations in name popularity over time using a large chart.

        Although \citet{byron2008} mathematically discussed the stream graph approach, their layer ordering and curve smoothing algorithms remained hard to understand and reproduce, with no fully disclosed solutions. \citet{DiBartolomeo2016} clarified those issues and, because of their contributions, they received the Best Paper Award at the 18th EG/VGTC Conference on Visualization (EuroVis), one of the most influential visualization conferences, highlighting the interest of the research community in ThemeRiver/stream graph design.

        However, stacked graphs and related variants are limited in handling multiple, multidimensional time series simultaneously. Although stacked graphs efficiently represent aggregated trends, they make it difficult to compare the individual contributions of each series. 
        Visual stacking is not informative for some data types or values, especially when handling negative values, which can often lead to misinterpretations of the space between curves~\cite{Heer2009}. 
        \citet{byron2008} recommend ordering layers to mitigate such effects; however, it does not fully resolve them. 
        To address these limitations, designers can use small multiples (see Figure~\ref{fig:space_share_split}) or horizon graphs~\cite{saito2005two} to encode positive and negative deviations. 
        Braided graphs~\cite{Javed2010Perception} can also be alternatives, although they become confusing visualizations in very high-dimensional scenarios.

        Another drawback of stacked charts is their static nature and limited interactivity. \citet{Shi2012} developed RankExplorer to mitigate such limitations by leveraging the abilities of stacked charts to handle temporal evolution, but introducing dynamic comparisons between different information groups. 
        In RankExplorer, multiple data items are ranked in a stacked graph, allowing users to compare the changing flow over time through markers identifying position changes, bar graphs showing classification oscillations, and lines that provide statistical summaries. However, the rank grouping can lead users to miss fine-grained patterns, in addition to not providing explanations for the ranking step.

        \citet{Dou2013} and \citet{Cui2014} integrated stream graphs in textual analysis tools. HierarchicalTopics~\cite{Dou2013} was demonstrated using news databases, where texts were organized into main categories that can branch into specific topics, creating a hierarchical tree-based structure~\cite{Burch2008}. Stream graphs were then used to describe the trend evolution of topics selected by the user, relying on topic spectrum timelines. However, the tool can suffer from clutter as the number of topics grows.

\section{Applications}
\label{sec6}
    This section presents a series of InfoVis applications in the context of time series, focusing on composed interfaces, i.e., multimodal visualization systems that integrate multiple coordinated techniques.
    Effective visualizations should be designed to support information acquisition, which means that graphical interfaces must guide users in data interpretation, according to their domain of knowledge~\cite{Federico2014}.
    Composed interfaces go further in mitigating the limitations of single-method visualizations by actively combining multiple visual and interactive components to leverage each individual technique's strengths. 
    The applications discussed in this section are grouped according to their application domains to illustrate how these composite systems address domain-specific challenges.
    
	\subsection{Textual Documents and Topics}
	\label{sec6:sub:text}
        Textual documents are often analyzed by correlating content similarity, citation networks, or explicit hyperlinks to reveal their relationships. 
        In the context of the Internet, websites are textual documents interconnected by hyperlinks, and a set of densely related pages (a cluster) represents a web topic. 
        These connections evolve over time as pages are created, removed, or modified.
        Text exploration on social networks presents additional challenges, combining relevance detection with the identification of trending topics. 
        When the temporal dependence of the publications is taken into account, the complexity of such analyses increases substantially. 
        Many researchers have developed visual metaphors for visualizing the spatiotemporal distribution of information~\cite{Cao2012}. 
        However, these methods are often limited in facilitating the visualization of relationships among multiple topics and their dynamic propagation. 
        
        \citet{Sun2014} introduced EvoRiver, an innovative approach that enables users to visualize the evolution of topic behaviors (cooperation and competition) over time. 
        The authors based the technique on a stream graph, extending and improving the original concept by incorporating support for positive (collaborative) and negative (competitive) topic interactions, avoiding the illogical overlaps that traditional stacked graphs can produce when negative values are present~\cite{Heer2009,Sun2014}.
        The EvoRiver is a visual composition comprising two parts: a top layer containing topic streams encoding positive (collaborative) behaviors and a bottom layer encoding negative (competitive) behaviors. Topics can migrate between layers when their behavior changes from competitive to collaborative power (or vice versa). 
        Time intervals can be interactively selected to investigate the relationships between topic pairs, which are visualized as linking edges between stream segments~\cite{Dang2016}.

        Specifically, the interface includes four main components: the EvoRiver chart itself, describing the flow of topics' evolution over time; a bivariate trend chart showing variations in topics' behaviors (positive/negative competition/collaboration power); edges linking related topics; and a Word Cloud~\cite{heimerl2014word} presenting keywords for a selected topic, with word size proportional to frequency. 
        \citet{Sun2014} demonstrated the interface using Twitter data to analyze the evolution of topics discussed during the 2012 US presidential election. They found that the topics ``Government'' and ``Politics'' were often closely related, but under the influence of ``International issues,'' ``Government,'' and ``Politics'' alternated between competitors and collaborators, thus allowing analysts to understand the complex dynamics of the main topics covered during that election process~\cite{Chen2017}. EvoRiver relies on a sentiment classifier based on statistical methods and expert labeling, which may introduce subjective bias to the technique.

    \subsection{Graph-structured Time Series}
	\label{sec6:sub:graphs}
        \citet{Toyoda2005} developed WebRelievo, a system for visualizing the structural evolution of websites. 
        The interface represents data as a graph~\cite{Herman2000}, with pages as nodes and hyperlinks as edges, thus modeling the site's time series as a sequence of graphs. Users can interactively analyze how site structures change over time, verifying when topics are clustered or split. 
        The interface displays temporal variations in separate graph views representing different time intervals, with the number of charts limited by the available screen area. However, WebRelievo's reliance on multiple static graph panels makes it hard to follow continuous evolution and suffers from occlusion when handling dense web communities. 
        
        \citet{Itoh2010} proposed an alternative with similar purposes but based on a 3D view to address the limitations. The evolution of elements is obtained by a ``slide planes'' technique that ``slices'' the time axis and displays the web graph according to the position of the plane at a given moment. Unfortunately, heavy edge crossings between planes still result in occlusion and cognitive overload.

        \citet{stopar2018streamstory} explored multidimensional climate measurements and GPS trajectories' time-varying data in a visualization system that first applies agglomerative clustering to reduce data dimensionality and then transforms the time series as a continuous-time Markov chain (states represent clusters, and transitions capture movement between them). The states and transitions are hierarchically aggregated and represented in a weighted graph, revealing the transition patterns over time.

        Visualizing graph-structured time series for large datasets remains challenging due to the computational demands of rendering their node-link structures and the tendency for occlusions. Therefore, choosing an efficient layout algorithm is critical to mitigate overlaps (force‑directed with edge bundling or hierarchical edge routing). Interactive tools such as zooming or fisheye distortion could be used to navigate between connections, in addition to color encodings to highlight clusters (see Section~\ref{sec4:sub:interacao}).

        \citet{costa2017studying} illustrate the use of probabilistic graphical modeling as a Dynamic Bayesian Network (a two-slice temporal Bayes net, 2TBN), which combines multivariate time series using a graph-based structure with a Kalman filter to estimate its dynamics. This study revealed the ability to obtain effective brain connections, as illustrated in resting-state and steady-state tasks of functional Magnetic Resonance Imaging (fMRI) data, where visual elements, such as time-varying graphs with dynamic arc lengths, represent the translation of the brain's Region of Interest (ROI) communication estimated across time. Although resting-state studies typically focus on functional connectivity, the authors hypothesized that the proposed 2TBN with time-varying parameters captures complementary aspects of intrinsic brain activity.

    \subsection{Financial Time Series}
	\label{sec6:sub:financial}
        The application of InfoVis methods to finance and business has been driven by the large numeric datasets generated by public and private companies. \citet{Lei2010} introduced a ringmap-based system for visually analyzing stock market data. Shares are clustered by a k-means-based algorithm and displayed on a spiral-like map (see Section~\ref{sec5:sub:spiral}). The design can be seen as an intersection of spiral graphs and heatmaps, where a variable number of concentric rings represents multiple shares over the same period. In addition, interactive zooming supports details-on-demand exploration.

        \citet{Lei2010} demonstrated their interface using Hang Seng Index data over a one-day business period. Each ring corresponds to the price evolution of a single stock, and the ring lengths encode market capitalization. The tool uses color encodings according to different sectors and performance levels. Black denotes the index, orange marks sectors that outperform the index, and gray indicates declining stocks during the trading session.

        \citet{Fu2005} used a specialized binary tree (SB-tree) to store time series containing the evolution of share prices. Financial data is volatile and highly dimensional, forcing visualization systems to process and update the visualized data periodically, increasing computational power demands. The approach processes data instances, calculates (ranks) their importance, and generates an SB-Tree with the most important data points. The representation incrementally updates the tree and thus prioritizes visualization of the most critical points without reprocessing the entire series.

        \citet{chang2007} developed WireVis, a multi-view system to analyze time series related to daily bank transactions and track fraud. The interface presents the relationships among accounts, time, and transaction keywords, through four coordinated views~\cite{Ko2016}. Specifically, it combines a heatmap, a word evaluation network, a sampling search tool, and a transaction descriptor to provide analysts with an overview of all transactions, which can be manually selected to handle specific details that update the visualization at each interaction. User studies demonstrated that WireVis enables the efficient detection of suspicious patterns in large datasets~\cite{chang2007}. 
        However, tracking multiple synchronized windows can be complex, as it imposes an increasing cognitive and perceptual load, requiring users to divide their attention between concurrent views (the multi-view trade-off)~\cite{Green2009}.
        \citet{chang2007} also conducted an extensive survey on visual analytics requirements in the context of the financial market to categorize tools, methodologies, and tasks.
        
        More recently, \cite{yuan2023tax} developed a similar multi-view interface to support tax auditors in exploring large-scale corporate financial time series, but augmented machine learning models to automate anomaly detection and forecast trends, which helps to reduce context switching and enables auditors to quickly detect suspicious points. The interface is interactive and scalable, allowing users to adjust time windows using sliders and filters for subset analysis. However, the underlying models remain ``black boxes'', lacking integration of explainability techniques~\cite{ortigossa2024explainable,xai_financial} to improve trust.

        There is a trend in current InfoVis solutions toward embedding machine learning models into visualization tools to support anomaly detection on a large scale and more quickly than relying only on human analysts. Machine learning models are able to bridge the gap between data visualization and scalability; that is, as data volumes exceed human capacity, automated models can pre-filter events, and InfoVis tools present humans with visual confirmations. We discuss this subject in more detail in Future Directions (Section~\ref{sec7:sub:future}).
        
    \subsection{Environmental Monitoring}
	\label{sec6:sub:air_qual}
        \citet{Li2013} developed interactive tools to visualize daily atmospheric pollution time series in Chinese cities, which were collected over a four-year monitoring period. The interface is based on a node-link tree structure~\cite{Burch2011} to organize the monitoring stations of each city hierarchically, allowing subdivision into districts. 
        Users can inspect and compare air quality features from multiple stations over time. The system was demonstrated in daily pollution data from Jinan (Shandong Province, China). Despite its effective layout, the visualization tool suffers from limited screen space utilization and scalability issues when representing large, multidimensional time-series datasets.
        
        \citet{Li2014} refined the previous design to handle data from more regions simultaneously. In this case, monitored cities were placed on a panel based on hyperbolic tree structures~\cite{Lamping1995}, where each city node can expand into temporal disc modules, which radially encode air quality time series for side-by-side comparison. 
        Both studies apply time series hierarchization to compare environmental datasets through layered views, applying InfoVis methods to present pollutant monitoring information and enhance the perception of variations and pollutant source behavior~\cite{Rodriguez2015}. The interfaces summarize large environmental datasets; however, they only support comparisons within the same city, i.e., cross-city analyses cannot be accomplished.

        \citet{Yoon2016} analyzed air pollution indices collected by the US Environmental Protection Agency (EPA) between 1998 and 2007. They investigated correlations between lung cancer incidence in particular regions of North America and the concentration levels of particulate matter in those regions, searching for relationships between lung cancer and exposure history to pollutants. Even highlighting that particulate matter is not the only risk factor for health, the authors used a simple visualization scheme linking time series and geographic views to confirm higher cancer risks in regions with elevated air pollutant levels.

        \citet{Alexandrina:2019} introduced an interactive web interface for air quality analysis based on monitoring particulate matter up to 10 \si{\micro\metre} (PM$_{10}$). The interface comprises two main groups of visualizations: a top panel with a scatter plot describing the PM$_{10}$ evolution over time, alongside four coordinated area charts (small multiples~\cite{ziegler2010}) showing related climate variables. The bottom panel mirrors the structure of the top layout, but presents a different interval of climate data for comparative exploration. 
        
        The authors designed a bivariate chart between the top and bottom groups to link their information and provide an intuition of the deviation between the two datasets. It is based on the statistical principle of bivariate analysis and visually contrasts two time series using an interpolation-based strategy that creates two concurrent curves that highlight the differences with contrasting colors. The tool is a coordinated multi-view interface that supports interactive brushing and zooming, and encodes multiple information channels simultaneously: seasonal color coding, tooltips, PM$_{10}$ concentration averages, and dynamically updated legends.

        Similarly, \citet{fonseca2021water} applied a symbolic representation to weather data from water-particle monitoring in Chile's Atacama Desert. Using a statistical process control approach, they proposed a bivariate SDA (see Section~\ref{sec4:sub:big_data}) control chart as a daily humidity monitoring tool.

    \subsection{Audio and Images}
	\label{sec6:sub:audio}
        \citet{Gomi2010} developed MIAOW, an image browser inspired by Tree-map approaches~\cite{Johnson1991}, a space-filling graphical metaphor. MIAOW is a 3D visualizer that hierarchically clusters and labels photographs according to their geographic location (latitude and longitude) and capture time. The visualization uses a space-time cube abstraction (see Section~\ref{sec5:sub:maps}) to display groups of images on the horizontal axis (delimited by rectangular nested areas), and time variation on the vertical axis. Interactive features include zooming, axis rotation, and clustered image browsing, but MIAOW has some limitations. For example, many cameras lack precise georeferencing, and clustering small thumbnails can produce misleading visual groupings.
        
        \citet{Javed2010} introduced stack zooming, an interactive technique to visualize audio time series. The approach hierarchically subdivides the information space, enabling users to focus on multiple areas of the time series through distinct detail levels~\cite{wang2017line}. 
        When a user selects a region of interest, a zoomed subchart is generated and stacked beneath the original, preserving the overall context while providing more details. Although it is not properly a multi-focus technique, stack zooming enables the creation of displays with reduced distortion while keeping the original representation of information, i.e., the general vision~\cite{Javed2010}. 
        However, stack zooming was designed for one-dimensional spaces; extending it to 2D introduces considerable layout and interaction complexity~\cite{Javed2012}.

        Continuous hierarchical zooming represents a significant advance in conventional step-by-step data navigation. Without it, users must select the information of interest one level at a time until the area of interest has been delimited, increasing the interaction count and cognitive load~\cite{Andrews2002}.
        Following this principle, \citet{Walker2016} developed TimeNotes, a system that supports interactive hierarchical exploration of multiple time series segments. Users can compare selected periods with previously selected intervals, allowing for the clustering and labeling of segments that contain trends or similar behaviors~\cite{Andrienko2018}. 
        \citet{Walker2016} demonstrated TimeNotes to $15$-minute audio recordings of animal monitoring collected by sensors mounted on birds, claiming that their solution is efficient, since analysts can isolate the information space and focus only on relevant data segments at desired granularities. The resulting views maintain a focus+context history of the analytical steps performed.

        \citet{itoh2024singdistvis} introduced SingDistVis, a focus+context design to provide overall views of the distribution of fundamental frequency trajectories in audio tapes by using heatmaps and details of selected data in polyline charts. 
        The heatmap highlights the vocal range trajectories of different voices, which can be further analyzed by drawing a selection region to render polyline charts. 
        The SingDistVis design enables simultaneous comparison of multiple audio time series, making it easier to detect differences in vocal range between speakers or instruments.

    \subsection{Event Sequences}
	\label{sec6:sub:event}
        ViDX~\cite{xu2016vidx} is an interface for monitoring industrial processes. 
        It interconnects multiple coordinated views centered on a timeline based on Marey's graph, a type of polyline visualization originally used for train and bus schedules~\cite{tufte2007}. 
        The timeline encodes processing times for each part of a particular workstation, grouping normal cycles into shaded bands rather than individual lines to reduce clutter while highlighting anomalous processes. 
        This primary view comprises a calendar for historical aggregate statistics, a radial graph for real-time failure detection, and small multiple histograms to display cycle times per workstation, providing users with a situational overview of the assembly line. 
        The ViDX system exemplifies a robust visual solution combining event-sequence timelines with multiple linked scenarios.

        Recently, \citet{bernard2024ivesa} designed IVESA, an interface for exploring event sequences in large multidimensional time series. 
        The system computes a set of event-based metrics (such as frequency, duration, and co-occurrence) and presents them on demand through multiple coordinated views.
        Although effective in handling complex data from multiple domains, IVESA requires substantial computational resources for real-time analysis and may exhibit some latency when processing high-dimensional time series.

\section{Discussions and Future Research on Time Series InfoVis}
\label{sec7}

    This review addresses modern information visualization (InfoVis) approaches for time series, with a particular emphasis on multidimensional time series. It demonstrates that representing time-oriented data is a multifaceted challenge, where different strategies are employed according to the context and the interdependence behavior of the data under analysis~\cite{Fua1999}. Modern and comprehensive tools usually incorporate user interaction mechanisms designed to help reduce the limitations imposed by high dimensionality. Some limitations include the inability to track information or conduct a systematic analysis of pattern detection, as many techniques primarily aim on visualizing the relationships between data records~\cite{Shi2014}.

    The larger the data dimensionality, the greater the number of challenges in developing graphical interfaces that can concisely and intuitively summarize the evolution of multiple features, even when employing a combination of different graphical methods and exploratory techniques. Consequently, this research differs from previous studies in several aspects, such as a review and discussion of several interactive-based tools, methods, and approaches specifically designed to handle multidimensional time series.

    \subsection{Software Tools and Libraries for InfoVis}
	\label{sec7:sub:software}
        The creation of InfoVis solutions has become increasingly accessible due to the availability of high-quality open-source libraries in multiple programming languages. Matplotlib (\url{https://matplotlib.org/}) is a Python library that provides a wide range of highly customizable plotting functions, supported by extensive documentation and numerous implementation examples suitable for both beginners and experienced developers. Similarly, Seaborn (\url{https://seaborn.pydata.org/}) and Plotly (\url{https://plotly.com/python/}) are valuable Python libraries used to generate complex, publication-quality visualizations. Datashader (\url{https://datashader.org/}) is an open‐source Python library specifically designed to visualize very large datasets.

        $\mathbf{R}$ is a statistical programming language well-known for its powerful data processing capabilities. For producing visualizations in $\mathbf{R}$, we refer to the ggplot2 grammar-based package~\cite{ggplot2}. JavaScript is the standard programming language for web applications and enables the creation of interactive and responsive HTML5 objects~\cite{Hales2012}. In this context, D3.js~\cite{Bostock2011} is a well-documented JavaScript library that enables dynamic data mapping and interactive visualizations using canvas and SVG graphical objects~\cite{murray2013}. 
        D3.js allows visualization developers to create highly customizable and interactive interfaces; however, it demands a solid experience in web development with JavaScript.
        NetworkD3 (\url{https://christophergandrud.github.io/networkD3/}) is an $\mathbf{R}$ package that leverages D3.js to create interactive graphs rendered with HTML compatibility, bridging $\mathbf{R}$ programming with web applications. 
        Shiny (\url{https://shiny.posit.co/}) is an open source package that integrates with ggplot2 from $\mathbf{R}$ and Seaborn/Plotly from Python, allowing users to create high-quality web applications for time series visualization based on a reactive programming model, which supports real-time updates and dynamic user interactions.

        Tableau (\url{https://www.tableau.com/}) and Microsoft Power BI (\url{https://powerbi.microsoft.com}) are advanced business intelligence (BI) software tools used for data analysis and the creation of interactive dashboards using a variety of interactive visualization techniques. However, full-featured versions of these tools require proprietary licensing and can be expensive. 

        We are currently experiencing an intense evolution in advanced generative approaches, such as ChatGPT~\cite{achiam2023gpt}, Llama~\cite{grattafiori2024llama}, and many other large language models (LLMs) that can generate increasingly accurate textual content and code. These models can be useful tools for InfoVis developers because of their increasing ability to generate and review code in various languages (e.g., Python, $\mathbf{R}$, JavaScript), which can reduce development time and minimize errors. In addition to general-purpose LLMs, specialized generative models focusing on visualization tasks have also emerged. Recent studies demonstrate how natural language processing can be integrated with visualization code synthesis~\cite{yang2024matplotagent} and chart understanding~\cite{NEURIPS2024_cdf6f8e9}. These approaches assist in creating interactive visualizations and also provide mechanisms to verify and optimize the generated code.

        Table~\ref{tab:packs} summarizes the strengths and limitations of each tool described in this section. Several other solutions, ranging from advanced platforms to straightforward and easy-to-use options, can be found online for chart generation. Here, we focus on widely used and well-documented tools and libraries.

        \begin{table*}[!t]
            \vspace{4pt}
            \caption{Summary of Tools and Libraries for InfoVis.}
            \vspace{2pt}
            \label{tab:packs}
            \footnotesize
            \centering
            \begin{tabular}{>{\raggedright}p{1.6cm}|p{3.6cm}|p{3.6cm}|p{3.6cm}|p{2.6cm}}
                \toprule[0.8pt]
                \textbf{Tool} & \textbf{Description} & \textbf{Strengths} & \textbf{Limitations} & \textbf{Interactivity} \\
                \midrule[0.8pt]
                Matplotlib & Python library for customizable plotting functions. & Highly customizable; suitable for publication-quality figures. & Complex coding for advanced features; not optimized for very large data. & Low (static by default). \\
                \midrule
                Seaborn & Python library built on Matplotlib for statistical visualizations. & Simplifies creation of complex plots compared to Matplotlib. & Similar performance limitations of Matplotlib with large data. & Low to moderate (static focus). \\
                \midrule
                Plotly & Python library for interactive, web-ready visualizations. & Enables dynamic and shareable visuals; web integration. & Complex coding for advanced features. & High (provides zoom and click events). \\
                \midrule
                Datashader & Python library for visualizing very large data. & Efficient for massive volumes; produces high-fidelity images. & Less intuitive for small data. & Moderate (can integrate with interactive tools). \\
                \midrule
                ggplot2 & Grammar-based $\mathbf{R}$ package. & High-capable for statistical graphics; extensible. & $\mathbf{R}$-specific. & Low (static by default). \\
                \midrule
                D3.js & JavaScript library for dynamic data mapping. & SVG-native; flexible for custom interfaces; supports animations. & Requires strong skills in web development. & Very high (focus on interactivity). \\
                \midrule
                NetworkD3 & $\mathbf{R}$ package leveraging D3.js for interactive graphs. & Bridges $\mathbf{R}$ coding with web interactivity. & Small number of visualizations; requires $\mathbf{R}$ and JavaScript skills. & High (D3.js-based). \\
                \midrule
                Shiny & Web app for data analysis and interactive dashboards. & User-friendly; integrates with $\mathbf{R}$ or Python ecosystems. & Complex coding for advanced features; free version is limited. & High (supports brushing, zooming, sliders). \\
                \midrule
                Tableau & BI software for data analysis and interactive dashboards. & User-friendly; dashboards with real-time updates. & Full features are expensive; proprietary licensing. & Very high (built-in dashboards). \\
                \midrule
                Power BI & BI software for data analysis and interactive dashboards. & User-friendly; supports collaboration. & Full features are expensive; tied to Microsoft environment. & Very high (built-in dashboards). \\
                \midrule
                General-Purpose LLMs & Generative models for textual content and code. & Accelerates prototyping; helps with error checking in coding. & Outputs require strong verification; potential inaccuracies. & Text-based interaction for queries and answers. \\
                \midrule
                Specialized LLMs & Generative models focused on visualization code synthesis. & Automates development from natural language. & Lack maturity (emerging field). & High (interactive code generation). \\
                \bottomrule[0.8pt]
            \end{tabular}
        \end{table*}

	\subsection{Challenges and Open Questions}
	\label{sec7:sub:desafio}
        Current visual exploration tools remain strongly based on traditional visualization models, such as line charts, histograms, pie charts, and bar charts~\cite{Heer2010}. Line graphs, in particular, are among the simplest and most popular methods for visualizing time-varying data, demonstrating a great capacity to synthesize information. However, while conventional line metaphors (even ordinary ones) effectively represent a few time series in several experiments, multidimensional time series demand the simultaneous observation of multiple features. Exploring the temporal evolution of multiple series is a common, yet challenging task in the context of multidimensional data~\cite{Javed2010Perception}.
        
        The main challenges in contemporary time series InfoVis relate to dealing with the ``curse of dimensionality'' and visualizing big data~\cite{Nonato2018}, which has become increasingly common in data analysis. Addressing these challenges demands innovative techniques and tools for summarizing and visualizing large datasets. In a systematic review of time-oriented data visualization, \citet{Aigner2007} argued that modern visualization tools must enable the perception of data's temporal evolution by highlighting patterns and trends, generating information, and deepening the understanding of the underlying phenomena.

        Another important aspect of time-oriented data is the dynamic flow of information, which may require periodic updates of the visualization~\cite{Milos2011}. Large datasets imply longer processing and rendering times. A robust visualization system must also be computationally scalable, i.e., its tools must be responsive even when large amounts of data are handled. Therefore, developing real-time visualization tools is challenging, opening fundamental questions regarding the construction of new interfaces that can integrate time-series representation with fluid control of the exploratory process while transmitting a clear vision of the information.
        
        The detection of anomalies is an issue that attracts the interest of data analysts. According to \citet{Lin2004}, anomalous behavior in data mining refers to a substantial deviation from the expected or ``normal'' pattern within a data sequence. The authors further argue that definitions of anomalies can vary significantly depending on the context, making the detection of unknown patterns particularly challenging. For a precise definition of anomalous behaviors in time series mining, refer to \citet{keogh2002finding}. 

        We noticed a lack of proposals for entirely new graphical metaphors. 
        For example, the stream graph, presented in Section~\ref{sec5:sub:stacked}, represents one of the last groundbreaking novelties proposed as a standalone chart. 
        On the other hand, visualization designers continue to innovate by integrating multiple metaphors and interactive tools to provide data scientists with renewed and powerful exploratory resources.

	\subsection{Future Directions}
	\label{sec7:sub:future}
        According to Gestalt theory, the brain has a cognitive tendency to complete missing information based on prior visual memory when its form is implicit~\cite{Wagemans2012}. This process depends on content stored in visual memory, which can differ substantially among viewers and is highly affected by their experiences and knowledge. However, a good visualization system must exploit graphical mechanisms that stimulate users' attention capabilities in order to attenuate the perceptual ``noise'' between the emitter (data source) and observer, actively engaging users in the task of detecting patterns and anomalies. When coupled with user interaction paradigms, the visual exploration of time series can promote efficient hypothesis testing and problem solving that include the discovery of patterns and anomalies~\cite{Lin2004}. Time-series clustering via Bayesian modeling is a promising area, given its flexibility and the inclusion of expert knowledge-based information. \citet{costa2024evaluating} presented a class from the Dynamic Bayesian Network (DBN), which enables time series clustering using multilevel information and decomposing different network-based structures, allowing causal InfoVis, through a probabilistic graphical model.

        When data dimensionality scales beyond the capabilities of conventional tools to track many attributes simultaneously, other visualization strategies must be considered. In this scenario, \citet{Nonato2018} conducted extensive research on multidimensional projection (MDP) methods from the perspective of visual analytics. MDP techniques are effective in reducing data dimensionality, transforming it into visual patterns that preserve and reflect similarities from the original data~\cite{eu2022projections}.
        Modern InfoVis solutions for time series can also be improved by incorporating techniques in periodization~\cite{andrienko2023s}, aggregation~\cite{nguyen2020exploring}, clustering~\cite{aghabozorgi2015time}, and by using methods for detecting correlations~\cite{sun2024task} and anomalies~\cite{xu2016vidx}.

        Moreover, recent advances in machine learning have led to models with remarkable performance rates in pattern discovery within large datasets. Despite the high power of such models in processing data from a wide range of domains, advanced machine learning models are non-linear ``black boxes'' that are complex to explain~\cite{longo2024explainable,ortigossa2024texplainer}. In response, significant research efforts have been directed toward developing comprehensive explainable artificial intelligence (XAI) methods~\cite{ortigossa2024explainable}. The integration of machine learning and visualization techniques has the potential to significantly enhance the insights derived from data analysis, since information visualization development is closely tied to XAI goals. 
        
        Such a valuable fusion of information systems can be seen in the works of \citet{xu2021mtseer}, who designed a multi-view interactive interface for evaluating machine learning models in multidimensional time-series forecasts using XAI, and \citet{shap_lundberg2018explainable}, which integrated XAI and multidimensional time-series visualization in the critical context of surgery monitoring. Another example is \cite{caruana2015intelligible}, where additive models were used to predict hospital readmission in large time series with the support of visualization grids to explore the additive components.

        In summary, designing visualization interfaces that integrate multiple and coordinated tools for data handling, processing, and presentation is a process that needs to root the design on a careful study of domain experts' requirements and aesthetic properties and represents a promising direction for constructing powerful information analytical tools capable of handling large and complex amounts of time series data we are currently generating.

\section{Conclusion}
\label{sec8}
    The analysis of temporal dependencies in large-scale, multidimensional time series datasets presents significant challenges in many application domains, as evidenced by the references discussed in this review. Visualization has the power to transform abstract phenomena into understandable visual elements~\cite{Shneiderman1996}. Therefore, well-designed visualization is fundamental in transforming abstract, time-oriented phenomena into perceptually intuitive representations, enabling analysts to uncover hidden insights with clarity. In Sections~\ref{sec5} and~\ref{sec6}, we explored the role of visual approaches and methods that address the complexity of time series data to make implicit information explicit and comprehensive.

    Several factors can impact the effectiveness of visualization tools in communicating information, including the selection of appropriate graphical metaphors (see Section~\ref{sec5}), the implementation of interactive manipulation techniques (see Section~\ref{sec4:sub:interacao}), user-centric designs that align with cognitive capacities (see Section~\ref{sec6}), and even the influence of device quality and dimensions on rendering fidelity~\cite{Thomas2005}. Addressing these factors is essential for enhancing the potential of information visualization in data interpretation. Table~\ref{tab:challenges} synthesizes visualization concepts exploited to mitigate the challenges related to visual data exploration, while Sections~\ref{sec7:sub:desafio} and~\ref{sec7:sub:future} outline open questions and future directions. Finally, we compared visualization software and environments to create InfoVis tools, highlighting their strengths for diverse applications (see Table~\ref{tab:packs}). In particular, recent breakthroughs, such as generative visualization models, have achieved prominent results in improving the understanding, synthesis, and optimization of visual representations (see Section~\ref{sec6}). 
    Therefore, developing visualization tools that provide clear and comprehensible layouts in an era of ever-growing data complexity is fundamental to allow analysts to extract meaningful information.

\begin{ack}
    This work was supported in part by the Coordenação de Aperfeiçoamento de Pessoal de Nível Superior--Brasil (CAPES)--Finance Code 001, in part by the National Council for Scientific and Technological Development (CNPq), and in part by The Paulo Pinheiro de Andrade Fellowship.
    The authors also acknowledge Angela C. P. Giampedro, from the University of São Paulo, for her valuable help.
    Figures~\ref{fig:data_cube}, \ref{fig:vis_stages}, and \ref{fig:gestalt} were created with BioRender.com.
    The opinions, hypotheses, conclusions, or recommendations expressed in this material are the authors' responsibility and do not necessarily reflect the views of the funding agencies.
\end{ack}

\small
\bibliography{references}

\begin{thebibliography}{219}
\providecommand{\natexlab}[1]{#1}
\providecommand{\url}[1]{\texttt{#1}}
\expandafter\ifx\csname urlstyle\endcsname\relax
  \providecommand{\doi}[1]{doi: #1}\else
  \providecommand{\doi}{doi: \begingroup \urlstyle{rm}\Url}\fi

\bibitem[Abbaspourazad et~al.(2023)Abbaspourazad, Elachqar, Miller, Emrani, Nallasamy, and Shapiro]{abbaspourazad2023large}
Salar Abbaspourazad, Oussama Elachqar, Andrew~C Miller, Saba Emrani, Udhyakumar Nallasamy, and Ian Shapiro.
\newblock Large-scale training of foundation models for wearable biosignals.
\newblock \emph{Preprint arXiv:2312.05409}, 2023.

\bibitem[Achiam et~al.(2023)Achiam, Adler, Agarwal, Ahmad, Akkaya, Aleman, Almeida, Altenschmidt, Altman, Anadkat, et~al.]{achiam2023gpt}
Josh Achiam, Steven Adler, Sandhini Agarwal, Lama Ahmad, Ilge Akkaya, Florencia~Leoni Aleman, Diogo Almeida, Janko Altenschmidt, Sam Altman, Shyamal Anadkat, et~al.
\newblock {GPT-4} technical report.
\newblock \emph{Preprint arXiv:2303.08774}, 2023.

\bibitem[Adnan et~al.(2016)Adnan, Just, and Baillie]{Adnan2016}
Muhammad Adnan, Mike Just, and Lynne Baillie.
\newblock Investigating time series visualisations to improve the user experience.
\newblock In \emph{Proceedings of the 2016 CHI Conference on Human Factors in Computing Systems}, CHI '16, pages 5444--5455, New York, NY, USA, 2016. ACM.

\bibitem[Aghabozorgi et~al.(2015)Aghabozorgi, Shirkhorshidi, and Wah]{aghabozorgi2015time}
Saeed Aghabozorgi, Ali~Seyed Shirkhorshidi, and Teh~Ying Wah.
\newblock Time-series clustering -- {A} decade review.
\newblock \emph{Information systems}, 53:\penalty0 16--38, 2015.

\bibitem[Aigner et~al.(2012)Aigner, Rind, and Hoffmann]{Aigner2012}
W.~Aigner, A.~Rind, and S.~Hoffmann.
\newblock Comparative evaluation of an interactive time-series visualization that combines quantitative data with qualitative abstractions.
\newblock \emph{Computer Graphics Forum}, 31\penalty0 (3pt2):\penalty0 995--1004, June 2012.
\newblock ISSN 0167-7055.

\bibitem[Aigner et~al.(2007)Aigner, Miksch, M{\"u}ller, Schumann, and Tominski]{Aigner2007}
Wolfgang Aigner, Silvia Miksch, Wolfgang M{\"u}ller, Heidrun Schumann, and Christian Tominski.
\newblock Visualizing time-oriented data--{A} systematic view.
\newblock \emph{Computers \& Graphics}, 31\penalty0 (3):\penalty0 401--409, 2007.

\bibitem[Aigner et~al.(2008)Aigner, Miksch, M{\"u}ller, Schumann, and Tominski]{aigner2008}
Wolfgang Aigner, Silvia Miksch, Wolfgang M{\"u}ller, Heidrun Schumann, and Christian Tominski.
\newblock Visual methods for analyzing time-oriented data.
\newblock \emph{IEEE Transactions on Visualization and Computer Graphics}, 14\penalty0 (1):\penalty0 47--60, 2008.

\bibitem[Aigner et~al.(2011)Aigner, Miksch, Schumann, and Tominski]{Aigner2011}
Wolfgang Aigner, Silvia Miksch, Heidrun Schumann, and Christian Tominski.
\newblock \emph{Visualization of Time-Oriented Data}.
\newblock Springer Publishing Company, Incorporated, New York, NY, USA, 1st edition, 2011.

\bibitem[Albers et~al.(2014)Albers, Correll, and Gleicher]{Albers2014}
Danielle Albers, Michael Correll, and Michael Gleicher.
\newblock Task-driven evaluation of aggregation in time series visualization.
\newblock In \emph{Proceedings of the SIGCHI Conference on Human Factors in Computing Systems}, CHI '14, pages 551--560, New York, NY, USA, 2014. ACM.

\bibitem[Alexandrina et~al.(2019)Alexandrina, Ortigossa, Lui, Gon{\c{c}}alves, Correa, Nonato, and Aguiar]{Alexandrina:2019}
Eduardo~Carlos Alexandrina, Evandro~S Ortigossa, Elaine~Schornobay Lui, Jos{\'e} Ant{\^o}nio~Silveira Gon{\c{c}}alves, Nivaldo~Aparecido Correa, Luis~Gustavo Nonato, and Monica~Lopes Aguiar.
\newblock Analysis and visualization of multidimensional time series: Particulate matter ({PM}10) from {S}{\~a}o {C}arlos-{SP} ({B}razil).
\newblock \emph{Atmospheric Pollution Research}, 10\penalty0 (4):\penalty0 1299--1311, 2019.

\bibitem[Ali et~al.(2019)Ali, Alqahtani, Jones, and Xie]{clustering2019}
Mohammed Ali, Ali Alqahtani, Mark~W. Jones, and Xianghua Xie.
\newblock Clustering and classification for time series data in visual analytics: A survey.
\newblock \emph{IEEE Access}, 7:\penalty0 181314--181338, 2019.

\bibitem[Andrews et~al.(2002)Andrews, Kienreich, Sabol, Becker, Droschl, Kappe, Granitzer, Auer, and Tochtermann]{Andrews2002}
Keith Andrews, Wolfgang Kienreich, Vedran Sabol, Jutta Becker, Georg Droschl, Frank Kappe, Michael Granitzer, Peter Auer, and Klaus Tochtermann.
\newblock The {I}nfo{S}ky visual explorer: {Exploiting} hierarchical structure and document similarities.
\newblock \emph{Information Visualization}, 1\penalty0 (3/4):\penalty0 166--181, December 2002.
\newblock ISSN 1473-8716.

\bibitem[Andrienko et~al.(2010)Andrienko, Andrienko, Demsar, Dransch, Dykes, Fabrikant, Jern, Kraak, Schumann, and Tominski]{Andrienko2010}
Gennady Andrienko, Natalia Andrienko, Urska Demsar, Doris Dransch, Jason Dykes, Sara~Irina Fabrikant, Mikael Jern, Menno-Jan Kraak, Heidrun Schumann, and Christian Tominski.
\newblock Space, time and visual analytics.
\newblock \emph{International Journal of Geographical Information Science}, 24\penalty0 (10):\penalty0 1577--1600, October 2010.
\newblock ISSN 1365-8816.

\bibitem[Andrienko et~al.(2017)Andrienko, Andrienko, Chen, Maciejewski, and Zhao]{andrienko2017visual}
Gennady Andrienko, Natalia Andrienko, Wei Chen, Ross Maciejewski, and Ye~Zhao.
\newblock Visual analytics of mobility and transportation: {State} of the art and further research directions.
\newblock \emph{IEEE Transactions on Intelligent Transportation Systems}, 18\penalty0 (8):\penalty0 2232--2249, 2017.

\bibitem[Andrienko and Andrienko(2013)]{Andrienko2013}
Natalia Andrienko and Gennady Andrienko.
\newblock Visual analytics of movement: {An} overview of methods, tools and procedures.
\newblock \emph{Information Visualization}, 12\penalty0 (1):\penalty0 3--24, January 2013.
\newblock ISSN 1473-8716.

\bibitem[Andrienko and Andrienko(2023)]{andrienko2023s}
Natalia Andrienko and Gennady Andrienko.
\newblock It's about time: {Analytical} time periodization.
\newblock \emph{Computer Graphics Forum}, 42\penalty0 (6):\penalty0 e14845, 2023.

\bibitem[Andrienko et~al.(2003)Andrienko, Andrienko, and Gatalsky]{ANDRIENKO2003}
Natalia Andrienko, Gennady Andrienko, and Peter Gatalsky.
\newblock Exploratory spatio-temporal visualization: an analytical review.
\newblock \emph{Journal of Visual Languages \& Computing}, 14\penalty0 (6):\penalty0 503--541, 2003.
\newblock ISSN 1045-926X.
\newblock Visual Data Mining.

\bibitem[Andrienko et~al.(2018)Andrienko, Lammarsch, Andrienko, Fuchs, Keim, Miksch, and Rind]{Andrienko2018}
Natalia Andrienko, Tim Lammarsch, Gennady Andrienko, Georg Fuchs, Daniel~A. Keim, Silvia Miksch, and Alexander Rind.
\newblock Viewing visual analytics as model building.
\newblock \emph{Computer Graphics Forum}, Early View\penalty0 (0):\penalty0 1--25, 2018.

\bibitem[Andrienko et~al.(2023)Andrienko, Andrienko, and Shirato]{andrienko2023episodes}
Natalia Andrienko, Gennady Andrienko, and Gota Shirato.
\newblock Episodes and topics in multivariate temporal data.
\newblock \emph{Computer Graphics Forum}, 42\penalty0 (6):\penalty0 e14926, 2023.

\bibitem[Arsenault et~al.(2025)Arsenault, Wang, and Patenaude]{xai_financial}
Pierre-Daniel Arsenault, Shengrui Wang, and Jean-Marc Patenaude.
\newblock A survey of explainable artificial intelligence ({XAI}) in financial time series forecasting.
\newblock \emph{ACM Comput. Surv.}, 57\penalty0 (10), May 2025.

\bibitem[Bach et~al.(2017)Bach, Dragicevic, Archambault, Hurter, and Carpendale]{Bach2017}
B.~Bach, P.~Dragicevic, D.~Archambault, C.~Hurter, and S.~Carpendale.
\newblock A descriptive framework for temporal data visualizations based on generalized space-time cubes.
\newblock \emph{Computer Graphics Forum}, 36\penalty0 (6):\penalty0 36--61, September 2017.
\newblock ISSN 0167-7055.

\bibitem[Bach et~al.(2014)Bach, Dragicevic, Archambault, Hurter, and Carpendale]{Bach2014}
Benjamin Bach, Pierre Dragicevic, Daniel Archambault, Christophe Hurter, and Sheelagh Carpendale.
\newblock A review of temporal data visualizations based on space-time cube operations.
\newblock In R.~Borgo, R.~Maciejewski, and I.~Viola, editors, \emph{STAR -- State of The Art Report, Eurographics Conference on Visualization (EuroVis)}, pages 23--41, Swansea, Wales, United Kingdom, 2014. Eurographics Association.

\bibitem[Badam et~al.(2016)Badam, Zhao, Sen, Elmqvist, and Ebert]{Badam2016}
Sriram~Karthik Badam, Jieqiong Zhao, Shivalik Sen, Niklas Elmqvist, and David Ebert.
\newblock Time{F}ork: {Interactive} prediction of time series.
\newblock In \emph{Proceedings of the 2016 CHI Conference on Human Factors in Computing Systems}, CHI '16, pages 5409--5420, New York, NY, USA, 2016. ACM.

\bibitem[Bai et~al.(2020)Bai, Tao, and Lin]{bai2020time}
Zhihui Bai, Yubo Tao, and Hai Lin.
\newblock Time-varying volume visualization: {A} survey.
\newblock \emph{Journal of Visualization}, 23:\penalty0 745--761, 2020.

\bibitem[Baldonado et~al.(2000)Baldonado, Woodruff, and Kuchinsky]{Baldonado2000}
Michelle Q.~Wang Baldonado, Allison Woodruff, and Allan Kuchinsky.
\newblock Guidelines for using multiple views in information visualization.
\newblock In \emph{Proceedings of the Working Conference on Advanced Visual Interfaces}, AVI '00, pages 110--119, New York, NY, USA, 2000. ACM.

\bibitem[Bartolomeo and Hu(2016)]{DiBartolomeo2016}
Marco~Di Bartolomeo and Yifan Hu.
\newblock There is more to streamgraphs than movies: {Better} aesthetics via ordering and lassoing.
\newblock In \emph{Proceedings of the Eurographics / IEEE VGTC Conference on Visualization}, EuroVis '16, pages 341--350, Goslar Germany, Germany, 2016. Eurographics Association.

\bibitem[Becker and Cleveland(1987)]{becker1987}
Richard~A. Becker and William~S. Cleveland.
\newblock Brushing scatterplots.
\newblock \emph{Technometrics}, 29\penalty0 (2):\penalty0 127--142, May 1987.

\bibitem[Bernard et~al.(2024)Bernard, Barth, Cuba, Meier, Peiris, and Shneiderman]{bernard2024ivesa}
J{\"u}rgen Bernard, Clara-Maria Barth, Eduard Cuba, Andrea Meier, Yasara Peiris, and Ben Shneiderman.
\newblock {IVESA} -- visual analysis of time-stamped event sequences.
\newblock \emph{IEEE Transactions on Visualization and Computer Graphics}, 2024.

\bibitem[Bloch et~al.(2008)Bloch, Byron, Carter, and Cox]{Bloch2008}
M.~Bloch, L.~Byron, S.~Carter, and A.~Cox.
\newblock The ebb and flow of movies: {Box} office receipts 1986-2007.
\newblock \url{http://www.nytimes.com/interactive/2008/02/23/movies/20080223_REVENUE_GRAPHIC.html}, 2008.
\newblock Online; acessado em 3 de março de 2018.

\bibitem[Borland and Taylor~II(2007)]{borland2007rainbow}
David Borland and Russell~M Taylor~II.
\newblock Rainbow color map (still) considered harmful.
\newblock \emph{IEEE Computer Graphics and Applications}, 27\penalty0 (2):\penalty0 14--17, 2007.

\bibitem[Bostock et~al.(2011)Bostock, Ogievetsky, and Heer]{Bostock2011}
Michael Bostock, Vadim Ogievetsky, and Jeffrey Heer.
\newblock D3: {Data}-driven documents.
\newblock \emph{IEEE Transactions on Visualization and Computer Graphics}, 17\penalty0 (12):\penalty0 2301--2309, December 2011.
\newblock ISSN 1077-2626.

\bibitem[Boyandin et~al.(2011)Boyandin, Bertini, Bak, and Lalanne]{Boyandin2011}
Ilya Boyandin, Enrico Bertini, Peter Bak, and Denis Lalanne.
\newblock Flowstrates: {An} approach for visual exploration of temporal origin-destination data.
\newblock In \emph{Proceedings of the 13th Eurographics / IEEE - VGTC Conference on Visualization}, EuroVis '11, pages 971--980, Chichester, UK, 2011. The Eurographs Association and John Wiley \& Sons, Ltd.

\bibitem[Burch and Weiskopf(2011)]{Burch2011}
Michael Burch and Daniel Weiskopf.
\newblock Visualizing dynamic quantitative data in hierarchies--{TimeEdgeTrees}: {Attaching} dynamic weights to tree edges.
\newblock In \emph{Proceedings of International Conference on Visualization Theory and Applications}, pages 177--186, New York, NY, USA, 2011. Springer.

\bibitem[Burch et~al.(2008)Burch, Beck, and Diehl]{Burch2008}
Michael Burch, Fabian Beck, and Stephan Diehl.
\newblock Timeline trees: {Visualizing} sequences of transactions in information hierarchies.
\newblock In \emph{Proceedings of the Working Conference on Advanced Visual Interfaces}, AVI '08, pages 75--82, New York, NY, USA, 2008. ACM.

\bibitem[Byron and Wattenberg(2008)]{byron2008}
Lee Byron and Martin Wattenberg.
\newblock Stacked graphs -- {Geometry} \& aesthetics.
\newblock \emph{IEEE Transactions on Visualization and Computer Graphics}, 14\penalty0 (6):\penalty0 1245--1252, November 2008.
\newblock ISSN 1077-2626.

\bibitem[Cao et~al.(2012)Cao, Lin, Sun, Lazer, Liu, and Qu]{Cao2012}
N.~Cao, Y.~R. Lin, X.~Sun, D.~Lazer, S.~Liu, and H.~Qu.
\newblock Whisper: {Tracing} the spatiotemporal process of information diffusion in real time.
\newblock \emph{IEEE Transactions on Visualization and Computer Graphics}, 18\penalty0 (12):\penalty0 2649--2658, December 2012.
\newblock ISSN 1077-2626.

\bibitem[Cao et~al.(2018)Cao, Wang, Li, Zhou, Li, and Li]{cao2018brits}
Wei Cao, Dong Wang, Jian Li, Hao Zhou, Lei Li, and Yitan Li.
\newblock {BRITS}: Bidirectional recurrent imputation for time series.
\newblock \emph{Advances in neural information processing systems}, 31, 2018.

\bibitem[Card et~al.(1999)Card, Mackinlay, and Shneiderman]{Card1999}
Stuart~K. Card, Jock~D. Mackinlay, and Ben Shneiderman, editors.
\newblock \emph{Readings in Information Visualization: {Using} Vision to Think}.
\newblock Morgan Kaufmann Publishers Inc., San Francisco, CA, USA, 1st edition, 1999.

\bibitem[Caruana et~al.(2015)Caruana, Lou, Gehrke, Koch, Sturm, and Elhadad]{caruana2015intelligible}
Rich Caruana, Yin Lou, Johannes Gehrke, Paul Koch, Marc Sturm, and Noemie Elhadad.
\newblock Intelligible models for healthcare: {Predicting} pneumonia risk and hospital 30-day readmission.
\newblock In \emph{Proceedings of the 21th ACM SIGKDD International Conference on Knowledge Discovery and Data Mining}, KDD '15, page 1721–1730, New York, NY, USA, 2015. Association for Computing Machinery.

\bibitem[Cazelles et~al.(2007)Cazelles, Chavez, Magny, Gu{\'e}gan, and Hales]{cazelles2007time}
Bernard Cazelles, Mario Chavez, Guillaume Constantin~de Magny, Jean-Francois Gu{\'e}gan, and Simon Hales.
\newblock Time-dependent spectral analysis of epidemiological time-series with wavelets.
\newblock \emph{Journal of the Royal Society Interface}, 4\penalty0 (15):\penalty0 625--636, 2007.

\bibitem[Chalmers(1996)]{Chalmers1996}
Matthew Chalmers.
\newblock A linear iteration time layout algorithm for visualising high-dimensional data.
\newblock In \emph{Proceedings of the 7th Conference on Visualization '96}, VIS '96, pages 127--ff., Los Alamitos, CA, USA, 1996. IEEE Computer Society Press.

\bibitem[Chan et~al.(2025)Chan, Nonato, Palpanas, Silva, and Freire]{chan2025tivy}
Gromit Yeuk-Yin Chan, Luis~Gustavo Nonato, Themis Palpanas, Cl{\'a}udio~T Silva, and Juliana Freire.
\newblock Tivy: Time series visual summary for scalable visualization.
\newblock \emph{arXiv preprint arXiv:2507.18972}, 2025.

\bibitem[Chang et~al.(2007{\natexlab{a}})Chang, Nesbitt, and Wilkins]{ChangD2007}
Dempsey Chang, Keith~V. Nesbitt, and Kevin Wilkins.
\newblock The {G}estalt principles of similarity and proximity apply to both the haptic and visual grouping of elements.
\newblock In \emph{Proceedings of the Eight Australasian Conference on User Interface - Volume 64}, AUIC '07, pages 79--86, Darlinghurst, Australia, Australia, 2007{\natexlab{a}}. Australian Computer Society, Inc.

\bibitem[Chang et~al.(2007{\natexlab{b}})Chang, Ghoniem, Kosara, Ribarsky, Yang, Suma, Ziemkiewicz, Kern, and Sudjianto]{chang2007}
Remco Chang, Mohammad Ghoniem, Robert Kosara, William Ribarsky, Jing Yang, Evan Suma, Caroline Ziemkiewicz, Daniel Kern, and Agus Sudjianto.
\newblock Wire{V}is: {Visualization} of categorical, time-varying data from financial transactions.
\newblock In \emph{Proceedings of the 2007 IEEE Symposium on Visual Analytics Science and Technology}, VAST '07, pages 155--162, Washington, DC, USA, 2007{\natexlab{b}}. IEEE Computer Society.

\bibitem[Chen et~al.(2017)Chen, Lin, and Yuan]{Chen2017}
Siming Chen, Lijing Lin, and Xiaoru Yuan.
\newblock Social media visual analytics.
\newblock \emph{Computer Graphics Forum}, 36\penalty0 (3):\penalty0 563--587, June 2017.
\newblock ISSN 0167-7055.

\bibitem[Chen and Guestrin(2016)]{XGBoost}
Tianqi Chen and Carlos Guestrin.
\newblock {XGBoost}: A scalable tree boosting system.
\newblock In \emph{Proceedings of the 22nd ACM SIGKDD International Conference on Knowledge Discovery and Data Mining}, KDD '16, pages 785--794, New York, NY, USA, 2016. Association for Computing Machinery.

\bibitem[Chu et~al.(2023)Chu, Ai, Wen, Shi, Ma, and Feng]{chu2023spatio}
Lele Chu, Bo~Ai, Yubo Wen, Qingtong Shi, Huadong Ma, and Wenjun Feng.
\newblock A spatio-temporal dynamic visualization method of time-varying wind fields based on particle system.
\newblock \emph{ISPRS International Journal of Geo-Information}, 12\penalty0 (4):\penalty0 146, 2023.

\bibitem[Cleveland and McGill(1984)]{Cleveland1984}
William~S. Cleveland and Robert McGill.
\newblock Graphical perception: {Theory}, experimentation, and application to the development of graphical methods.
\newblock \emph{Journal of the American Statistical Association}, 79\penalty0 (387):\penalty0 531--554, 1984.

\bibitem[Cockburn et~al.(2009)Cockburn, Karlson, and Bederson]{cockburn2009review}
Andy Cockburn, Amy Karlson, and Benjamin~B Bederson.
\newblock A review of overview+detail, zooming, and focus+context interfaces.
\newblock \emph{ACM Computing Surveys (CSUR)}, 41\penalty0 (1):\penalty0 1--31, 2009.

\bibitem[Costa et~al.(2017)Costa, Nichols, and Smith]{costa2017studying}
Lilia Costa, Thomas Nichols, and Jim~Q Smith.
\newblock Studying the effective brain connectivity using multiregression dynamic models.
\newblock \emph{Brazilian Journal of Probability and Statistics}, 31\penalty0 (4):\penalty0 765--800, 2017.

\bibitem[Costa et~al.(2024)Costa, Anacleto, Nascimento, Smith, Queen, Louzada, and Nichols]{costa2024evaluating}
Lilia Costa, Osvaldo Anacleto, Diego~C Nascimento, James~Q Smith, Catriona~M Queen, Francisco Louzada, and Thomas Nichols.
\newblock Evaluating brain group structure methods using hierarchical dynamic models.
\newblock \emph{Pattern Recognition}, 155:\penalty0 110687, 2024.

\bibitem[Cowan(2010)]{cowan2010magical}
Nelson Cowan.
\newblock The magical mystery four: {How} is working memory capacity limited, and why?
\newblock \emph{Current Directions in Psychological Science}, 19\penalty0 (1):\penalty0 51--57, 2010.

\bibitem[Craft and Cairns(2005)]{Craft2005}
Brock Craft and Paul Cairns.
\newblock Beyond guidelines: {What} can we learn from the visual information seeking mantra?
\newblock In \emph{Proceedings of the Ninth International Conference on Information Visualisation}, IV '05, pages 110--118, Washington, DC, USA, 2005. IEEE Computer Society.

\bibitem[Cuba(2015)]{cuba2015research}
Nicholas Cuba.
\newblock Research note: {Sankey} diagrams for visualizing land cover dynamics.
\newblock \emph{Landscape and Urban Planning}, 139:\penalty0 163--167, 2015.

\bibitem[Cui et~al.(2014)Cui, Liu, Wu, and Wei]{Cui2014}
Weiwei Cui, Shixia Liu, Zhuofeng Wu, and Hao Wei.
\newblock How hierarchical topics evolve in large text corpora.
\newblock \emph{IEEE Transactions on Visualization and Computer Graphics}, 20\penalty0 (12):\penalty0 2281--2290, December 2014.
\newblock ISSN 1077-2626.

\bibitem[Cui(2019)]{cui2019visual}
Wenqiang Cui.
\newblock Visual analytics: {A} comprehensive overview.
\newblock \emph{IEEE Access}, 7:\penalty0 81555--81573, 2019.

\bibitem[Dang et~al.(2016)Dang, Pendar, and Forbes]{Dang2016}
T.~N. Dang, N.~Pendar, and A.~G. Forbes.
\newblock Time{A}rcs: {Visualizing} fluctuations in dynamic networks.
\newblock In \emph{Proceedings of the Eurographics / IEEE VGTC Conference on Visualization}, EuroVis '16, pages 61--69, Goslar Germany, Germany, 2016. Eurographics Association.

\bibitem[De~Livera et~al.(2011)De~Livera, Hyndman, and Snyder]{de2011forecasting}
Alysha~M De~Livera, Rob~J Hyndman, and Ralph~D Snyder.
\newblock Forecasting time series with complex seasonal patterns using exponential smoothing.
\newblock \emph{Journal of the American statistical association}, 106\penalty0 (496):\penalty0 1513--1527, 2011.

\bibitem[Donders et~al.(2006)Donders, Van Der~Heijden, Stijnen, and Moons]{donders2006gentle}
A~Rogier~T Donders, Geert~JMG Van Der~Heijden, Theo Stijnen, and Karel~GM Moons.
\newblock A gentle introduction to imputation of missing values.
\newblock \emph{Journal of clinical epidemiology}, 59\penalty0 (10):\penalty0 1087--1091, 2006.

\bibitem[Dou et~al.(2013)Dou, Yu, Wang, Ma, and Ribarsky]{Dou2013}
Wenwen Dou, Li~Yu, Xiaoyu Wang, Zhiqiang Ma, and William Ribarsky.
\newblock Hierarchical{T}opics: {Visually} exploring large text collections using topic hierarchies.
\newblock \emph{IEEE Transactions on Visualization and Computer Graphics}, 19\penalty0 (12):\penalty0 2002--2011, December 2013.
\newblock ISSN 1077-2626.

\bibitem[Eick and Karr(2002)]{Eick2002}
S.~G. Eick and A.~F Karr.
\newblock Visual scalability.
\newblock \emph{Journal of Computational and Graphical Statistics}, 11\penalty0 (1):\penalty0 22--43, 2002.

\bibitem[Elmqvist et~al.(2011)Elmqvist, Vande~Moere, Jetter, Cernea, Reiterer, and Jankun-Kelly]{Elmqvist2011}
Niklas Elmqvist, Andrew Vande~Moere, Hans-Christian Jetter, Daniel Cernea, Harald Reiterer, and T.~J. Jankun-Kelly.
\newblock Fluid interaction for information visualization.
\newblock \emph{Information Visualization}, 10\penalty0 (4):\penalty0 327--340, October 2011.
\newblock ISSN 1473-8716.

\bibitem[Esling and Agon(2012)]{esling2012time}
Philippe Esling and Carlos Agon.
\newblock Time-series data mining.
\newblock \emph{ACM Computing Surveys (CSUR)}, 45\penalty0 (1):\penalty0 1--34, 2012.

\bibitem[Facebook(2023)]{prophet}
Facebook.
\newblock Prophet: Automatic forecasting procedure.
\newblock \url{https://github.com/facebook/prophet}, 2023.
\newblock Online, visited on June 2025.

\bibitem[Fang et~al.(2020)Fang, Xu, and Jiang]{Fang2020}
Yujie Fang, Hui Xu, and Jie Jiang.
\newblock A survey of time series data visualization research.
\newblock \emph{IOP Conference Series: Materials Science and Engineering}, 782\penalty0 (2):\penalty0 022013, mar 2020.

\bibitem[Federico et~al.(2014)Federico, Hoffmann, Rind, Aigner, and Miksch]{Federico2014}
Paolo Federico, Stephan Hoffmann, Alexander Rind, Wolfgang Aigner, and Silvia Miksch.
\newblock Qualizon graphs: {Space}-efficient time-series visualization with qualitative abstractions.
\newblock In \emph{Proceedings of the 2014 International Working Conference on Advanced Visual Interfaces}, AVI '14, pages 273--280, New York, NY, USA, 2014. ACM.

\bibitem[Feng et~al.(2010)Feng, Kwock, Lee, and Taylor]{feng2010matching}
David Feng, Lester Kwock, Yueh Lee, and Russell Taylor.
\newblock Matching visual saliency to confidence in plots of uncertain data.
\newblock \emph{IEEE Transactions on Visualization and Computer Graphics}, 16\penalty0 (6):\penalty0 980--989, 2010.

\bibitem[Fonseca et~al.(2021)Fonseca, Ferreira, Nascimento, Fiaccone, Ulloa-Correa, Garcia-Pina, and Louzada]{fonseca2021water}
Anderson Fonseca, Paulo~Henrique Ferreira, Diego Carvalho~do Nascimento, Rosemeire Fiaccone, Christopher Ulloa-Correa, Ayon Garcia-Pina, and Francisco Louzada.
\newblock Water particles monitoring in the atacama desert: Spc approach based on proportional data.
\newblock \emph{Axioms}, 10\penalty0 (3):\penalty0 154, 2021.

\bibitem[Fu(2011)]{Fu2011}
Tak-chung Fu.
\newblock A review on time series data mining.
\newblock \emph{Engineering Applications of Artificial Intelligence}, 24\penalty0 (1):\penalty0 164--181, February 2011.
\newblock ISSN 0952-1976.

\bibitem[Fu et~al.(2005)Fu, Chung, Lam, Luk, and Ng]{Fu2005}
Tak-chung Fu, Fu-lai Chung, Chun-fai Lam, Robert Luk, and Chak-man Ng.
\newblock Adaptive data delivery framework for financial time series visualization.
\newblock In \emph{Proceedings of the 2005 International Conference on Mobile Business}, ICMB '05, pages 267--273, Washington, DC, USA, 2005. IEEE Computer Society.

\bibitem[Fua et~al.(1999)Fua, Ward, and Rundensteiner]{Fua1999}
Ying-Huey Fua, Matthew~O. Ward, and Elke~A. Rundensteiner.
\newblock Hierarchical parallel coordinates for exploration of large datasets.
\newblock In \emph{Proceedings of the Conference on Visualization '99: Celebrating Ten Years}, VIS '99, pages 43--50, Los Alamitos, CA, USA, 1999. IEEE Computer Society Press.
\newblock ISBN 0-7803-5897-X.

\bibitem[Furnas(2006)]{furnas2006fisheye}
George~W Furnas.
\newblock A fisheye follow-up: {Further} reflections on focus + context.
\newblock In \emph{Proceedings of the SIGCHI Conference on Human Factors in Computing Systems}, pages 999--1008, 2006.

\bibitem[Gilch and M{\"u}ller(2018)]{gilch2018elo}
Lorenz~A Gilch and Sebastian M{\"u}ller.
\newblock On {Elo} based prediction models for the {FIFA} worldcup 2018.
\newblock \emph{Preprint arXiv:1806.01930}, 2018.

\bibitem[Goethem et~al.(2017)Goethem, Staals, Loffler, Dykes, and Speckmann]{Goethem2017}
Arthur~Van Goethem, Frank Staals, Maarten Loffler, Jason Dykes, and Bettina Speckmann.
\newblock Multi-granular trend detection for time-series analysis.
\newblock \emph{IEEE Transactions on Visualization and Computer Graphics}, 23\penalty0 (1):\penalty0 661--670, 1 2017.
\newblock ISSN 1077-2626.

\bibitem[Gogolou et~al.(2018)Gogolou, Tsandilas, Palpanas, and Bezerianos]{gogolou2018comparing}
Anna Gogolou, Theophanis Tsandilas, Themis Palpanas, and Anastasia Bezerianos.
\newblock Comparing similarity perception in time series visualizations.
\newblock \emph{IEEE Transactions on Visualization and Computer Graphics}, 25\penalty0 (1):\penalty0 523--533, 2018.

\bibitem[Gomi and Itoh(2010)]{Gomi2010}
Ai~Gomi and Takayuki Itoh.
\newblock {MIAOW}: {A} 3{D} image browser applying a location- and time-based hierarchical data visualization technique.
\newblock In \emph{Proceedings of the International Conference on Advanced Visual Interfaces}, AVI '10, pages 225--232, New York, NY, USA, 2010. ACM.

\bibitem[Gong et~al.(2021)Gong, Chung, and Glass]{gong2021ast}
Yuan Gong, Yu-An Chung, and James Glass.
\newblock {AST}: Audio spectrogram transformer.
\newblock \emph{Preprint arXiv:2104.01778}, 2021.

\bibitem[Grattafiori et~al.(2024)Grattafiori, Dubey, Jauhri, Pandey, Kadian, Al-Dahle, Letman, Mathur, Schelten, Vaughan, et~al.]{grattafiori2024llama}
Aaron Grattafiori, Abhimanyu Dubey, Abhinav Jauhri, Abhinav Pandey, Abhishek Kadian, Ahmad Al-Dahle, Aiesha Letman, Akhil Mathur, Alan Schelten, Alex Vaughan, et~al.
\newblock The llama 3 herd of models.
\newblock \emph{Preprint arXiv:2407.21783}, 2024.

\bibitem[Green et~al.(2009)Green, Ribarsky, and Fisher]{Green2009}
Tera~Marie Green, William Ribarsky, and Brian Fisher.
\newblock Building and applying a human cognition model for visual analytics.
\newblock \emph{Information Visualization}, 8\penalty0 (1):\penalty0 1--13, January 2009.
\newblock ISSN 1473-8716.

\bibitem[Grinstein et~al.(2001)Grinstein, Trutschl, and Cvek]{Grinstein2001}
Georges Grinstein, Marjan Trutschl, and Urska Cvek.
\newblock High-dimensional visualizations.
\newblock In \emph{Proceedings of the 7th Data Mining Conference}, KDD Workshop, pages 7--19, San Francisco, California, USA, 2001. ACM.

\bibitem[Gruendl et~al.(2016)Gruendl, Riehmann, Pausch, and Froehlich]{Gruendl2016}
Henning Gruendl, Patrick Riehmann, Yves Pausch, and Bernd Froehlich.
\newblock Time-series plots integrated in parallel-coordinates displays.
\newblock In \emph{Proceedings of the Eurographics / IEEE VGTC Conference on Visualization}, EuroVis '16, pages 321--330, Goslar Germany, Germany, 2016. Eurographics Association.

\bibitem[Hales(2012)]{Hales2012}
Wesley Hales.
\newblock \emph{HTML5 and JavaScript Web Apps}.
\newblock O'Reilly Media, Inc., Sebastopol, CA, EUA, 2012.

\bibitem[Hao et~al.(2007)Hao, Dayal, Keim, and Schreck]{Hao2007}
Ming Hao, Umeshwar Dayal, Daniel Keim, and Tobias Schreck.
\newblock Multi-resolution techniques for visual exploration of large time-series data.
\newblock In \emph{Proceedings of the 9th Joint Eurographics / IEEE VGTC Conference on Visualization}, EUROVIS'07, pages 27--34, Aire-la-Ville, Switzerland, Switzerland, 2007. Eurographics Association.

\bibitem[Hao et~al.(2005)Hao, Dayal, Keim, and Schreck]{Hao2005}
Ming~C. Hao, Umeshwar Dayal, Daniel~A. Keim, and Tobias Schreck.
\newblock Importance-driven visualization layouts for large time series data.
\newblock In \emph{Proceedings of the Proceedings of the 2005 IEEE Symposium on Information Visualization}, INFOVIS '05, pages 27--, Washington, DC, USA, 2005. IEEE Computer Society.

\bibitem[Harrower and Brewer(2003)]{harrower2003colorbrewer}
Mark Harrower and Cynthia~A Brewer.
\newblock {ColorBrewer.org: An} online tool for selecting colour schemes for maps.
\newblock \emph{The Cartographic Journal}, 40\penalty0 (1):\penalty0 27--37, 2003.

\bibitem[Havre et~al.(2000)Havre, Hetzler, and Nowell]{Havre2000}
Susan Havre, Beth Hetzler, and Lucy Nowell.
\newblock Theme{R}iver: {Visualizing} theme changes over time.
\newblock In \emph{Proceedings of the IEEE Symposium on Information Visualization}, INFOVIS '00, pages 115--, Washington, DC, USA, 2000. IEEE Computer Society.

\bibitem[Havre et~al.(2002)Havre, Hetzler, Whitney, and Nowell]{Havre2002}
Susan Havre, Elizabeth Hetzler, Paul Whitney, and Lucy Nowell.
\newblock Theme{R}iver: {Visualizing} thematic changes in large document collections.
\newblock \emph{IEEE Transactions on Visualization and Computer Graphics}, 8\penalty0 (1):\penalty0 9--20, January 2002.
\newblock ISSN 1077-2626.

\bibitem[Heer and Robertson(2007)]{Heer2007}
Jeffrey Heer and George Robertson.
\newblock Animated transitions in statistical data graphics.
\newblock \emph{IEEE Transactions on Visualization and Computer Graphics}, 13\penalty0 (6):\penalty0 1240--1247, November 2007.
\newblock ISSN 1077-2626.

\bibitem[Heer and Shneiderman(2012)]{Heer2012}
Jeffrey Heer and Ben Shneiderman.
\newblock Interactive dynamics for visual analysis.
\newblock \emph{Communications of the ACM}, 55\penalty0 (4):\penalty0 45--54, April 2012.
\newblock ISSN 0001-0782.

\bibitem[Heer et~al.(2009)Heer, Kong, and Agrawala]{Heer2009}
Jeffrey Heer, Nicholas Kong, and Maneesh Agrawala.
\newblock Sizing the horizon: {The} effects of chart size and layering on the graphical perception of time series visualizations.
\newblock In \emph{Proceedings of the SIGCHI Conference on Human Factors in Computing Systems}, CHI '09, pages 1303--1312, New York, NY, USA, 2009. ACM.

\bibitem[Heer et~al.(2010)Heer, Bostock, and Ogievetsky]{Heer2010}
Jeffrey Heer, Michael Bostock, and Vadim Ogievetsky.
\newblock A tour through the visualization zoo.
\newblock \emph{Communications of the ACM}, 53\penalty0 (6):\penalty0 59--67, June 2010.
\newblock ISSN 0001-0782.

\bibitem[Heimerl et~al.(2014)Heimerl, Lohmann, Lange, and Ertl]{heimerl2014word}
Florian Heimerl, Steffen Lohmann, Simon Lange, and Thomas Ertl.
\newblock Word cloud explorer: {Text} analytics based on word clouds.
\newblock In \emph{2014 47th Hawaii International Conference on System Science}, pages 1833--1842. IEEE, 2014.

\bibitem[Heinrich and Weiskopf(2013)]{heinrich2013state}
Julian Heinrich and Daniel Weiskopf.
\newblock State of the art of parallel coordinates.
\newblock \emph{Eurographics State-of-the-Art Reports (STARs)}, pages 95--116, 2013.

\bibitem[Herman et~al.(2000)Herman, Melancon, and Marshall]{Herman2000}
Ivan Herman, Guy Melancon, and M.~Scott Marshall.
\newblock Graph visualization and navigation in information visualization: {A} survey.
\newblock \emph{IEEE Transactions on Visualization and Computer Graphics}, 6\penalty0 (1):\penalty0 24--43, January 2000.
\newblock ISSN 1077-2626.

\bibitem[Hoffman(1998)]{hoffman1998}
D.~D. Hoffman.
\newblock \emph{Visual Intelligence: {How} We Create What We See}.
\newblock W. W. Norton and Company, Inc., New York, NY, USA, 1st edition, 1998.

\bibitem[Hullman(2019)]{hullman2019authors}
Jessica Hullman.
\newblock Why authors don't visualize uncertainty.
\newblock \emph{IEEE Transactions on Visualization and Computer Graphics}, 26\penalty0 (1):\penalty0 130--139, 2019.

\bibitem[Hyndman and Athanasopoulos(2018)]{hyndman2018forecasting}
Rob~J Hyndman and George Athanasopoulos.
\newblock \emph{Forecasting: principles and practice}.
\newblock OTexts, 2018.

\bibitem[Inselberg and Dimsdale(1990)]{Inselberg1990}
Alfred Inselberg and Bernard Dimsdale.
\newblock Parallel coordinates: {A} tool for visualizing multi-dimensional geometry.
\newblock In \emph{Proceedings of the 1st Conference on Visualization}, VIS '90, pages 361--378, Los Alamitos, CA, USA, 1990. IEEE Computer Society Press.

\bibitem[Iram et~al.(2019)Iram, Fernando, and Hill]{iram2019connecting}
Shamaila Iram, Terrence Fernando, and Richard Hill.
\newblock Connecting to smart cities: {Analyzing} energy times series to visualize monthly electricity peak load in residential buildings.
\newblock In \emph{Proceedings of the Future Technologies Conference (FTC) 2018: Volume 1}, pages 333--342. Springer, 2019.

\bibitem[Itoh et~al.(2010)Itoh, Toyoda, and Kitsuregawa]{Itoh2010}
Masahiko Itoh, Masashi Toyoda, and Masaru Kitsuregawa.
\newblock An interactive visualization framework for time-series of web graphs in a 3{D} environment.
\newblock In \emph{Proceedings of the 2010 14th International Conference Information Visualisation}, IV '10, pages 54--60, Washington, DC, USA, 2010. IEEE Computer Society.

\bibitem[Itoh et~al.(2024)Itoh, Nakano, Fukayama, Hamasaki, and Goto]{itoh2024singdistvis}
Takayuki Itoh, Tomoyasu Nakano, Satoru Fukayama, Masahiro Hamasaki, and Masataka Goto.
\newblock {SingDistVis}: interactive overview+ detail visualization for {F0} trajectories of numerous singers singing the same song.
\newblock \emph{Multimedia Tools and Applications}, pages 1--21, 2024.

\bibitem[Javed and Elmqvist(2010)]{Javed2010}
Waqas Javed and Niklas Elmqvist.
\newblock Stack zooming for multi-focus interaction in time-series data visualization.
\newblock In \emph{2010 IEEE Pacific Visualization Symposium (PacificVis)}, pages 33--40, Taipei, Taiwan, 2010. IEEE Computer Society.

\bibitem[Javed et~al.(2010)Javed, McDonnel, and Elmqvist]{Javed2010Perception}
Waqas Javed, Bryan McDonnel, and Niklas Elmqvist.
\newblock Graphical perception of multiple time series.
\newblock \emph{IEEE Transactions on Visualization and Computer Graphics}, 16\penalty0 (6):\penalty0 927--934, November 2010.
\newblock ISSN 1077-2626.

\bibitem[Javed et~al.(2012)Javed, Ghani, and Elmqvist]{Javed2012}
Waqas Javed, Sohaib Ghani, and Niklas Elmqvist.
\newblock Polyzoom: {Multiscale} and multifocus exploration in 2{D} visual spaces.
\newblock In \emph{Proceedings of the SIGCHI Conference on Human Factors in Computing Systems}, CHI '12, pages 287--296, New York, NY, USA, 2012. ACM.

\bibitem[Johnson and Shneiderman(1991)]{Johnson1991}
Brian Johnson and Ben Shneiderman.
\newblock Tree-{M}aps: {A} space-filling approach to the visualization of hierarchical information structures.
\newblock In \emph{Proceedings of the 2nd Conference on Visualization '91}, VIS '91, pages 284--291, Los Alamitos, CA, USA, 1991. IEEE Computer Society Press.

\bibitem[Jugel et~al.(2014)Jugel, Jerzak, Hackenbroich, and Markl]{jugel2014m4}
Uwe Jugel, Zbigniew Jerzak, Gregor Hackenbroich, and Volker Markl.
\newblock M4: a visualization-oriented time series data aggregation.
\newblock \emph{Proceedings of the VLDB Endowment}, 7\penalty0 (10):\penalty0 797--808, 2014.

\bibitem[kaggle.com(2017{\natexlab{a}})]{seattle_data}
kaggle.com.
\newblock Did it rain in {Seattle}? (1948-2017).
\newblock \url{https://www.kaggle.com/datasets/rtatman/did-it-rain-in-seattle-19482017}, 2017{\natexlab{a}}.
\newblock Online, visited on June 2024.

\bibitem[kaggle.com(2017{\natexlab{b}})]{usnames_data}
kaggle.com.
\newblock {US} baby names.
\newblock \url{https://www.kaggle.com/datasets/kaggle/us-baby-names}, 2017{\natexlab{b}}.
\newblock Online, visited on June 2024.

\bibitem[kaggle.com(2022)]{fifa_data}
kaggle.com.
\newblock {FIFA} world cup 2022: {Complete} dataset.
\newblock \url{https://www.kaggle.com/datasets/die9origephit/fifa-world-cup-2022-complete-dataset}, 2022.
\newblock Online, visited on June 2024.

\bibitem[Kehrer and Hauser(2013)]{Kehrer2013}
Johannes Kehrer and Helwig Hauser.
\newblock Visualization and visual analysis of multifaceted scientific data: {A} survey.
\newblock \emph{IEEE Transactions on Visualization and Computer Graphics}, 19\penalty0 (3):\penalty0 495--513, March 2013.
\newblock ISSN 1077-2626.

\bibitem[Keim(2002)]{Keim:2002}
Daniel~A Keim.
\newblock Information visualization and visual data mining.
\newblock \emph{IEEE Transactions on Visualization and Computer Graphics}, 8\penalty0 (1):\penalty0 1--8, 2002.

\bibitem[Keogh et~al.(2002)Keogh, Lonardi, and Chiu]{keogh2002finding}
Eamonn Keogh, Stefano Lonardi, and Bill'Yuan-chi' Chiu.
\newblock Finding surprising patterns in a time series database in linear time and space.
\newblock In \emph{Proceedings of the eighth ACM SIGKDD international conference on Knowledge discovery and data mining}, pages 550--556, 2002.

\bibitem[Ko et~al.(2016)Ko, Chor, Afzal, Yau, Chae, Malik, Beck, Jang, Ribarsky, and Eber]{Ko2016}
S.~Ko, I.~Chor, S.~Afzal, C.~Yau, J.~Chae, A.~Malik, K.~Beck, Y.~Jang, W.~Ribarsky, and D.~S. Eber.
\newblock A survey on visual analysis approaches for financial data.
\newblock In \emph{Proceedings of the Eurographics / IEEE VGTC Conference on Visualization: State of the Art Reports}, EuroVis '16, pages 599--617, Goslar, Germany, 2016. Eurographics Association.

\bibitem[Konyha et~al.(2012)Konyha, Lez, Matković, Jelović, and Hauser]{Konyha2012}
Zoltan Konyha, Alan Lez, Kresimir Matković, Mario Jelović, and Helwig Hauser.
\newblock Interactive visual analysis of families of curves using data aggregation and derivation.
\newblock In \emph{Proceedings of the 12th International Conference on Knowledge Management and Knowledge Technologies}, i-KNOW '12, pages 24:1--24:8, New York, NY, USA, 2012. ACM.
\newblock ISBN 978-1-4503-1242-4.

\bibitem[Krstajic et~al.(2011)Krstajic, Bertini, and Keim]{Milos2011}
Milos Krstajic, Enrico Bertini, and Daniel~A. Keim.
\newblock Cloud{L}ines: {Compact} display of event episodes in multiple time-series.
\newblock \emph{IEEE Transactions on Visualization and Computer Graphics}, 17:\penalty0 2432--2439, 2011.

\bibitem[Kumar et~al.(2005)Kumar, Lolla, Keogh, Lonardi, Chotirat, and Wei]{Kumar2005}
Nitin Kumar, Venkata~Nishanth Lolla, Eamonn Keogh, Stefano Lonardi, Ann~Ratanamahatana Chotirat, and Li~Wei.
\newblock Time-series bitmaps: {A} practical visualization tool for working with large time series databases.
\newblock In \emph{Proceedings of the 2005 SIAM International Conference on Data Mining}, pages 531--535, Philadelphia, PA, USA, 2005. SIAM: Society for Industrial and Applied Mathematics.

\bibitem[Köthur et~al.(2015)Köthur, Witt, Sips, Marwan, Schinkel, and Dransch]{Kothur2015}
P.~Köthur, C.~Witt, M.~Sips, N.~Marwan, S.~Schinkel, and D.~Dransch.
\newblock Visual analytics for correlation-based comparison of time series ensembles.
\newblock \emph{Computer Graphics Forum}, 34\penalty0 (3):\penalty0 411--420, June 2015.
\newblock ISSN 0167-7055.

\bibitem[Lamping et~al.(1995)Lamping, Rao, and Pirolli]{Lamping1995}
John Lamping, Ramana Rao, and Peter Pirolli.
\newblock A focus+context technique based on hyperbolic geometry for visualizing large hierarchies.
\newblock In \emph{Proceedings of the SIGCHI Conference on Human Factors in Computing Systems}, CHI '95, pages 401--408, New York, NY, USA, 1995. ACM Press/Addison-Wesley Publishing Co.

\bibitem[Lei and Zhang(2010)]{Lei2010}
Su~Te Lei and Kang Zhang.
\newblock A visual analytics system for financial time-series data.
\newblock In \emph{Proceedings of the 3rd International Symposium on Visual Information Communication}, VINCI '10, pages 20:1--20:9, New York, NY, USA, 2010. ACM.
\newblock ISBN 978-1-4503-0436-8.

\bibitem[Lepot et~al.(2017)Lepot, Aubin, and Clemens]{lepot2017interpolation}
Mathieu Lepot, Jean-Baptiste Aubin, and Fran{\c{c}}ois~HLR Clemens.
\newblock Interpolation in time series: An introductive overview of existing methods, their performance criteria and uncertainty assessment.
\newblock \emph{Water}, 9\penalty0 (10):\penalty0 796, 2017.

\bibitem[Li et~al.(2013)Li, Jiang, Liu, and Meng]{Li2013}
Ning Li, Zhifang Jiang, Zixiang Liu, and Xiangxu Meng.
\newblock A method of hierarchical time-series data visualization.
\newblock In \emph{Proceedings of the 6th International Symposium on Visual Information Communication and Interaction}, VINCI '13, pages 113--114, New York, NY, USA, 2013. ACM.

\bibitem[Li et~al.(2014)Li, Jiang, Liu, and Sun]{Li2014}
Ning Li, Zhifang Jiang, Zixiang Liu, and Haoxin Sun.
\newblock Hyperbolic tree + time disc: {Visualizing} hierarchical time-series data.
\newblock In \emph{Proceedings of the 7th International Symposium on Visual Information Communication and Interaction}, VINCI '14, pages 188--191, New York, NY, USA, 2014. ACM.

\bibitem[Lim et~al.(2021)Lim, Ar{\i}k, Loeff, and Pfister]{lim2021temporal}
Bryan Lim, Sercan~{\"O} Ar{\i}k, Nicolas Loeff, and Tomas Pfister.
\newblock Temporal fusion transformers for interpretable multi-horizon time series forecasting.
\newblock \emph{International Journal of Forecasting}, 37\penalty0 (4):\penalty0 1748--1764, 2021.

\bibitem[Lin et~al.(2004)Lin, Keogh, Lonardi, Lankford, and Nystrom]{Lin2004}
Jessica Lin, Eamonn Keogh, Stefano Lonardi, Jeffrey~P. Lankford, and Donna~M. Nystrom.
\newblock Visually mining and monitoring massive time series.
\newblock In \emph{Proceedings of the Tenth ACM SIGKDD International Conference on Knowledge Discovery and Data Mining}, KDD '04, pages 460--469, New York, NY, USA, 2004. ACM.

\bibitem[Lin et~al.(2005)Lin, Keogh, and Lonardi]{Lin2005}
Jessica Lin, Eamonn Keogh, and Stefano Lonardi.
\newblock Visualizing and discovering non-trivial patterns in large time series databases.
\newblock \emph{Information Visualization}, 4\penalty0 (2):\penalty0 61--82, July 2005.
\newblock ISSN 1473-8716.

\bibitem[Liu et~al.(2018)Liu, Xu, and Ren]{liu2018tpflow}
Dongyu Liu, Panpan Xu, and Liu Ren.
\newblock {TPFlow}: Progressive partition and multidimensional pattern extraction for large-scale spatio-temporal data analysis.
\newblock \emph{IEEE Transactions on Visualization and Computer Graphics}, 25\penalty0 (1):\penalty0 1--11, 2018.

\bibitem[Liu et~al.(2014)Liu, Cui, Wu, and Liu]{liu2014survey}
Shixia Liu, Weiwei Cui, Yingcai Wu, and Mengchen Liu.
\newblock A survey on information visualization: {Recent} advances and challenges.
\newblock \emph{The Visual Computer}, 30:\penalty0 1373--1393, 2014.

\bibitem[Liu et~al.(2016)Liu, Maljovec, Wang, Bremer, and Pascucci]{liu2016visualizing}
Shusen Liu, Dan Maljovec, Bei Wang, Peer-Timo Bremer, and Valerio Pascucci.
\newblock Visualizing high-dimensional data: {Advances} in the past decade.
\newblock \emph{IEEE Transactions on Visualization and Computer Graphics}, 23\penalty0 (3):\penalty0 1249--1268, 2016.

\bibitem[Liu and Heer(2018)]{liu2018somewhere}
Yang Liu and Jeffrey Heer.
\newblock Somewhere over the rainbow: {An} empirical assessment of quantitative colormaps.
\newblock In \emph{Proceedings of the 2018 CHI Conference on Human Factors in Computing Systems}, pages 1--12, 2018.

\bibitem[Liu et~al.(2013)Liu, Jiang, and Heer]{liu2013immens}
Zhicheng Liu, Biye Jiang, and Jeffrey Heer.
\newblock {imMens}: Real-time visual querying of big data.
\newblock In \emph{Computer graphics forum}, volume~32, pages 421--430. Wiley Online Library, 2013.

\bibitem[Lohse(1991)]{Lohse1991}
Jerry Lohse.
\newblock A cognitive model for the perception and understanding of graphs.
\newblock In \emph{Proceedings of the SIGCHI Conference on Human Factors in Computing Systems}, CHI '91, pages 137--144, New York, NY, USA, 1991. ACM.

\bibitem[Longo et~al.(2024)Longo, Brcic, Cabitza, Choi, Confalonieri, Del~Ser, Guidotti, Hayashi, Herrera, Holzinger, et~al.]{longo2024explainable}
Luca Longo, Mario Brcic, Federico Cabitza, Jaesik Choi, Roberto Confalonieri, Javier Del~Ser, Riccardo Guidotti, Yoichi Hayashi, Francisco Herrera, Andreas Holzinger, et~al.
\newblock Explainable artificial intelligence ({XAI}) 2.0: {A} manifesto of open challenges and interdisciplinary research directions.
\newblock \emph{Information Fusion}, 106:\penalty0 102301, 2024.

\bibitem[Lundberg et~al.(2018)Lundberg, Nair, Vavilala, Horibe, Eisses, Adams, Liston, Low, Newman, Kim, et~al.]{shap_lundberg2018explainable}
Scott~M Lundberg, Bala Nair, Monica~S Vavilala, Mayumi Horibe, Michael~J Eisses, Trevor Adams, David~E Liston, Daniel King-Wai Low, Shu-Fang Newman, Jerry Kim, et~al.
\newblock Explainable machine-learning predictions for the prevention of hypoxaemia during surgery.
\newblock \emph{Nature Biomedical Engineering}, 2\penalty0 (10):\penalty0 749--760, 2018.

\bibitem[Lupton and Allwood(2017)]{lupton2017hybrid}
Richard~C Lupton and Julian~M Allwood.
\newblock Hybrid sankey diagrams: {Visual} analysis of multidimensional data for understanding resource use.
\newblock \emph{Resources, Conservation and Recycling}, 124:\penalty0 141--151, 2017.

\bibitem[Maaten and Hinton(2008)]{Maaten2008}
Laurens Van~Der Maaten and Geoffrey Hinton.
\newblock Visualizing data using t-{SNE}.
\newblock \emph{Journal of Machine Learning Research}, 9\penalty0 (1):\penalty0 2579--2605, 2008.

\bibitem[McLachlan et~al.(2008)McLachlan, Munzner, Koutsofios, and North]{McLachlan2008}
Peter McLachlan, Tamara Munzner, Eleftherios Koutsofios, and Stephen North.
\newblock Live{RAC}: {Interactive} visual exploration of system management time-series data.
\newblock In \emph{Proceedings of the SIGCHI Conference on Human Factors in Computing Systems}, CHI '08, pages 1483--1492, New York, NY, USA, 2008. ACM.

\bibitem[Miller(1956)]{miller1956magical}
George~A Miller.
\newblock The magical number seven, plus or minus two: {Some} limits on our capacity for processing information.
\newblock \emph{Psychological Review}, 63\penalty0 (2):\penalty0 81, 1956.

\bibitem[Moher et~al.(2009)Moher, Liberati, Tetzlaff, and Altman]{Moher:2009}
David Moher, Alessandro Liberati, Jennifer Tetzlaff, and Douglas~G Altman.
\newblock Preferred reporting items for systematic reviews and meta-analyses: {The PRISMA} statement.
\newblock \emph{Annals of internal medicine}, 151\penalty0 (4):\penalty0 264--269, 2009.

\bibitem[Munzner(2014)]{Munzner:2014}
Tamara Munzner.
\newblock \emph{Visualization Analysis and Design}.
\newblock CRC Press, 2014.

\bibitem[Murray(2013)]{murray2013}
S.~Murray.
\newblock \emph{Interactive Data Visualization for the Web}.
\newblock O'Reilly Media, Inc., 1005 Gravenstein Highway North, Sebastopol, CA, USA, 1st edition, 2013.

\bibitem[Narayanswamy et~al.(2024)Narayanswamy, Liu, Ayush, Yang, Xu, Liao, Garrison, Tailor, Sunshine, Liu, et~al.]{narayanswamy2024scaling}
Girish Narayanswamy, Xin Liu, Kumar Ayush, Yuzhe Yang, Xuhai Xu, Shun Liao, Jake Garrison, Shyam Tailor, Jake Sunshine, Yun Liu, et~al.
\newblock Scaling wearable foundation models.
\newblock \emph{Preprint arXiv:2410.13638}, 2024.

\bibitem[Nascimento et~al.(2020)Nascimento, Pimentel, Souza, Leite, Edwards, Santos, and Louzada]{nascimento2020dynamic}
Diego~C Nascimento, Bruno Pimentel, Renata Souza, Joao~P Leite, Dylan~J Edwards, Taiza~EG Santos, and Francisco Louzada.
\newblock Dynamic time series smoothing for symbolic interval data applied to neuroscience.
\newblock \emph{Information Sciences}, 517:\penalty0 415--426, 2020.

\bibitem[Nguyen et~al.(2020)Nguyen, Qiao, Heer, and Hullman]{nguyen2020exploring}
Francis Nguyen, Xiaoli Qiao, Jeffrey Heer, and Jessica Hullman.
\newblock Exploring the effects of aggregation choices on untrained visualization users' generalizations from data.
\newblock \emph{Computer Graphics Forum}, 39\penalty0 (6):\penalty0 33--48, 2020.

\bibitem[Nonato and Aupetit(2018)]{Nonato2018}
L.~G. Nonato and M.~Aupetit.
\newblock Multidimensional projection for visual analytics: {Linking} techniques with distortions, tasks, and layout enrichment.
\newblock \emph{IEEE Transactions on Visualization and Computer Graphics}, 25\penalty0 (8):\penalty0 2650--2673, June 2018.
\newblock URL \url{http://10.1109/TVCG.2018.2846735}.

\bibitem[North and Shneiderman(1997)]{North1997}
Chris North and Ben Shneiderman.
\newblock A taxonomy of multiple window coordinations.
\newblock \emph{Technical Report CS-TR-3854}, pages 1--, 1997.

\bibitem[NYT(2008)]{nyt_stream}
NYT.
\newblock {The New York Times} -- {Ebb} and flow at the box office.
\newblock \url{https://archive.nytimes.com/www.nytimes.com/imagepages/2008/02/24/business/24METRIC_GRAPHIC.html}, 2008.
\newblock Online, visited on June 2024.

\bibitem[Oliveira and Levkowitz(2003)]{Oliveira2003}
M.~C.~F. Oliveira and H.~Levkowitz.
\newblock From visual data exploration to visual data mining: {A} survey.
\newblock \emph{IEEE Transactions on Visualization and Computer Graphics}, 9\penalty0 (3):\penalty0 378--394, jul 2003.
\newblock ISSN 1077-2626.

\bibitem[Oreshkin et~al.(2019)Oreshkin, Carpov, Chapados, and Bengio]{oreshkin2019n}
Boris~N Oreshkin, Dmitri Carpov, Nicolas Chapados, and Yoshua Bengio.
\newblock {N-BEATS}: Neural basis expansion analysis for interpretable time series forecasting.
\newblock \emph{Preprint arXiv:1905.10437}, 2019.

\bibitem[Ortigossa et~al.(2022)Ortigossa, Dias, and Nascimento]{eu2022projections}
Evandro~S. Ortigossa, Fábio~Felix Dias, and Diego Carvalho~do Nascimento.
\newblock Getting over high-dimensionality: {How} multidimensional projection methods can assist data science.
\newblock \emph{Applied Sciences}, 12\penalty0 (13), 2022.

\bibitem[Ortigossa et~al.(2024)Ortigossa, Gon{\c{c}}alves, and Nonato]{ortigossa2024explainable}
Evandro~S Ortigossa, Thales Gon{\c{c}}alves, and Luis~Gustavo Nonato.
\newblock {EXplainable} artificial intelligence ({XAI})--{From} theory to methods and applications.
\newblock \emph{IEEE Access}, 12:\penalty0 80799--80846, 2024.

\bibitem[Ortigossa et~al.(2025)Ortigossa, Dias, Barr, Silva, and Nonato]{ortigossa2024texplainer}
Evandro~S Ortigossa, F{\'a}bio~F Dias, Brian Barr, Claudio~T Silva, and Luis~Gustavo Nonato.
\newblock {T-Explainer}: {A} model-agnostic explainability framework based on gradients.
\newblock \emph{IEEE Intelligent Systems}, pages 1--11, 2025.
\newblock \doi{10.1109/MIS.2025.3564330}.

\bibitem[Palmeiro et~al.(2022)Palmeiro, Malveiro, Costa, Polido, Moreira, and Bizarro]{palmeiro2022data}
Jo{\~a}o Palmeiro, Beatriz Malveiro, Rita Costa, David Polido, Ricardo Moreira, and Pedro Bizarro.
\newblock {Data+Shift: Supporting} visual investigation of data distribution shifts by data scientists.
\newblock \emph{Preprint arXiv:2204.14025}, 2022.

\bibitem[Perin et~al.(2013)Perin, Vernier, and Fekete]{Perin2013}
Charles Perin, Frederic Vernier, and Jean-Daniel Fekete.
\newblock Interactive horizon graphs: {Improving} the compact visualization of multiple time series.
\newblock In \emph{Proceedings of the SIGCHI Conference on Human Factors in Computing Systems}, CHI '13, pages 3217--3226, New York, NY, USA, 2013. ACM.

\bibitem[Pham et~al.(2020)Pham, Nguyen, and Dang]{pham2020contimap}
Vung Pham, Ngan Nguyen, and Tommy Dang.
\newblock {ContiMap: Continuous} heatmap for large time series data.
\newblock In \emph{2020 Visualization in Data Science (VDS)}, pages 42--51. IEEE, 2020.

\bibitem[Pike et~al.(2009)Pike, Stasko, Chang, and O'Connell]{Pike2009}
William~A. Pike, John Stasko, Remco Chang, and Theresa~A. O'Connell.
\newblock The science of interaction.
\newblock \emph{Information Visualization}, 8\penalty0 (4):\penalty0 263--274, December 2009.
\newblock ISSN 1473-8716.

\bibitem[Pratama et~al.(2016)Pratama, Permanasari, Ardiyanto, and Indrayani]{pratama2016review}
Irfan Pratama, Adhistya~Erna Permanasari, Igi Ardiyanto, and Rini Indrayani.
\newblock A review of missing values handling methods on time-series data.
\newblock In \emph{2016 International Conference on Information Technology Systems and Innovation (ICITSI)}, pages 1--6. IEEE, 2016.

\bibitem[Priyadarshini et~al.(2025)Priyadarshini, Bajaj, and Zaitsev]{priyadarshini2025energy}
MS~Priyadarshini, Mohit Bajaj, and Ievgen Zaitsev.
\newblock Energy feature extraction and visualization of voltage sags using wavelet packet analysis for enhanced power quality monitoring.
\newblock \emph{Scientific Reports}, 15\penalty0 (1):\penalty0 2226, 2025.

\bibitem[Resnikoff(1989)]{Resnikoff1989}
Howard~L. Resnikoff.
\newblock \emph{The Illusion of Reality}.
\newblock Springer-Verlag New York, Inc., New York, NY, USA, 1989.

\bibitem[Roberts(2007)]{Roberts2007}
Jonathan~C. Roberts.
\newblock State of the art: {Coordinated} \& multiple views in exploratory visualization.
\newblock In \emph{Proceedings of the Fifth International Conference on Coordinated and Multiple Views in Exploratory Visualization}, CMV '07, pages 61--71, Washington, DC, USA, 2007. IEEE Computer Society.

\bibitem[Roberts et~al.(2018)Roberts, Laramee, Smith, Brookes, and D'Cruze]{roberts2018smart}
Richard~C Roberts, Robert~S Laramee, Gary~A Smith, Paul Brookes, and Tony D'Cruze.
\newblock Smart brushing for parallel coordinates.
\newblock \emph{IEEE Transactions on Visualization and Computer Graphics}, 25\penalty0 (3):\penalty0 1575--1590, 2018.

\bibitem[Rodriguez and Mendez(2015)]{Rodriguez2015}
L.~Rodriguez and D.~Mendez.
\newblock Pollution spots: {A} novel method for air pollution monitoring.
\newblock In \emph{Air Pollution XXIII}, volume 198 of \emph{WIT Transactions on Ecology and the Environment}, pages 129--140, Southampton, UK, 2015. WIT Press.

\bibitem[Rodriguez-Fernandez et~al.(2023)Rodriguez-Fernandez, Montalvo-Garcia, Piccialli, Nalepa, and Camacho]{rodriguez2023deepvats}
Victor Rodriguez-Fernandez, David Montalvo-Garcia, Francesco Piccialli, Grzegorz~J Nalepa, and David Camacho.
\newblock {DeepVATS}: Deep visual analytics for time series.
\newblock \emph{Knowledge-Based Systems}, 277:\penalty0 110793, 2023.

\bibitem[Sadana and Stasko(2016)]{Sadana2016}
Ramik Sadana and John Stasko.
\newblock Multiple coordinated visualizations for tablets.
\newblock \emph{Computer Graphics Forum}, 35\penalty0 (3):\penalty0 261--270, June 2016.

\bibitem[Saito et~al.(2005)Saito, Miyamura, Yamamoto, Saito, Hoshiya, and Kaseda]{saito2005two}
Takafumi Saito, Hiroko~Nakamura Miyamura, Mitsuyoshi Yamamoto, Hiroki Saito, Yuka Hoshiya, and Takumi Kaseda.
\newblock Two-tone pseudo coloring: {Compact} visualization for one-dimensional data.
\newblock In \emph{IEEE Symposium on Information Visualization, 2005. INFOVIS 2005.}, pages 173--180. IEEE, 2005.

\bibitem[Salinas et~al.(2020)Salinas, Flunkert, Gasthaus, and Januschowski]{salinas2020deepar}
David Salinas, Valentin Flunkert, Jan Gasthaus, and Tim Januschowski.
\newblock {DeepAR}: Probabilistic forecasting with autoregressive recurrent networks.
\newblock \emph{International journal of forecasting}, 36\penalty0 (3):\penalty0 1181--1191, 2020.

\bibitem[Satriadi et~al.(2020)Satriadi, Ens, Cordeil, Czauderna, and Jenny]{satriadi2020maps}
Kadek~Ananta Satriadi, Barrett Ens, Maxime Cordeil, Tobias Czauderna, and Bernhard Jenny.
\newblock Maps around me: {3D} multiview layouts in immersive spaces.
\newblock \emph{Proceedings of the ACM on Human-Computer Interaction}, 4\penalty0 (ISS):\penalty0 1--20, 2020.

\bibitem[Schmidt(2008)]{schmidt2008sankey}
Mario Schmidt.
\newblock The sankey diagram in energy and material flow management--{Part} ii: {Methodology} and current applications.
\newblock \emph{Journal of Industrial Ecology}, 12\penalty0 (2):\penalty0 173--185, 2008.

\bibitem[Shakeel et~al.(2022)Shakeel, Iram, Al-Aqrabi, Alsboui, and Hill]{shakeel2022comprehensive}
Hafiz~Muhammad Shakeel, Shamaila Iram, Hussain Al-Aqrabi, Tariq Alsboui, and Richard Hill.
\newblock A comprehensive state-of-the-art survey on data visualization tools: {Research} developments, challenges and future domain specific visualization framework.
\newblock \emph{IEEE Access}, 10:\penalty0 96581--96601, 2022.

\bibitem[Shi et~al.(2012)Shi, Cui, Liu, Xu, Chen, and Qu]{Shi2012}
Conglei Shi, Weiwei Cui, Shixia Liu, Panpan Xu, Wei Chen, and Huamin Qu.
\newblock Rank{E}xplorer: {Visualization} of ranking changes in large time series data.
\newblock \emph{IEEE Transactions on Visualization and Computer Graphics}, 18\penalty0 (12):\penalty0 2669--2678, October 2012.

\bibitem[Shi et~al.(2014)Shi, Wu, Liu, Zhou, and Qu]{Shi2014}
Conglei Shi, Yingcai Wu, Shixia Liu, Hong Zhou, and Huamin Qu.
\newblock Loyal{T}racker: {Visualizing} loyalty dynamics in search engines.
\newblock \emph{IEEE Transactions on Visualization and Computer Graphics}, 20\penalty0 (12):\penalty0 1733--1742, 2014.

\bibitem[Shneiderman(1996)]{Shneiderman1996}
Ben Shneiderman.
\newblock The eyes have it: {A} task by data type taxonomy for information visualizations.
\newblock In \emph{Proceedings of the 1996 IEEE Symposium on Visual Languages}, VL '96, pages 336--, Washington, DC, USA, 1996. IEEE Computer Society.

\bibitem[Shneiderman(2008)]{Shneiderman2008}
Ben Shneiderman.
\newblock Extreme visualization: {Squeezing} a billion records into a million pixels.
\newblock In \emph{Proceedings of the 2008 ACM SIGMOD International Conference on Management of Data}, SIGMOD '08, pages 3--12, New York, NY, USA, 2008. ACM.

\bibitem[Shumway and Stoffer(2011)]{shumway2006}
Robert~H. Shumway and David~S. Stoffer.
\newblock \emph{Time series analysis and its applications: {With} {R} examples}.
\newblock Springer Publishing Company, Incorporated, New York, NY, USA, 3rd edition, 2011.

\bibitem[Silva and Catarci(2000)]{Silva2000}
Sônia~Fernandes Silva and Tiziana Catarci.
\newblock Visualization of linear time-oriented data: {A} survey.
\newblock In \emph{Proceedings of the First International Conference on Web Information Systems Engineering}, volume~1 of \emph{WISE '00}, pages 310--, Washington, DC, USA, 2000. IEEE Computer Society.

\bibitem[Song and Szafir(2018)]{song2018s}
Hayeong Song and Danielle~Albers Szafir.
\newblock Where's my data? evaluating visualizations with missing data.
\newblock \emph{IEEE transactions on visualization and computer graphics}, 25\penalty0 (1):\penalty0 914--924, 2018.

\bibitem[Stasko and Zhang(2000)]{Stasko2000}
John Stasko and Eugene Zhang.
\newblock Focus+context display and navigation techniques for enhancing radial, space-filling hierarchy visualizations.
\newblock In \emph{Proceedings of the IEEE Symposium on Information Visualization 2000}, INFOVIS '00, pages 57--, Washington, DC, USA, 2000. IEEE Computer Society.

\bibitem[Stopar et~al.(2018)Stopar, Skraba, Grobelnik, and Mladenic]{stopar2018streamstory}
Luka Stopar, Primoz Skraba, Marko Grobelnik, and Dunja Mladenic.
\newblock {StreamStory: Exploring} multivariate time series on multiple scales.
\newblock \emph{IEEE Transactions on Visualization and Computer Graphics}, 25\penalty0 (4):\penalty0 1788--1802, 2018.

\bibitem[Sun et~al.(2014)Sun, Wu, Liu, Peng, Zhu, and Liang]{Sun2014}
G.~Sun, Y.~Wu, S.~Liu, T.~Q. Peng, J.~J.~H. Zhu, and R.~Liang.
\newblock Evo{R}iver: {Visual} analysis of topic coopetition on social media.
\newblock \emph{IEEE Transactions on Visualization and Computer Graphics}, 20\penalty0 (12):\penalty0 1753--1762, December 2014.
\newblock ISSN 1077-2626.

\bibitem[Sun et~al.(2024)Sun, Li, Jin, Dai, Peng, and Chen]{sun2024task}
Jiancheng Sun, Xiaohe Li, Yongnu Jin, Liyun Dai, Xiangdong Peng, and Chunlin Chen.
\newblock Task-oriented analysis and visualization of correlation patterns in multi-sensor time series.
\newblock \emph{Knowledge-Based Systems}, page 111525, 2024.

\bibitem[Thakur and Hanson(2010)]{Thakur2010}
Sidharth Thakur and Andrew~J. Hanson.
\newblock A {3D} visualization of multiple time series on maps.
\newblock In \emph{Proceedings of the 2010 14th International Conference Information Visualisation}, IV '10, pages 336--343, Washington, DC, USA, 2010. IEEE Computer Society.

\bibitem[Thomas and Cook(2005)]{Thomas2005}
James~J. Thomas and Kristin~A. Cook.
\newblock \emph{Illuminating the Path: {The} Research and Development Agenda for Visual Analytics}.
\newblock IEEE Computer Society Press, Los Alamitos, CA, USA, 2005.

\bibitem[Tominski et~al.(2017)Tominski, Gladisch, Kister, Dachselt, and Schumann]{Tominski2017}
C.~Tominski, S.~Gladisch, U.~Kister, R.~Dachselt, and H.~Schumann.
\newblock Interactive lenses for visualization: {An} extended survey.
\newblock \emph{Computer Graphics Forum}, 36\penalty0 (6):\penalty0 173--200, September 2017.
\newblock ISSN 0167-7055.

\bibitem[Tominski et~al.(2004)Tominski, Abello, and Schumann]{Tominski2004}
Christian Tominski, James Abello, and Heidrun Schumann.
\newblock Axes-based visualizations with radial layouts.
\newblock In \emph{Proceedings of the ACM Symposium on Applied Computing}, SAC '04, pages 1242--1247, New York, NY, USA, 2004. ACM.
\newblock ISBN 1-58113-812-1.

\bibitem[Toyoda and Kitsuregawa(2005)]{Toyoda2005}
Masashi Toyoda and Masaru Kitsuregawa.
\newblock A system for visualizing and analyzing the evolution of the web with a time series of graphs.
\newblock In \emph{Proceedings of the Sixteenth ACM Conference on Hypertext and Hypermedia}, HYPERTEXT '05, pages 151--160, New York, NY, USA, 2005. ACM.

\bibitem[Tufte(2007)]{tufte2007}
E.~R. Tufte.
\newblock \emph{The Visual Display of Quantitative Information}.
\newblock Graphics Press LLC, Cheshire, Connecticut, USA, 2nd edition, 2007.

\bibitem[Ulmer et~al.(2023)Ulmer, Angelini, Fekete, Kohlhammer, and May]{ulmer2023survey}
Alex Ulmer, Marco Angelini, Jean-Daniel Fekete, J{\"o}rn Kohlhammer, and Thorsten May.
\newblock A survey on progressive visualization.
\newblock \emph{IEEE Transactions on Visualization and Computer Graphics}, 30\penalty0 (9):\penalty0 6447--6467, 2023.

\bibitem[Veras and Collins(2019)]{veras2019discriminability}
Rafael Veras and Christopher Collins.
\newblock Discriminability tests for visualization effectiveness and scalability.
\newblock \emph{IEEE Transactions on Visualization and Computer Graphics}, 26\penalty0 (1):\penalty0 749--758, 2019.

\bibitem[Wagemans et~al.(2012{\natexlab{a}})Wagemans, Elder, Kubovy, Palmer, Peterson, Singh, and Heydt]{Wagemans2012}
J.~Wagemans, J.~H. Elder, M.~Kubovy, S.~E. Palmer, M.~A. Peterson, M.~Singh, and R.~Von~Der Heydt.
\newblock A century of {G}estalt psychology in visual perception: {I}. perceptual grouping and figure-ground organization.
\newblock \emph{Psychological Bulletin}, 138\penalty0 (6):\penalty0 1172--1217, 2012{\natexlab{a}}.

\bibitem[Wagemans et~al.(2012{\natexlab{b}})Wagemans, Feldman, Gepshtein, Kimchi, Pomerantz, Helm, and Leeuwen]{Wagemans2012b}
J.~Wagemans, J.~Feldman, S.~Gepshtein, R.~Kimchi, J.~R. Pomerantz, P.~A. Van~Der Helm, and C.~Van Leeuwen.
\newblock A century of {G}estalt psychology in visual perception: {II}. conceptual and theoretical foundations.
\newblock \emph{Psychological Bulletin}, 138\penalty0 (6):\penalty0 1218--1252, 2012{\natexlab{b}}.

\bibitem[Walker et~al.(2016)Walker, Borgo, and Jones]{Walker2016}
James Walker, Rita Borgo, and Mark~W. Jones.
\newblock Time{N}otes: {A} study on effective chart visualization and interaction techniques for time-series data.
\newblock \emph{IEEE Transactions on Visualization and Computer Graphics}, 22\penalty0 (1):\penalty0 549--558, Janaury 2016.
\newblock ISSN 1077-2626.

\bibitem[Wang et~al.(2017{\natexlab{a}})Wang, Guo, and Zhang]{wang2017spatial}
Deqiang Wang, Danhuai Guo, and Hui Zhang.
\newblock Spatial temporal data visualization in emergency management: {A} view from data-driven decision.
\newblock In \emph{Proceedings of the 3rd ACM SIGSPATIAL International Workshop on the Use of GIS in Emergency Management}, pages 1--7, 2017{\natexlab{a}}.

\bibitem[Wang et~al.(2024{\natexlab{a}})Wang, Du, Yang, Qian, Cao, Zhang, Wang, Liang, and Wen]{wang2024deep}
Jun Wang, Wenjie Du, Yiyuan Yang, Linglong Qian, Wei Cao, Keli Zhang, Wenjia Wang, Yuxuan Liang, and Qingsong Wen.
\newblock Deep learning for multivariate time series imputation: A survey.
\newblock \emph{arXiv preprint arXiv:2402.04059}, 2024{\natexlab{a}}.

\bibitem[Wang et~al.(2016)Wang, Wu, Chen, Luo, and Qu]{Wang2016}
Yun Wang, Tongshuang Wu, Zhutian Chen, Qiong Luo, and Huamin Qu.
\newblock {STAC}: {Enhancing} stacked graphs for time series analysis.
\newblock In Chuck Hansen, Ivan Viola, and Xiaoru Yuan, editors, \emph{2016 IEEE Pacific Visualization Symposium (PacificVis)}, pages 234--238, Taipei, Taiwan, 2016. IEEE Computer Society.

\bibitem[Wang et~al.(2017{\natexlab{b}})Wang, Han, Zhu, Deussen, and Chen]{wang2017line}
Yunhai Wang, Fubo Han, Lifeng Zhu, Oliver Deussen, and Baoquan Chen.
\newblock Line graph or scatter plot? {Automatic} selection of methods for visualizing trends in time series.
\newblock \emph{IEEE Transactions on Visualization and Computer Graphics}, 24\penalty0 (2):\penalty0 1141--1154, 2017{\natexlab{b}}.

\bibitem[Wang et~al.(2024{\natexlab{b}})Wang, Xia, He, Chen, Liu, Zhu, Liang, Wu, Liu, Malladi, Chevalier, Arora, and Chen]{NEURIPS2024_cdf6f8e9}
Zirui Wang, Mengzhou Xia, Luxi He, Howard Chen, Yitao Liu, Richard Zhu, Kaiqu Liang, Xindi Wu, Haotian Liu, Sadhika Malladi, Alexis Chevalier, Sanjeev Arora, and Danqi Chen.
\newblock {CharXiv}: Charting gaps in realistic chart understanding in multimodal {LLMs}.
\newblock In \emph{Advances in Neural Information Processing Systems}, volume~37, pages 113569--113697. Curran Associates, Inc., 2024{\natexlab{b}}.

\bibitem[Ward et~al.(2010)Ward, Grinstein, and Keim]{Ward2010}
Matthew Ward, Georges Grinstein, and Daniel Keim.
\newblock \emph{Interactive Data Visualization: {Foundations}, Techniques, and Applications}.
\newblock A. K. Peters, Ltd., Natick, MA, USA, 1st edition, 2010.

\bibitem[Ward et~al.(2015)Ward, Grinstein, and Keim]{Ward:2015}
Matthew~O. Ward, Georges Grinstein, and Daniel Keim.
\newblock \emph{Interactive data visualization: {Foundations}, techniques, and applications}.
\newblock CRC Press, 2 edition, 2015.
\newblock ISBN 978-1482257373.

\bibitem[Ware(2019)]{Ware2004}
Colin Ware.
\newblock \emph{Information Visualization: {Perception} for Design}.
\newblock Morgan Kaufmann Publishers Inc., San Francisco, CA, USA, 4th edition, 2019.

\bibitem[Ware et~al.(2023)Ware, Stone, and Szafir]{ware2023rainbow}
Colin Ware, Maureen Stone, and Danielle~Albers Szafir.
\newblock Rainbow colormaps are not all bad.
\newblock \emph{IEEE Computer Graphics and Applications}, 43\penalty0 (3):\penalty0 88--93, 2023.

\bibitem[Wattenberg(2005{\natexlab{a}})]{Wattenberg2005}
Martin Wattenberg.
\newblock A note on space-filling visualizations and space-filling curves.
\newblock In \emph{Proceedings of the 2005 IEEE Symposium on Information Visualization}, INFOVIS '05, pages 24--, Washington, DC, USA, 2005{\natexlab{a}}. IEEE Computer Society.

\bibitem[Wattenberg(2005{\natexlab{b}})]{Wattenberg2005b}
Martin Wattenberg.
\newblock Baby names, visualization, and social data analysis.
\newblock In \emph{Proceedings of the 2005 IEEE Symposium on Information Visualization}, INFOVIS '05, pages 1--, Washington, DC, USA, 2005{\natexlab{b}}. IEEE Computer Society.

\bibitem[Weber et~al.(2001)Weber, Alexa, and Müller]{Weber01}
Marc Weber, Marc Alexa, and Wolfgang Müller.
\newblock Visualizing time-series on spirals.
\newblock In \emph{Proceedings of the IEEE Symposium on Information Visualization}, INFOVIS '01, pages 7--, Washington, DC, USA, 2001. IEEE Computer Society.

\bibitem[Wickham(2016)]{ggplot2}
Hadley Wickham.
\newblock \emph{ggplot2: {Elegant} Graphics for Data Analysis}.
\newblock Springer-Verlag New York, 2016.
\newblock URL \url{https://ggplot2.tidyverse.org}.

\bibitem[Wijk and Selow(1999)]{VanWijk1999}
Jarke J.~Van Wijk and Edward R.~Van Selow.
\newblock Cluster and calendar based visualization of time series data.
\newblock In \emph{Proceedings of the 1999 IEEE Symposium on Information Visualization}, INFOVIS '99, pages 4--, Washington, DC, USA, 1999. IEEE Computer Society.

\bibitem[Wong(1999)]{Wong1999}
P.~C. Wong.
\newblock Guest editor's introduction: {Visual} data mining.
\newblock \emph{IEEE Computer Graphics and Applications}, 19\penalty0 (5):\penalty0 20--21, 09 1999.
\newblock ISSN 0272-1716.

\bibitem[Wu et~al.(2021)Wu, Chang, Hellerstein, Satyanarayan, and Wu]{wu2021diel}
Yifan Wu, Remco Chang, Joseph~M Hellerstein, Arvind Satyanarayan, and Eugene Wu.
\newblock {DIEL}: Interactive visualization beyond the here and now.
\newblock \emph{IEEE Transactions on Visualization and Computer Graphics}, 28\penalty0 (1):\penalty0 737--746, 2021.

\bibitem[Xu et~al.(2021)Xu, Yuan, Wang, Silva, and Bertini]{xu2021mtseer}
Ke~Xu, Jun Yuan, Yifang Wang, Claudio Silva, and Enrico Bertini.
\newblock {mTSeer}: {Interactive} visual exploration of models on multivariate time-series forecast.
\newblock In \emph{Proceedings of the 2021 CHI Conference on Human Factors in Computing Systems}, pages 1--15, 2021.

\bibitem[Xu et~al.(2016)Xu, Mei, Ren, and Chen]{xu2016vidx}
Panpan Xu, Honghui Mei, Liu Ren, and Wei Chen.
\newblock {ViDX: Visual} diagnostics of assembly line performance in smart factories.
\newblock \emph{IEEE Transactions on Visualization and Computer Graphics}, 23\penalty0 (1):\penalty0 291--300, 2016.

\bibitem[Yang et~al.(2024)Yang, Zhou, Wang, Cong, Han, Yan, Liu, Tan, Liu, Yu, et~al.]{yang2024matplotagent}
Zhiyu Yang, Zihan Zhou, Shuo Wang, Xin Cong, Xu~Han, Yukun Yan, Zhenghao Liu, Zhixing Tan, Pengyuan Liu, Dong Yu, et~al.
\newblock {MatPlotAgent}: Method and evaluation for {LLM}-based agentic scientific data visualization.
\newblock \emph{Preprint arXiv:2402.11453}, 2024.

\bibitem[Yi et~al.(2007)Yi, Kang, Stasko, and Jacko]{Yi2007}
Ji~Soo Yi, Youn~Ah Kang, John Stasko, and Julie Jacko.
\newblock Toward a deeper understanding of the role of interaction in information visualization.
\newblock \emph{IEEE Transactions on Visualization and Computer Graphics}, 13\penalty0 (6):\penalty0 1224--1231, November 2007.
\newblock ISSN 1077-2626.

\bibitem[Yi et~al.(2023)Yi, Zhang, Cao, Wang, Long, Hu, He, Niu, Fan, and Xiong]{yi2023survey}
Kun Yi, Qi~Zhang, Longbing Cao, Shoujin Wang, Guodong Long, Liang Hu, Hui He, Zhendong Niu, Wei Fan, and Hui Xiong.
\newblock A survey on deep learning based time series analysis with frequency transformation.
\newblock \emph{Preprint arXiv:2302.02173}, 2023.

\bibitem[Yoon et~al.(2016)Yoon, Xu, and Tourassi]{Yoon2016}
Hong-Jun Yoon, Songhua Xu, and Georgia~D. Tourassi.
\newblock Predicting lung cancer incidence from air pollution exposures using shapelet-based time series analysis.
\newblock In \emph{{IEEE-EMBS} International Conference on Biomedical and Health Informatics ({BHI})}, pages 565--568, Piscataway, NJ, USA, 2016. IEEE.

\bibitem[Yuan et~al.(2023)Yuan, Li, Li, Wong, Zhang, and Qu]{yuan2023tax}
Linping Yuan, Boyu Li, Siqi Li, Kam~Kwai Wong, Rong Zhang, and Huamin Qu.
\newblock Tax-scheduler: {An} interactive visualization system for staff shifting and scheduling at tax authorities.
\newblock \emph{Visual Informatics}, 7\penalty0 (2):\penalty0 30--40, 2023.

\bibitem[Zeng et~al.(2014)Zeng, Fu, Arisona, Erath, and Qu]{zeng2014visualizing}
Wei Zeng, Chi-Wing Fu, Stefan~M{\"u}ller Arisona, Alexander Erath, and Huamin Qu.
\newblock Visualizing mobility of public transportation system.
\newblock \emph{IEEE Transactions on Visualization and Computer Graphics}, 20\penalty0 (12):\penalty0 1833--1842, 2014.

\bibitem[Zhang et~al.(2021)Zhang, Bales, and Fleyeh]{zhang2021time}
Fan Zhang, Chris Bales, and Hasan Fleyeh.
\newblock From time series to image analysis: {A} transfer learning approach for night setback identification of district heating substations.
\newblock \emph{Journal of Building Engineering}, 43:\penalty0 102537, 2021.

\bibitem[Zhao et~al.(2011{\natexlab{a}})Zhao, Chevalier, and Balakrishnan]{Zhao2011}
Jian Zhao, Fanny Chevalier, and Ravin Balakrishnan.
\newblock Krono{M}iner: {Using} multi-foci navigation for the visual exploration of time-series data.
\newblock In \emph{Proceedings of the SIGCHI Conference on Human Factors in Computing Systems}, CHI '11, pages 1737--1746, New York, NY, USA, 2011{\natexlab{a}}. ACM.

\bibitem[Zhao et~al.(2011{\natexlab{b}})Zhao, Chevalier, Pietriga, and Balakrishnan]{Zhao2011b}
Jian Zhao, Fanny Chevalier, Emmanuel Pietriga, and Ravin Balakrishnan.
\newblock Exploratory analysis of time-series with {C}hrono{L}enses.
\newblock \emph{IEEE Transactions on Visualization and Computer Graphics}, 17\penalty0 (12):\penalty0 2422--2431, December 2011{\natexlab{b}}.
\newblock ISSN 1077-2626.

\bibitem[Zhao et~al.(2008)Zhao, Forer, and Harvey]{Zhao2008}
Jinfeng Zhao, Pip Forer, and Andrew~S. Harvey.
\newblock Activities, ringmaps and geovisualization of large human movement fields.
\newblock \emph{Information Visualization}, 7\penalty0 (3):\penalty0 198--209, June 2008.
\newblock ISSN 1473-8716.

\bibitem[Ziegler et~al.(2010)Ziegler, Jenny, Gruse, Keim, et~al.]{ziegler2010}
Hartmut Ziegler, Marco Jenny, Tino Gruse, Daniel Keim, et~al.
\newblock Visual market sector analysis for financial time series data.
\newblock In \emph{Visual Analytics Science and Technology (VAST)}, pages 83--90, Salt Lake City, UT, USA, Oct 2010. IEEE.

\end{thebibliography}

\end{document}